\documentclass[twocolumn,preprintnumbers,prd,superscriptaddress,nofootinbib,floatfix,showpacs,showkeys,aps,10pt]{revtex4-2}

\usepackage{amsmath}
\usepackage{amsfonts}
\usepackage{amssymb}
\usepackage{graphicx}
\usepackage{color}
\usepackage{xcolor}
\usepackage{hyperref}
\usepackage{booktabs}
\usepackage{hyperref}
\usepackage{cleveref}
\usepackage{braket}
\usepackage{mathtools}

\definecolor{vividviolet}{rgb}{0.62, 0.0, 1.0}
\definecolor{amaranth}{rgb}{0.9, 0.17, 0.31}
\definecolor{palatinateblue}{rgb}{0.15, 0.23, 0.89}
\definecolor{brightpink}{rgb}{1.0, 0.0, 0.5}
\definecolor{cornflowerblue}{rgb}{0.39, 0.58, 0.93}
\definecolor{deepcarminepink}{rgb}{0.94, 0.19, 0.22}
\definecolor{radicalred}{rgb}{1.0, 0.21, 0.37}
\hypersetup{ linktoc=all,
    colorlinks, linkcolor={palatinateblue},
    citecolor={brightpink}, urlcolor={amaranth}
}

\newcommand{\be}{\begin{equation}}
\newcommand{\ee}{\end{equation}}
\newcommand{\bs}{\begin{split}}
\newcommand{\bea}{\begin{eqnarray}}
\newcommand{\eea}{\end{eqnarray}}

\newcommand{\bes}{\begin{subequations}}
\newcommand{\ees}{\end{subequations}}

\newcommand{\x}{\mathsf{x}}

\usepackage{tikz,xcolor}
\definecolor{lime}{HTML}{A6CE39}
\DeclareRobustCommand{\orcidicon}{%
	\begin{tikzpicture}
	\draw[lime, fill=lime] (0,0) 
	circle [radius=0.16] 
	node[white] {{\fontfamily{qag}\selectfont \tiny ID}};
	\draw[white, fill=white] (-0.0625,0.095) 
	circle [radius=0.007];
	\end{tikzpicture}
	\hspace{-2mm}
}
\foreach \x in {A, ..., Z}{%
	\expandafter\xdef\csname orcid\x\endcsname{\noexpand\href{https://orcid.org/\csname orcidauthor\x\endcsname}{\noexpand\orcidicon}}
}

\begin{document}

\title{Making two particle detectors in flat spacetime communicate quantumly}

\author{Alessio Lapponi\orcidA{}}
\email{alessio.lapponi-ssm@unina.it}
\affiliation{Scuola Superiore Meridionale, Largo San Marcellino 10, 80138 Napoli, Italy.}
\affiliation{
Istituto Nazionale di Fisica Nucleare (INFN), Sezione di Napoli,
Complesso Universitario di Monte S. Angelo, Via Cintia Edificio 6, 80126 Napoli, Italy.}

\author{Jorma Louko\orcidC{}}
\email{jorma.louko@nottingham.ac.uk}
\affiliation{School of Mathematical Sciences, University of Nottingham, University Park, Nottingham, NG7 2RD, UK}

\author{Stefano Mancini\orcidD{}}
\email{stefano.mancini@unicam.it}
\affiliation{School of Science and Technology, University of Camerino, Via Madonna delle Carceri 9, 62032 Camerino, Italy.}
\affiliation{Istituto Nazionale di Fisica Nucleare (INFN), Sezione di Perugia, Via A.~Pascoli, 06123 Perugia, Italy.}

\date{April 2024; updated July 2024.\\ aaPublished in Phys.\ Rev.\ D \textbf{110}, 025018 (2024), doi.org/10.1103/PhysRevD.110.025018.\\ aaFor Open Access purposes, this Author Accepted Manuscript is made available under CC BY public copyright.}

\begin{abstract}
    A communication protocol with non-zero quantum capacity is found when the two communicating parts are particle detector models in
    $(3+1)$-dimensional spacetime. In particular, as detectors, we consider two harmonic oscillators interacting with a scalar field, whose evolution is generalized for whatever background spacetime and whatever spacetime smearing of the detectors. We then specialize to Minkowski spacetime and an initial Minkowski vacuum, considering a rapid interaction between the field and the two detectors, studying the case where the receiver is static and the sender is moving. The possibility to have a quantum capacity greater than zero stems from a relative acceleration between the detectors. Indeed, no reliable quantum communication is possible when the two detectors are static or moving inertially with respect to each other, but a reliable quantum communication can be achieved between a uniformly accelerated sender and an inertial receiver. 
\end{abstract}
\maketitle

\section{Introduction}
An intriguing intersection between two pillars of modern physics, i.e. quantum mechanics and Einstein's theory of relativity, is provided by the theory of \textit{relativistic quantum information} (RQI) \cite{MannRQI}. Each quantum communication protocol relies on a composite quantum system wherein its components exchange information via a quantum channel \cite{ManciniBook}. However, when a quantum system is subjected to relativistic effects, such as high velocities or strong gravitational fields, significant modifications of it are expected \cite{Wald_1999}. For this reason, the study of RQI becomes indispensable as we contemplate the extension of quantum communication and computation protocols to relativistic regimes, especially in the context of space-based quantum technologies \cite{Belenchia_2022,KRELINA2023101563} and relativistic quantum cryptography \cite{Radchenko_2014}.

Notable effects of the spacetime curvature could be seen in the context of quantum field theory. In fact, the framework known as \textit{quantum field theory in curved spacetimes} \cite{Birrell:1982ix,Hollands_2015} predicts a non-unique definition of the particle number operator, meaning that the amount of particles measured by observers in different frames could be different. This effect occurs, in particular, when observers undergo a non-inertial motion \cite{PCWDavies_1975,Unruh1976}, or lie in a spacetime with a horizon or a time-dependent gravitational field \cite{Hawking:1975vcx,Ford1987}. Because of this mismatch of measured particles, the communication capabilities of quantum channels were recently proven to decrease in these contexts \cite{Mancini_2014,Good_2021}.

The concept of particles produced has no meaning without a second quantum system measuring the presence of those particles. To this aim, \textit{Unruh-DeWitt detectors} (or, \textit{particle detector models}) play a pivotal role on understanding the physics of particle production in gravitational contexts \cite{Unruh1976,Unruh1984,Hu_2012}. In general, they consist on a localized quantum system interacting with an observable of the field. The detection of a particle by an Unruh-DeWitt detector is related to the transition between its ground state to a whatever excited state. For example, if uniformly linearly accelerated, the probability of transition has a thermal probability distribution, proving that the particles produced by the Unruh effect can be effectively detected \cite{Unruh1984,1991RSPSA.435..205R}. Applications of particle detectors go beyond the thermal acceleration, giving insights on the nature of quantum fields in several spacetime contexts, such as cosmological expansions \cite{Gibbons1977,Mart_n_Mart_nez_2014} and black holes \cite{Hodkinson2014,NG2014,Bueley_2022}. 

In this context, an outstanding result is that the vacuum of a quantum field presents an entanglement between space-like separated points, which can be harvested by moving particle detectors \cite{Reznik_2003,Reznik_2005,Salton_2015}. Because of this, one can exploit the classical and quantum correlations of the field state to communicate messages. Hence, recent studies have developed communication protocols between two distant particle detectors interacting with a mediator field \cite{Cliche2010,Jonsson_2018,Lapponi_2023}. These schemes can exploit qubit systems \cite{Cliche2010} (with two-level detectors) or bosonic systems (with harmonic oscillator detectors) \cite{Brown_2013,Lapponi_2023}.

The prominent problem, when dealing with the interaction between the field and the particle detectors, is the lack of exact solutions for the evolution of the system beyond perturbative regime. In fact, when studying the probability of transition of the detectors - since relativistic effects on quantum systems are expected to be perturbations of them - in a communication context a perturbative regime implies a negligible amount of signal communicated \cite{Simidjia2020}. In case of communication of qubits, with two-level particle detectors, this limit can be overcome by using the algebraic approach for quantum field theory \cite{Landulfo_2016,Tjoa_2022}. In case we communicate bosons - with harmonic oscillator detectors - it was shown that the evolution of the covariance matrix always allows a non-perturbative approach for the evolution of the system \cite{Brown_2013}. In particular, Ref.~\cite{Lapponi_2023} considers the Heisenberg evolution of the detectors' moment operator, following a quantum Langevin equation. With this method, the quantum channel properties can be found exactly also in the strong coupling regime. 

Motivated by this fact, in this paper we study the Heisenberg evolution for two harmonic oscillator detectors interacting with a scalar field in a general $(3+1)$D spacetime. In particular, the protocol in Ref.~\cite{Lapponi_2023} is generalized for whatever detectors' smearings and trajectories. The aim is to find a particular protocol allowing a reliable communication of quantum messages i.e. a quantum capacity of the channel greater than zero. Indeed, for a channel involving communicating particle detectors in a $(3+1)D$ spacetime, while a classical capacity greater than zero is easily obtainable (see e.g. Refs.~\cite{Tjoa_2022,Lapponi_2023}), a quantum capacity greater than zero was never obtained so far - unless one considers entanglement assistance \cite{Landulfo_2016,Barcellos2021}, detectors operating in bounded regions of space \cite{Jonsson_2018} or detectors interacting with a finite number of modes of the field \cite{chen2008capacity,Bruschi_2013}.

For this reason, we wonder if a reliable communication of quantum messages in an open $(3+1)$D spacetime is even possible or if there is some limit preventing this kind of communication. 
Ref.~\cite{Jonsson_2018} pointed out the role of the no-cloning theorem. The theorem proves that quantum states cannot be \textit{cloned} without errors, meaning that a quantum message cannot be sent reliably to two different receivers. Then, if the sender's detector interacts with the field in each direction, in an isotropic spacetime, there is potentially more than one receiver achieving the same input message. The input message is then \textit{cloned} and the no-cloning theorem violated. Henceforth, this theorem should prevent a quantum capacity greater than zero in each isotropic spacetime. 
As a consequence, in case of a $(3+1)$D spacetime, a quantum capacity greater than zero is expected to occur in very anisotropic situations. This could be reached when the two detectors move at relativistic speeds with respect to each other.

Then, to explore this possibility, we consider three different situations with the detectors in a Minkowski background spacetime: 1) the detectors are static; 2) the detectors move inertially w.r.t. each other; 3) the sender's detector is Rindler accelerated and the receiver is static. In case both the detectors are static, because the situation is fully isotropic, the no-cloning theorem should prevent any reliable quantum communication. The same reasoning does not apply for inertial detectors. However, we show that, in case the detectors travel inertially, not only the quantum capacity is still zero, but also the classical capacity is expected to decrease. Finally, in the third case, where the sender is Rindler accelerated, we prove that the quantum capacity can be greater than zero. In particular, this is possible if the sender, after preparing the state, waits enough time before sending it to the receiver. This is due to the fact that, from the receiver's perspective, the state to be communicated gets amplified more and more during the time the sender waits. This amplification could overcome physical limits given by the uncertainty principle that the sender would have in the static case.

The paper is structured as follows. In Sec.~\ref{sec: Hamiltonian} we specify the Hamiltonian of the system, keeping an eye on the prescriptions needed in case the two detectors are not static w.r.t. each other. In Sec.~\ref{sec: QLE}, we study the Heisenberg evolution of the detectors' moment operator. In Sec.~\ref{sec communication protocol} we build a general communication protocol using the aforementioned Heisenberg evolution, describing the state of each detector as a one-mode Gaussian state and the system of the two detectors as a two-modes Gaussian state. In Sec.~\ref{sec: OMGC} we recognize the channel arising from the general protocol as a one-mode Gaussian channel. The properties and quantum capacity of this class of channels are defined and discussed. In Sec.~\ref{sec: specific protocol} we consider a rapid interaction between field and detectors. The properties of the channel are studied when the two detectors are static \ref{ssec: static detectors}, inertially moving \ref{ssec: inertial detectors} and when the sender is Rindler accelerated \ref{ssec: accelerating detectors}. The results and the possible perspectives for future works are discussed in Sec.~\ref{sec: conclusions}. 

Throughout this paper, we work in natural units $\hbar=c=1$.

\section{Hamiltonian of the system}\label{sec: Hamiltonian}
We consider two non-relativistic quantum systems, labelled with $A$ and $B$, whose Hamiltonian, in their proper frame, is that of a $1D$ quantum harmonic oscillator i.e.
\begin{equation}\label{harmonic oscillator hamiltonian}
    \hat{H}_i=\omega_i \left(a_i^\dagger a_i+\frac{1}{2}\right)\,,
\end{equation}
where $\omega_i$ is the frequency of the oscillator $i=A,B$. Each of these harmonic oscillators travel with a general trajectory in a $(3+1)D$ spacetime. These oscillators can be thought as an infinite level Unruh-deWitt detector whose energy gap is $\omega_i$ \cite{Brown_2013}. 

The detectors interact with a massless scalar field, namely
\begin{equation}\label{massless scalar field}
    \hat{\Phi}(t,\mathbf{x})=\int\frac{d^3\mathbf{k}}{\sqrt{(2\pi)^3 2|\mathbf{k}|}}\left(\hat{a}_{\mathbf{k}}e^{-i(|\mathbf{k}|t-i\mathbf{k}\cdot\mathbf{x})}+\text{H.c.}\right),
\end{equation}
The interaction between the detector $i=A,B$ and the field can be modeled via the Hamiltonian density
\begin{equation}\label{single interaction}
\hat{h}_{i,\Phi}=f_i(\mathbf{x},t)\hat{q}_i(t)\otimes\hat{\Phi}(\mathbf{x},t)\,,
\end{equation}
where the moment operator $\hat{q}_i$ is chosen to be the position operator of the $1D$ quantum harmonic oscillator, i.e.
\begin{equation}
    \hat{q}_i=\frac{1}{\sqrt{2m_i\omega_i}}(a^\dagger_i+a_i)\,,
\end{equation}
where $m_i$ is the mass of the oscillator $i$. The function $f_i(\mathbf{x},t)$ in Eq.~\eqref{single interaction} is the spacetime smearing function of the detector $i$. In other words, $f_i$ indicates how the field-detector interaction is distributed in space and time. Usually, in the detector proper frame $f_i(\mathbf{x},t)$ is defined as the product between:
\begin{itemize}
    \item a space dependent function $\tilde{f}_i(\mathbf{x})$, indicating the position of the detector in space and its ``shape" around its center of mass;
    \item a time dependent function $\lambda_i(t)$ called the \textit{switching-in function}, indicating how the field-detector interaction is turned on and off in time. 
\end{itemize}
Considering the interaction of the field with both the detectors $A$ and $B$, the complete interaction Hamiltonian density is given by
\begin{equation}\label{complete interaction hamiltonian density}
    \hat{h}_I=\left(f_A\hat{q}_A+f_B\hat{q}_B\right)\otimes \hat{\Phi}\,.
\end{equation}
The operator $\hat{h}_I$ from Eq.~\eqref{complete interaction hamiltonian density} is a scalar and then it is independent from the coordinates chosen \cite{Mart_n_Mart_nez_2021,Perche_2021}. The Heisenberg evolution of the system, however, depends on the Hamiltonian of the detector \eqref{harmonic oscillator hamiltonian} and on the interaction Hamiltonian $\hat{H}_I$, obtained by integrating Eq.~\eqref{complete interaction hamiltonian density} in space. Then, $\hat{H}_I$ is observer-dependent and so is the Heisenberg evolution of the involved operators. To study the evolution of the system, we need to define the observer's frame and its coordinates.

To account for the most general case, we consider the two detectors lying in a general background spacetime and following general trajectories. Each detector $i$ has a proper observer positioned at the center of mass of the detector $i$. Since each detector is represented by a non-relativistic quantum system, the coordinates used by each observer should be locally non-relativistic. For this reason, we consider the proper observer comoving with the detector $i$ to use the Fermi-normal coordinates associated with their trajectory. Namely, these coordinates could be written as $(t_i,x_i,y_i,z_i)$, where $t_i$ is the proper time of the observer, and the space coordinates $(x_i,y_i,z_i)$ are defined such that the basis generating them is made by vectors always orthogonal to the proper velocity of the detector $i$ (see Refs.~\cite{PhysRevD.101.045017} and \cite{Mart_n_Mart_nez_2021}, for further details). Moreover, since we work with detectors having a spatial extension, the Fermi-normal coordinates must be well defined along the detectors' shape, to be considered as non-relativistic quantum systems. This is true in general only if the detector is small enough, as shown in Ref.~\cite{Perche_2022}. 

The interaction Hamiltonian $\hat{H}_I$, for an observer $i$ working in the Fermi-normal coordinates $t_i,\mathbf{x}_i$, is obtained through the integration of Eq.~\eqref{complete interaction hamiltonian density} in $d\mathbf{x}_i$, i.e.\small
\begin{equation}
    \hat{H}_I^i(t_i)=\int_{\Sigma_{t_i}}\sum_{j=A,B}f_j(t_i,\mathbf{x}_i)\hat{q}_j(t_i)\otimes \hat{\Phi}(t_i,\mathbf{x}_i)\sqrt{-g_i(t_i,\mathbf{x}_i)}d\mathbf{x}_i\,,
\end{equation}\normalsize
where $g_i$ is determinant of the metric tensor of the spacetime where $i$ lies and $\Sigma_{t_i}$ is the Cauchy surface $t_i=\text{const}$. For simplicity, we define the \textit{smeared field operator} as
\begin{equation}\label{smeared field external observer}
    \hat{\varphi}_i^j(t_j)=\int_{\Sigma_{t_j}}f_i(\mathbf{x}_j,t_j)\hat{\Phi}(\mathbf{x}_j,t_j)\sqrt{-g_j(\mathbf{x}_j,t_j)}d\mathbf{x}_j\,,
\end{equation}
so that the interaction Hamiltonian for the observer $i=A,B$ can be written as
\begin{align}\label{interaction hamiltonian generic observer}
    \hat{H}_I^i(t_i)=\hat{q}_A(t_i)\otimes\hat{\varphi}_A^i(t_i)+\hat{q}_B(t_i)\otimes\hat{\varphi}_B^i(t_i)\,.
\end{align}
Notice that, when we write the smeared field operator $\hat{\varphi}_a^b$, the label $a$ refers to the smearing function used, while the label $b$ refers to the observer performing the integration.

\section{Quantum Langevin equation}\label{sec: QLE}
We want to study the Heisenberg evolution of the operators $\hat{q}_A$ and $\hat{q}_B$. The evolution of $\hat{q}_A$ is governed by the sum of the Hamiltonian of the harmonic oscillator $A$, given by Eq.~\eqref{harmonic oscillator hamiltonian} with $i=A$ and the interaction Hamiltonian given by Eq.~\eqref{interaction hamiltonian generic observer} with $i=A$. At this point, we can follow the same procedure done in Ref.~\cite{Lapponi_2023}, recognizing the interaction Hamiltonian \eqref{interaction hamiltonian generic observer} as the one occurring in the Caldeira-Legett model for the quantum Brownian motion, when the smeared field plays the role of an Ohmic environment \cite{CALDEIRA1983587,Breuer2006}. The Heisenberg evolution of the moment operator $\hat{q}_A$ is then determined by the following quantum Langevin equation
\begin{align}
   & m_A\frac{d^2}{(dt_A)^2}q_A(t_A)+m_A\omega_A^2q_A\nonumber\\&-\sum_{j=A,B}\int_{-\infty}^{t_A}\chi^A_{Aj}(t_A,s_A)q_j(s_A)ds_A=\varphi_A^A(t_A)\,,\label{quantum langevin equation for Alice}
\end{align}
where we defined the \textit{dissipation kernel}
\begin{equation}\label{dissipation kernel}
    \chi_{ij}^i(t_i,s_i)\coloneqq i\theta(t_i-s_i)\bra{\Phi}\left[\varphi_i^i(t_i),\varphi_j^i(s_i)\right]\ket{\Phi}\,,
\end{equation}
for $i,j=A,B$ and denoted with $\ket{\Phi}$ the initial state of the scalar field. 
If the two detectors are not causally correlated, then the commutators between the field operators with $i\ne j$ vanish and so do the off-diagonal elements of the dissipation kernel \eqref{dissipation kernel}.

Analogously, for the moment operator of the oscillator $B$ we have
\begin{align}
    &m_B\frac{d^2}{(dt_B)^2}q_B(t_B)+m_B\omega_B^2q_B\nonumber\\&-\sum_{j=A,B}\int_{-\infty}^{t_B}\chi_{Bj}^B(t_B,s_B)q_j(s_B)ds_B=\varphi_B^B(t_B)\,.\label{quantum langevin equation for Bob}
\end{align}
The aim is to solve the two coupled differential equations \eqref{quantum langevin equation for Alice} and \eqref{quantum langevin equation for Bob}. In the communication protocol we have in mind, the detector $A$ wants to communicate its state to the detector $B$. For this reason, we need to calculate the coupled Langevin equations in the proper coordinates of the detector $B$, i.e. $t_B$. Then equation \eqref{quantum langevin equation for Alice}, in terms of $t_B$, becomes
\begin{align}\label{QLE for Alice seen by Bob}
    &m_A\frac{1}{\dot{t}_A^2}\ddot{q}_A-m_A\frac{\ddot{t}_A}{\dot{t}_A^3}\dot{q}_A+m_A\omega_A^2q_A\nonumber\\&-\sum_{j=A,B}\int_{-\infty}^{t_B}\chi_{Aj}^A(t_B,s_B)q_j(s_B)\dot{t}_A(s_B)ds_B=\varphi_A^A(t_B)\,,
\end{align}
where we denoted with the upper dot the derivative w.r.t. $t_B$. Finally, we can write Eqs.~\eqref{quantum langevin equation for Bob} and \eqref{QLE for Alice seen by Bob} together in the following compact form
\begin{widetext}
\footnotesize
\begin{equation}\label{vectorial form of langevin equation}
    \left(\begin{matrix}
        \frac{d^2}{dt_B^2}-\frac{\ddot{t}_A}{\dot{t}_A}\frac{d}{dt_B}+\dot{t}_A^2\omega_A^2&0\\0&\frac{d^2}{dt_B^2}+\omega_B^2
    \end{matrix}\right)\left(\begin{matrix}
        q_A\\q_B
    \end{matrix}\right)-\int_{-\infty}^{t_B}\left(\begin{matrix}
        \frac{\dot{t}_A(t_B)^2\dot{t}_A(s_B)}{m_A}&0\\0&\frac{1}{m_B}
    \end{matrix}\right)\left(\begin{matrix}
        \chi_{AA}^A(t_B,s_B)&\chi_{AB}^A(t_B,s_B)\\\chi_{BA}^B(t_B,s_B)&\chi_{BB}^B(t_B,s_B)
    \end{matrix}\right)\left(\begin{matrix}
        q_A(s_B)\\q_B(s_B)
    \end{matrix}\right)ds_B=\left(\begin{matrix}
        \frac{\dot{t}_A^2(t_B)\varphi_A^A(t_B)}{m_A}\\\frac{\varphi_B^B(t_B)}{m_B}
    \end{matrix}\right)\,,
\end{equation}\normalsize
\end{widetext}
where we multiplied Eq.~\eqref{QLE for Alice seen by Bob} by $m_A^{-1}\dot{t}_A^2(t_B)$ and Eq.~\eqref{quantum langevin equation for Bob} by $m_B^{-1}$. 
If the detectors are not causally correlated, i.e. $\chi_{AB}^A=\chi_{BA}^B=0$, then the off-diagonal terms of Eq.~\eqref{vectorial form of langevin equation} disappear and the evolution of $\hat{q}_A$ becomes completely independent from $\hat{q}_B$ and viceversa.

We now define the Green function matrix 
\begin{equation}
    \mathbb{G}(t_B,s_B)\coloneqq\left(\begin{matrix}G_{AA}(t_B,s_B)&G_{AB}(t_B,s_B)\\G_{BA}(t_B,s_B)&G_{BB}(t_B,s_B)\end{matrix}\right)\,,
\end{equation}
solution of the homogeneous form of the Langevin equation \eqref{vectorial form of langevin equation}, which is reported in Eq.~\eqref{homogeneous quantum langevin equation} of Appendix \ref{appendix Green function matrix}. Imposing the causality condition $\mathbb{G}(t<s)=0$, the Green function matrix follows the boundary conditions $\mathbb{G}(t=s,s)=0$ and $\dot{\mathbb{G}}(t\to s^+,s)=\mathbb{I}$. 

Then, the evolution of the operators $\hat{q}_i$ from a time $s_B$ to a time $t_B$ can be expressed through the Green function matrix as\small
\begin{align}
    &\left(\begin{matrix}
        q_A(t_B)\\q_B(t_B)
    \end{matrix}\right)=\dot{\mathbb{G}}(t_B,s_B)\left(\begin{matrix}
        q_A(s_B)\\q_B(s_B)
    \end{matrix}\right)+\mathbb{G}(t_B,s_B)\left(\begin{matrix}
        \dot{q}_A(s_B)\\\dot{q}_B(s_B)
    \end{matrix}\right)\nonumber\\&+\int_{s_B}^{t_B} \mathbb{G}(t_B,r_B)\mathbb{M}^{-1}\mathbb{F}^2(r_B)\left(\begin{matrix}
        \varphi_{A}^A(r_B)\\\varphi_{B}^B(r_B)
    \end{matrix}\right)dr_B\,,\label{moment operator evolution}
\end{align}\normalsize
where, for simplicity, we defined the matrix $\mathbb{M}\coloneqq\text{diag}(m_A,m_B)$ and the matrix $\mathbb{F}(t)\coloneqq\text{diag}(\dot{t}_A(t),1)$.

\section{Communication protocol}\label{sec communication protocol}
Tracing away the field, the system made by two harmonic oscillators can be seen as a two-modes bosonic system. Within multi-mode Bosonic systems, Gaussian states have a pivotal importance both in quantum optics and in quantum information theory \cite{serafini2017quantum}. Motivated by this fact, we consider the system of the two detectors to be initially in a \textit{two-modes bosonic Gaussian state}. The properties of these states
are defined by two elements: the \textit{covariance matrix}
\begin{equation}\label{covariance matrix calculation basis}
    \sigma=\left(\begin{array}{c|c}
       \sigma_{qq} & \sigma_{qp}\\\hline
        \sigma_{pq} & \sigma_{pp}
    \end{array}\right)\,,
\end{equation}
where, for $\alpha,\beta=q,p$, we have\footnotesize
\begin{equation}
    \sigma_{\alpha\beta}=\frac{1}{2}\left(\begin{matrix}
        \langle\left\{\hat{\alpha}_A,\hat{\beta}_A\right\}\rangle-2\langle\hat{\alpha}_A\rangle\langle\hat{\beta}_A\rangle&\langle\left\{\hat{\alpha}_A,\hat{\beta}_B\right\}\rangle-2\langle\hat{\alpha}_A\rangle\langle\hat{\beta}_B\rangle\\\langle\left\{\hat{\alpha}_B,\hat{\beta}_A\right\}\rangle-2\langle\hat{\alpha}_B\rangle\langle\hat{\beta}_A\rangle&\langle\left\{\hat{\alpha}_B,\hat{\beta}_B\right\}\rangle-2\langle\hat{\alpha}_B\rangle\langle\hat{\beta}_B\rangle
    \end{matrix}\right)\,,
\end{equation}
\normalsize
and the \textit{first momentum vector}
\begin{equation}\label{first momentum vector}
    \mathbf{d}=\left(\begin{matrix}
        \langle \hat{q}_A\rangle\\\langle \hat{q}_B\rangle\\\langle \hat{p}_A\rangle\\\langle \hat{p}_B\rangle
    \end{matrix}\right)\,.
\end{equation}

The operator $\hat{p}_i$ is canonically conjugate to $\hat{q}_i$, so that
\begin{equation}\label{algebra}
    \left[\hat{q}_i,\hat{p}_j\right]=i\delta_{ij}\,.
\end{equation}
If $\hat{q}_i=\frac{1}{\sqrt{2m_i\omega_i}}(a^\dagger_i+a_i)$, then $\hat{p}_i=i\sqrt{\frac{m_i\omega_i}{2}}(a_i^\dagger-a_i)$ from Eq.~\eqref{algebra}. Moreover, by applying the Heisenberg evolution to the operator $\hat{q}_i$, one obtains $\frac{d}{dt_i}\hat{q}_i=\hat{p}_i/m_i$. Since the evolution of the system is computed in Bob's frame, we have $\hat{p}_A=m_A\frac{dq_A}{dt_A}=m_A\dot{q}_A/\dot{t}_A$. At this point, Eq.~\eqref{moment operator evolution} can be rewritten in terms of the value of the operators $\hat{p}_i$ at the initial time $s_B$, i.e.:
\begin{align}
    &\left(\begin{matrix}
        q_A(t_B)\\q_B(t_B)
    \end{matrix}\right)=\dot{\mathbb{G}}(t_B,s_B)\left(\begin{matrix}
        q_A(s_B)\\q_B(s_B)
    \end{matrix}\right)\nonumber\\&+\mathbb{G}(t_B,s_B)\mathbb{M}^{-1}\mathbb{F}(s_B)\left(\begin{matrix}
        p_A(s_B)\\p_B(s_B)
    \end{matrix}\right)\nonumber\\&+\int_{s_B}^{t_B} \mathbb{G}(t_B,r_B)\mathbb{M}^{-1}\mathbb{F}^2(r_B)\left(\begin{matrix}
        \varphi_{A}^A(r_B)\\\varphi_{B}^B(r_B)
    \end{matrix}\right)dr_B\,,\label{moment operator evolution for momentum}
\end{align}\normalsize
where $\mathbb{F}$ and $\mathbb{M}$ are defined at the end of Sec.~\ref{sec: QLE}. By applying a time derivative to Eq.~\eqref{moment operator evolution for momentum} and multiplying it from the left by $\mathbb{F}^{-1}(t_B)\mathbb{M}$, we can write the evolved operators $\hat{p}_i$ as
\begin{widetext}\footnotesize
\begin{equation}\label{momentum evolution}
    \left(\begin{matrix}
        p_A(t_B)\\p_B(t_B)
    \end{matrix}\right)=\mathbb{F}^{-1}(t_B)\mathbb{M}\ddot{\mathbb{G}}(t_B,s_B)\left(\begin{matrix}
        q_A(s_B)\\q_B(s_B)
    \end{matrix}\right)+\mathbb{F}^{-1}(t_B)\mathbb{M}\dot{\mathbb{G}}(t_B,s_B)\mathbb{M}^{-1}\mathbb{F}(s_B)\left(\begin{matrix}
        p_A(s_B)\\p_B(s_B)
    \end{matrix}\right)+\int_{s_B}^{t_B}\mathbb{F}^{-1}(t_B)\mathbb{M}\dot{\mathbb{G}}(t_B,r_B)\mathbb{M}^{-1}\mathbb{F}(r_B)^2\left(\begin{matrix}
        \varphi_{A}^A(r_B)\\\varphi_{B}^B(r_B)
    \end{matrix}\right)dr_B\,.
\end{equation}\normalsize
\end{widetext}
The first momentum vector \eqref{first momentum vector} does not affect the entropy-related quantities of a Gaussian state. Being interested in them in the following, we can consider $\mathbf{d}=\mathbf{0}$ without loss of generality. Finally, using Eqs.~\eqref{moment operator evolution for momentum} and \eqref{momentum evolution}, we can write the evolution of the covariance matrix \eqref{covariance matrix calculation basis} from a time $s_B$ to a time $t_B$. Settting for simplicity $t\coloneqq t_B$ and $s\coloneqq s_B$, we have
\begin{widetext}
\begin{align}
    \sigma_{qq}(t)=&\dot{\mathbb{G}}(t,s)\sigma_{qq}(s)\dot{\mathbb{G}}^T(t,s)+\dot{\mathbb{G}}(t,s)\sigma_{qp}(s)\mathbb{F}(s)\mathbb{M}^{-1}\mathbb{G}^T(t,s)+\mathbb{G}(t,s)\mathbb{M}^{-1}\mathbb{F}(s)\sigma_{pq}(s)\dot{\mathbb{G}}^T(t,s)\nonumber\\&+\mathbb{G}(t,s)\mathbb{M}^{-1}\mathbb{F}(s)\sigma_{pp}(s)\mathbb{F}(s)\mathbb{M}^{-1}\mathbb{G}^T(t,s)+\int_s^t\int_s^t \mathbb{G}(t,r)\mathbb{M}^{-1}\mathbb{F}^2(r) \nu(r,r') \mathbb{F}^2(r')
\mathbb{M}^{-1}\mathbb{G}^{T}(t,r')drdr'\,;\label{cov matrix evo qq}
\end{align}\small
\begin{align}
   \sigma_{qp}(t)=&\dot{\mathbb{G}}(t,s)\sigma_{qq}(s)\ddot{\mathbb{G}}^T(t,s)\mathbb{M}\mathbb{F}^{-1}(t)+\dot{\mathbb{G}}(t,s)\sigma_{qp}(s)\mathbb{M}^{-1}\mathbb{F}(s)\dot{\mathbb{G}}^T(t,s)\mathbb{F}^{-1}(t)\mathbb{M}+\mathbb{G}(t,s)\mathbb{M}^{-1}\mathbb{F}(s)\sigma_{pq}(s)\ddot{\mathbb{G}}^T(t,s)\mathbb{M}\mathbb{F}^{-1}(t)\nonumber\\&+\mathbb{G}(t,s)\mathbb{M}^{-1}\mathbb{F}(s)\sigma_{pp}(s)\mathbb{F}(s)\mathbb{M}^{-1}\dot{\mathbb{G}}^T(t,s)\mathbb{M}\mathbb{F}^{-1}(t)+\int_s^t\int_s^t \mathbb{G}(t,r)\mathbb{M}^{-1}\mathbb{F}^2(r) \nu(r,r')\mathbb{F}^2(r') 
\mathbb{M}^{-1}\dot{\mathbb{G}}^{T}(t,r')\mathbb{M}\mathbb{F}^{-1}(t)drdr'\,;\label{cov matrix evo qp}
\end{align} \normalsize
\begin{equation}\label{cov matrix evo pq}
    \sigma_{pq}(t)=\sigma_{qp}^T(t)\,;
\end{equation}\small
\begin{align}
    \sigma_{pp}(t)=&\mathbb{F}^{-1}(t)\mathbb{M}\ddot{\mathbb{G}}(t,s)\sigma_{qq}(s)\ddot{\mathbb{G}}^T(t,s)\mathbb{M}\mathbb{F}^{-1}(t)+\mathbb{F}^{-1}(t)\mathbb{M}\ddot{\mathbb{G}}(t,s)\sigma_{qp}(s)\mathbb{F}(s)\mathbb{M}^{-1}\dot{\mathbb{G}}^T(t,0)\mathbb{M}\mathbb{F}^{-1}(t)\nonumber\\&+\mathbb{F}^{-1}(t)\mathbb{M}\dot{\mathbb{G}}(t,s)\mathbb{M}^{-1}\mathbb{F}(s)\sigma_{pq}(s)\ddot{\mathbb{G}}^T(t,s)\mathbb{M}\mathbb{F}^{-1}(t)+\mathbb{F}^{-1}(t)\mathbb{M}\dot{\mathbb{G}}(t,s)\mathbb{M}^{-1}\mathbb{F}(s)\sigma_{pp}(s)\mathbb{F}(s)\mathbb{M}^{-1}\dot{\mathbb{G}}^T(t,s)\mathbb{M}\mathbb{F}^{-1}(t)\nonumber\\&+\int_s^t\int_s^t \mathbb{F}^{-1}(t)\mathbb{M}\dot{\mathbb{G}}(t,r)\mathbb{M}^{-1}\mathbb{F}^2(r) \nu(r,r') 
\mathbb{F}^2(r')\mathbb{M}^{-1}\dot{\mathbb{G}}^{T}(t,r')\mathbb{M}\mathbb{F}^{-1}(t)drdr'\,,\label{cov matrix evo pp}
\end{align}\normalsize
\end{widetext}
where we defined the noise kernel
\begin{equation}\label{noise kernel}
    \nu(t,t')\coloneqq\frac{1}{2}\left(\begin{matrix}
        \langle\left\{\varphi_A^A(t),\varphi_A^A(t')\right\}   \rangle&\langle\left\{\varphi_B^B(t),\varphi_A^A(t')\right\}\rangle\\\langle\left\{\varphi_A^A(t),\varphi_B^B(t')\right\}\rangle&\langle\left\{\varphi_B^B(t),\varphi_B^B(t')\right\}\rangle
    \end{matrix}\right)\,.
\end{equation}
Eqs.~\eqref{cov matrix evo qq}, \eqref{cov matrix evo qp}, \eqref{cov matrix evo pq} and \eqref{cov matrix evo pp} could be rewritten in the following compact form
\begin{equation}\label{two-modes transformation calculation basis}
    \sigma(t)=\mathbf{T}\sigma(s)\mathbf{T}^T+\mathbf{N}\,,
\end{equation}
where
\begin{equation}\label{complete transmission matrix}
    \mathbf{T}=\left(\begin{array}{c|c}
       \dot{\mathbb{G}}(t,s) & \mathbb{G}(t,s)\mathbb{M}^{-1}\mathbb{F}(s)\\\hline
        \mathbb{F}^{-1}(t)\mathbb{M}\ddot{\mathbb{G}}(t,s) & \mathbb{F}^{-1}(t)\mathbb{M}\dot{\mathbb{G}}(t,s)\mathbb{M}^{-1}\mathbb{F}(s)
    \end{array}\right)\,,
\end{equation}
and\begin{widetext}
\scriptsize
\begin{equation}\label{complete noise matrix}
    \mathbf{N}=\left(\begin{array}{c|c}
    \int_s^t\int_s^t \mathbb{G}(t,r)\mathbb{M}^{-1}\mathbb{F}^2(r)\nu(r,r')\mathbb{F}^2(r')
\mathbb{M}^{-1}\mathbb{G}^{T}(t,r')drdr'&\int_s^t\int_s^t \mathbb{G}(t,r)\mathbb{M}^{-1}\mathbb{F}^2(r)\nu(r,r') \mathbb{F}^2(r')
\mathbb{M}^{-1}\dot{\mathbb{G}}^{T}(t,r')\mathbb{M}\mathbb{F}^{-1}(t)drdr'\\\hline\int_s^t\int_s^t \mathbb{F}^{-1}(t)\mathbb{M}\dot{\mathbb{G}}(t,r)\mathbb{M}^{-1}\mathbb{F}^2(r) \nu^T(r,r') \mathbb{F}^2(r')
\mathbb{M}^{-1}\mathbb{G}^{T}(t,r')drdr'&\int_s^t\int_s^t \mathbb{F}^{-1}(t)\mathbb{M}\dot{\mathbb{G}}(t,r)\mathbb{M}^{-1}\mathbb{F}^2(r)\nu(r,r')\mathbb{F}^2(r')
\mathbb{M}^{-1}\dot{\mathbb{G}}^{T}(t,r')\mathbb{M}\mathbb{F}^{-1}(t)drdr'\end{array}\right)\,.
\end{equation}\normalsize
\end{widetext}

The communication protocol consists of Alice sending information about her detector's state to Bob. In other words, we now define a \textit{quantum channel} - in general, a map whose input and output are quantum states - whose input is the state of Alice's detector at a time $s$ and the output is the state of Bob's detector at a time $t$ (the properties of this channel are studied in Sec.~\ref{sec: OMGC}). Namely, we wonder how much information about Alice's state at the time $s$ is achievable from Bob's state at the time $t$. To perform this study, it is convenient to rewrite the covariance matrix \eqref{covariance matrix calculation basis} in the form
\begin{equation}\label{covariance matrix physical basis}
     \sigma=\left(\begin{array}{c|c}
       \sigma_{AA} & \sigma_{AB}\\\hline
        \sigma_{BA} & \sigma_{BB}
    \end{array}\right)\,,
\end{equation}
where, for $i,j=A,B$
\begin{equation}\label{one-mode Gaussian state}
    \sigma_{ij}=\frac{1}{2}\left(\begin{matrix}
        \langle\left\{\hat{q}_i,\hat{q}_j\right\}\rangle&\langle\left\{\hat{q}_i,\hat{p}_j\right\}\rangle\\\langle\left\{\hat{p}_i,\hat{q}_j\right\}\rangle&\langle\left\{\hat{p}_i,\hat{p}_j\right\}\rangle
    \end{matrix}\right)\,.
\end{equation}
The state of Alice detector is represented by the one-mode Gaussian state $\sigma_{AA}$. Analogously $\sigma_{BB}$ is the one-mode Gaussian state representing Bob's detector. The off-diagonal term $\sigma_{AB}=\sigma_{BA}^T$ expresses the correlations between the two oscillators. If we suppose the two detectors to be initially uncorrelated, we need $\sigma_{AB}(s)=0$. The expectation value of the energy of the oscillator $i$, whose state is represented by $\sigma_{ii}$, depends on the observer. If the observer is moving alongside the oscillator, then
\begin{equation}
    \label{oscillator energy}
    \langle E_i\rangle=\text{Tr}(\mathbb{E}_i\sigma_{ii}\mathbb{E}_i)\,,
\end{equation} 
where $\mathbb{E}_i=\text{diag}\left(\sqrt{\frac{m_i\omega_i^2}{2}},\frac{1}{\sqrt{2m_i}}\right)$. For an external observer $j$, from the conservation of the action, the energy $\langle E_i \rangle$ from Eq.~\eqref{oscillator energy} is multiplied by a factor $\frac{dt_i}{dt_j}$. Hence, in Bob's frame, Alice's detector carries an energy $\dot{t}_A\langle E_A\rangle=\dot{t}_A\text{Tr}(\mathbb{E}_A\sigma_{AA}\mathbb{E}_A)$.

To obtain the covariance matrix in the form given by Eq.~\eqref{covariance matrix physical basis}, one can apply to the covariance matrix \eqref{covariance matrix calculation basis} the permutation
\begin{equation}
    P=\left(\begin{matrix}
        1&0&0&0\\0&0&1&0\\0&1&0&0\\0&0&0&1
    \end{matrix}\right)\,,
\end{equation}
which exchanges the second and third rows and columns of the matrix \eqref{covariance matrix calculation basis}. Since $PP=\mathbb{I}$, Eq.~\eqref{two-modes transformation calculation basis} still holds by applying the same transformation $P$ to the matrices $\mathbf{T}$ and $\mathbf{N}$ of Eqs.~\eqref{complete transmission matrix} and \eqref{complete noise matrix}, respectively. In this way, we can write
\begin{equation}\label{new transmission matrix}
    \mathbf{T}'\coloneqq P\mathbf{T}P=\left(\begin{array}{c|c}
       T_{AA} & T_{AB}\\\hline
    T_{BA}& T_{BB}
    \end{array}\right)\,;
\end{equation}
\begin{equation}\label{new noise matrix}
    \mathbf{N}'\coloneqq P\mathbf{N}P=\left(\begin{array}{c|c}
       N_{AA} & N_{AB}\\\hline
    N_{BA}& N_{BB}
    \end{array}\right)\,.
\end{equation}
Then, Eq.~\eqref{two-modes transformation calculation basis} can be rewritten as
\begin{align}
    &\left(\begin{array}{c|c}
       \sigma_{AA}(t) & \sigma_{AB}(t)\\\hline
    \sigma_{BA}(t)& \sigma_{BB}(t)
    \end{array}\right)=\left(\begin{array}{c|c}
       T_{AA} & T_{AB}\\\hline
    T_{BA}& T_{BB}
    \end{array}\right)\left(\begin{array}{c|c}
       \sigma_{AA}(s) & \sigma_{AB}(s)\\\hline
    \sigma_{BA}(s)& \sigma_{BB}(s)
    \end{array}\right)\nonumber\\&\times\left(\begin{array}{c|c}
       T_{AA}^T & T_{BA}^T\\\hline
    T_{AB}^T& T_{BB}^T
    \end{array}\right)+\left(\begin{array}{c|c}
       N_{AA} & N_{AB}\\\hline
    N_{BA}& N_{BB}
    \end{array}\right)\,.\label{complete cov matrix transformation phys basis}
\end{align}
The input state of the channel is Alice's detector state at the time $s$, namely $\sigma_{in}=\sigma_{AA}(s)$. The output state is the state of the detector $B$ at a time $t$, namely $\sigma_{out}=\sigma_{BB}(t)$. The latter, using Eq.~\eqref{complete cov matrix transformation phys basis} and $\sigma_{AB}(s)=\sigma_{BA}(s)=0$, can be written in terms of the former as
\begin{equation}\label{output in terms of the input}
    \sigma_{BB}(t)=T_{BA}\sigma_{AA}(s)T^T_{BA}+T_{BB}\sigma_{BB}(s)T_{BB}^T+N_{BB}\,,
\end{equation}
where, using Eqs.~\eqref{complete transmission matrix} and \eqref{complete noise matrix} alongside Eqs.~\eqref{new transmission matrix} and \eqref{new noise matrix}, we can find
\begin{equation}\label{transmissivity matrix}
    T_{BA}=\left(\begin{matrix}
        \dot{G}_{BA}(t,s)&G_{BA}(t,s)\frac{\dot{t}_A(s)}{m_A}\\\ddot{G}_{BA}(t,s)m_B&\dot{G}_{BA}(t,s)m_B\frac{\dot{t}_A(s)}{m_A}
    \end{matrix}\right)\,;
\end{equation}
\begin{equation}
    T_{BB}=\left(\begin{matrix}
    \dot{G}_{BB}(t,s)&G_{BB}(t,s)m_B^{-1}\\\ddot{G}_{BB}(t,s)m_B&\dot{G}_{BB}(t,s)
    \end{matrix}\right)\,;
\end{equation}
and $N_{BB}=\left(\begin{matrix}N_{11}&N_{12}\\N_{12}&N_{22}\end{matrix}\right)$, with
\begin{widetext}
    \begin{align}
    N_{11}=&m_A^{-2}\int_s^t\int_s^t\dot{t}^2_A(r)\dot{t}^2_A(r')G_{BA}(t,r)\nu_{AA}(r,r')G_{BA}(t,r')drdr'+m_A^{-1}m_B^{-1}\int_s^t\int_s^t\dot{t}_A^2(r')G_{BB}(t,r)\nu_{BA}(r,r')G_{BA}(t,r')drdr'\nonumber\\&+m_A^{-1}m_B^{-1}\int_s^t\int_s^t\dot{t}_A^2(r)G_{BA}(t,r)\nu_{AB}(r,r')G_{BB}(t,r')drdr'+m_B^{-2}\int_s^t\int_s^tG_{BB}(t,r)\nu_{BB}(r,r')G_{BB}(t,r')drdr'\label{N11}\,;
\end{align}
\begin{align}
    N_{12}=&m_A^{-2}m_B\int_s^t\int_s^t\dot{t}^2_A(r)\dot{t}^2_A(r')G_{BA}(t,r)\nu_{AA}(r,r')\dot{G}_{BA}(t,r')drdr'+m_A^{-1}\int_s^t\int_s^t\dot{t}_A^2(r')G_{BB}(t,r)\nu_{BA}(r,r')\dot{G}_{BA}(t,r')drdr'\nonumber\\&+m_A^{-1}\int_s^t\int_s^t\dot{t}_A^2(r)G_{BA}(t,r)\nu_{AB}(r,r')\dot{G}_{BB}(t,r')drdr'+m_B^{-1}\int_s^t\int_s^tG_{BB}(t,r)\nu_{BB}(r,r')\dot{G}_{BB}(t,r')drdr'\label{N12}\,;
\end{align}
\begin{align}
    N_{22}=&m_A^{-2}m_B^2\int_s^t\int_s^t\dot{t}^2_A(r)\dot{t}^2_A(r')\dot{G}_{BA}(t,r)\nu_{AA}(r,r')\dot{G}_{BA}(t,r')drdr'+m_A^{-1}m_B\int_s^t\int_s^t\dot{t}_A^2(r')\dot{G}_{BB}(t,r)\nu_{BA}(r,r')\dot{G}_{BA}(t,r')drdr'\nonumber\\&+m_A^{-1}m_B\int_s^t\int_s^t\dot{t}_A^2(r)\dot{G}_{BA}(t,r)\nu_{AB}(r,r')G_{BB}(t,r')drdr'+\int_s^t\int_s^t\dot{G}_{BB}(t,r)\nu_{BB}(r,r')\dot{G}_{BB}(t,r')drdr'\label{N22}\,.
\end{align}
\end{widetext}
\section{One-mode Gaussian channels}\label{sec: OMGC}
A quantum channel transforming each one-mode Gaussian state $\sigma_{in}$ into another one-mode Gaussian state $\sigma_{out}$ is called \textit{one-mode Gaussian channel} \cite{Caruso_2006}. In general, a one-mode Gaussian channel $\mathcal{N}$ acts on the input $\sigma_{in}$ through the following map
\begin{equation}\label{quantum channel mapping}
    \mathcal{N}:\sigma_{in}\mapsto\sigma_{out}=\mathbb{T}\sigma_{in}\mathbb{T}^T+\mathbb{N}\,.
\end{equation}
To ensure the complete positiveness of the channel, the following condition should be satisfied
\begin{equation}\label{CP condition}
    \det\mathbb{N}\ge\frac{1}{2}(1-\det\mathbb{T})\,.
\end{equation}

Comparing Eq.~\eqref{output in terms of the input} with Eq.~\eqref{quantum channel mapping}, since $\sigma_{in}=\sigma_{AA}(s)$ and $\sigma_{out}=\sigma_{BB}(t)$, we can recognize the channel defined in the communication protocol in Sec.~\ref{sec communication protocol} as a one-mode Gaussian channel characterized by the matrices
\begin{equation}
    \mathbb{T}=T_{BA}\,;
\end{equation}
\begin{equation}\label{noise matrix}
    \mathbb{N}=T_{BB}\sigma_{BB}(s)T^T_{BB}+N_{BB}\,.
\end{equation}
The matrices $\mathbb{T}$ and $\mathbb{N}$, therefore characterize the capacities of the channel $\mathcal{N}$. Moreover, for one-mode Gaussian channels, it is conjectured that the one-mode Gaussian states are the ones preserving more classical and quantum information under the application of a quantum channel \cite{Giovannetti_2014,Holevo2015}. Following this conjecture, to calculate the capacities of a channel, one can consider exclusively Gaussian states as inputs of the channel without loss of generality.

\subsection{Canonical form}\label{ssec: canonical form}
To evaluate the capacity of the channel $\mathcal{N}$, we need to reduce it to its \textit{canonical form} \cite{Caruso_2006}.
First of all, we apply a unitary transformation $U_{in}$ on the input Gaussian state, whose density matrix is $\rho_{in}$ and another unitary transformation $U_{out}$ on the output Gaussian state whose density matrix is $\rho_{out}$. $U_{in}$ and $U_{out}$ are called \textit{pre-processing} and \textit{post-processing} transformations, respectively. Let be $\rho_{in}$ and $\rho_{out}$ represented by the covariance matrices $\sigma_{in}$ and $\sigma_{out}$, respectively. Then, a unitary transformation acting on a density matrix corresponds to a symplectic transformation acting on a covariance matrix, so that the action of $U_{in}$ on $\rho_{in}$ (resp.~the action of $U_{out}$ on $\rho_{out}$) corresponds to the action of a symplectic transformation $S_{in}$ on $\sigma_{in}$ ($S_{out}$ on $\sigma_{out}$). Since the pre-processing and post-processing transformations are unitaries, the properties of the channel $\mathcal{N}$ are invariant up to their application. The mapping \eqref{quantum channel mapping} can be rewritten as
\begin{equation}\label{map canonical}
    U_{out}\circ\mathcal{N}\circ U_{in}:\sigma_{in}\mapsto \mathbb{T}_c\sigma_{in}\mathbb{T}^T_c+\mathbb{N}_c\,,
\end{equation}
where 
\begin{equation}\label{canonical T}
    \mathbb{T}_c=S_{in}\mathbb{T}S_{out}^T\,,
\end{equation}
\begin{equation}\label{canonical N}
    \mathbb{N}_c=S_{out}\mathbb{N}S_{out}^T\,,
\end{equation}
It is possible to choose $S_{in}$ and $S_{out}$ so that\footnote{In some particular cases, the matrices $\mathbb{T}$ and $\mathbb{N}$ cannot be reduced in this form. Instead, they reduces analogously to rank one matrices. However, this occurs in very singular cases, so that this possibility is not taken into account in this work.} $\mathbb{T}_c=\sqrt{|\tau|}\mathbb{I}$ and $\mathbb{N}_c=\sqrt{W}\mathbb{I}$. Calling the elements of $\mathbb{T}$ and $\mathbb{N}$ as $T_{ij}\coloneqq\left\{\mathbb{T}\right\}_{ij}$ and $N_{ij}\coloneqq\left\{\mathbb{N}\right\}_{ij}$, where $i,j=1,2$, the pre-processing $S_{in}$ and the post-processing $S_{out}$ reducing the channel to its canonical form are
\begin{equation}\label{preprocMatrix}
    S_{in}=\frac{\sqrt[4]{W}}{\sqrt{N_{11}|\tau|}}\left(\begin{matrix}
        \frac{N_{11}T_{22}-N_{12}T_{12}}{\sqrt{W}}&-T_{12}\\
        \frac{N_{12}T_{11}-N_{11}T_{21}}{\sqrt{W}}&T_{11}
    \end{matrix}\right)\,,
\end{equation}
\begin{equation}
    S_{out}=\frac{\sqrt[4]{W}}{\sqrt{N_{11}}}\left(\begin{matrix}
        1&0\\
        -\frac{N_{12}}{\sqrt{W}}&\frac{N_{11}}{\sqrt{W}}
    \end{matrix}\right).
\end{equation}
From Eq.~\eqref{map canonical}, we can finally write the output $\sigma_{out}$ in terms of the input $\sigma_{in}$ as
\begin{equation}\label{Final canonical map}
    \sigma_{out}=|\tau|\sigma_{in}+\sqrt{W}\mathbb{I}\,.
\end{equation}
From Eqs.~\eqref{canonical T} and \eqref{canonical N}, by applying the determinant on both sides, we have 
\begin{align}
    &\tau=\det\mathbb{T}\,;\label{tau definition}\\
    &W=\det\mathbb{N}\label{W definition}\,.
\end{align}

The meaning of the parameters $\tau$ and $W$ can be easily understood from Eq.~\eqref{Final canonical map}. Namely, $\tau$, called \textit{transmissivity} of the channel, indicates the fraction of input signal which is present in the output. The parameter $W$ refers to the amount of signal achieved by Bob which is not present into Alice's input. This is associated with the \textit{additive noise} achieved by Bob. In particular, with $W$ and $\tau$, one can evaluate the average number of noisy particles $\overline{n}$ that Bob achieves, as
\begin{equation}\label{additivenoise}
    \overline{n}=\begin{cases}&\frac{\sqrt{W}}{\left|1-\tau\right|}-\frac{1}{2}\quad\text{if}\quad \tau\ne1\,,\\&\sqrt{W}\quad\text{otherwise}\,.\end{cases}
\end{equation}
The complete positiveness condition \eqref{CP condition} reduces to $\overline{n}\ge0$.

\subsection{Capacities}\label{ssec: Capacities}
We now study the quality of the communication of a generic channel characterized by generic values of $\tau$ and $W$. The perfect quantum channel occurs when $\tau=1$ and $\overline{n}=0$. The further we go from this ideal situation, the worse would be the quality of the communication through the channel. 

The quantification of this ``quality" is well provided by the capacities of a quantum channel. In particular, the \textit{classical capacity} (\textit{quantum capacity}) of the quantum channel $\mathcal{N}$ quantifies the quality of the communication of classical messages (quantum messages) under the channel $\mathcal{N}$. By using the formal definition, the classical capacity (quantum capacity) of a quantum channel $\mathcal{N}$ is the maximum rate of classical information (quantum information) that the channel $\mathcal{N}$ can transmit reliably. In other words, if the capacity of a channel is zero, then the channel cannot transmit information reliably. Instead, as long as the capacity of a channel is positive, information can be transmitted with an arbitrarily low amount of error. However, the less is the magnitude of the capacity, the more are the uses of the quantum channel needed to transmit information reliably - in practice, a lower capacity requires more time for a reliable communication.

The possibility of a reliable communication of classical messages with harmonic oscillators detectors is guaranteed by the fact that we are using Bosonic channels. For such channels, the classical capacity is always greater than zero and can be arbitrarily high by increasing the energy of the channel input \cite{Lupo_2011, Pilyavets_2012}. Moreover, for static detectors always interacting with the field after the switching-in, the classical capacity was extensively studied in Ref.~\cite{Lapponi_2023}. For this reason, in this paper we focus more on the possibility of transmitting quantum messages reliably, by studying the quantum capacity within the protocol described in Sec.~\ref{sec communication protocol}.

Then, we try to evaluate the quantum capacity $Q$ of the channel $\mathcal{N}$, characterized by the parameters $\tau$ and $W$. This task is still an open problem if we consider several uses of the quantum channel. Indeed, in this case, different inputs of the channel uses could be entangled and this fact drastically complicates the evaluation (see Refs.~\cite{Cubitt_2015,Oskouei_2018} for more details). However, the problem simplifies if we consider input states separable over each channel use. In this case, we evaluate the so-called \textit{single-letter quantum capacity} $Q^{(1)}$. In general, $Q^{(1)}(\mathcal{N})\le Q(\mathcal{N})$, so that $Q^{(1)}>0$ is sufficient to prove that a reliable quantum communication is possible.

It is (see e.g.~\cite{ManciniBook})
\begin{equation}\label{single letter quantum capacity}
    Q^{(1)}(\mathcal{N})=\max\left\{0,I_c(\mathcal{N})\right\}\,.
\end{equation}
where $I_c$ is the \textit{maximized coherent information}, i.e. the coherent information of the channel $\mathcal{N}$ maximized over all the possible inputs. The latter, for a one-mode Gaussian channel characterized by the parameters $\tau$ and $\overline{n}$, results \cite{HolevoQC,Br_dler_2015}
\begin{equation}\label{maxcoherentinf2}
    I_\textrm{c}(\tau,W)=\theta(\tau)\log\frac{\tau}{|1-\tau|}-h\left(\frac{1}{2}+\overline{n}\right)\,,
\end{equation}
where $\log$ denotes base $2$ logarithm and $h:(1/2,+\infty)\to \mathbb{R}_+$ is the function
\begin{align}
    h(x)\coloneqq&\left(x+\frac{1}{2}\right)\log\left(x+\frac{1}{2}\right)-\left(x-\frac{1}{2}\right)\log\left(x-\frac{1}{2}\right)\,,\label{entropy function}
\end{align}
From Eq.~\eqref{single letter quantum capacity} we have that $Q^{(1)}>0$ if $I_c$ is positive. Since $h$ is definite positive, from Eq.~\eqref{maxcoherentinf2}, a necessary condition to have $I_c>0$ is that $\tau>1/2$. This is obviously a consequence of the no-cloning theorem \cite{NoCloning}.

\section{Implementing the communication protocol in different scenarios}\label{sec: specific protocol}
The features and properties of the communication channel built with a pair or harmonic oscillators, interacting with a field, were established in Secs.~\ref{sec: QLE} and \ref{sec communication protocol}. Now, we focus on a more specific protocol, defining a particular spacetime smearing for the detectors which is convenient to calculate the quantum capacity of the channel explicitly and to compare the results with the literature.

We take the detectors to travel on prescribed trajectories in Minkowski spacetime. Before specifying the trajectories, we define the spacetime smearing of the detectors. As mentioned in Sec.~\ref{sec: Hamiltonian}, in the detectors' proper frame, the smearing function $f_i$ is usually factorized into a space dependent function $\tilde{f}_i$, giving the shape of the detector - in other words, its spatial distribution - and a time dependent function $\lambda_i$ giving the switching-in function. Hence, by calling $(t_i,\mathbf{x}_i)$ the proper coordinates of the observer comoving with the detector $i$, we have
\begin{equation}
    f_i(\mathbf{x}_i,t_i)=\lambda_i(t_i)\tilde{f}_i(\mathbf{x}_i)\,.
\end{equation}
A finite size for the detector $i$, given by the function $\tilde{f}_i(\mathbf{x}_i)$ provides an ultraviolet cutoff for the modes of the field interacting with the detector. In particular, from Ref.~\cite{Schlicht_2004}, a shape of the detector $i$ following a Lorentzian distribution with effective size $\epsilon$, gives an exponential cutoff for the modes of the field $e^{-\epsilon\mathbf{k}}$. This cutoff is convenient since it usually allows analytical solutions for the correlation function.

Motivated by this fact, for both the detectors, we consider a Lorentzian shape
\begin{equation}\label{lorentzian smearing}
    \tilde{f}_{i}(\mathbf{x}_i)=\frac{1}{\pi^2}\frac{\epsilon}{\left(\mathbf{x}_i\cdot\mathbf{x}_i+\epsilon^2\right)^2}\,,
\end{equation}
Since $\epsilon^{-1}$ represents an ultraviolet cutoff for the energies of the modes the detector interacts with, the energy of the detector itself - computed through an average $\langle E_i\rangle$ from Eq.~\eqref{oscillator energy} - must satisfy
\begin{equation}\label{energy condition}
    \epsilon\langle E_i\rangle\ll1\,.
\end{equation}
Since the minimum energy of the oscillator $i$ occurs in its ground state, where $\langle E_i\rangle=\omega_i/2$, a necessary condition to satisfy Eq.~\eqref{energy condition} is
\begin{equation}\label{energy condition on the frequency}
    \epsilon\omega_i\ll1\,.
\end{equation}
We mostly use the proper coordinates of the receiver Bob. Then, for the sake of simplicity, we write $(t,\mathbf{x})\coloneqq(t_B,\mathbf{x}_B)$ from now on. The distance between Bob's detector and Alice's detector, measured in Bob's frame, is indicated with the function $d(t)$. Since the Lorentzian smearing function \eqref{lorentzian smearing} does not have compact support, to assume the two detectors uncorrelated when Alice prepares the state (i.e. $\sigma_{AB}(s)=0$) we need the two detectors to be far from each other at the time $s$, i.e. we assume $d(s)\gg\epsilon$. Moreover, we assume $d(t)\gg\epsilon$ for any time $t$ to ensure that the communication between the detectors occurs only because of the interaction with the field - not because the two detectors ``touch" each other at a certain time.\\

Regarding the switching in function, we resort to a rapid interaction between field and detector \cite{Simidjia2020,Barcellos2021,Tjoa_2022}. Namely we consider Alice detector to interact with the field only at a certain time $t_I^A$, so that
\begin{equation}\label{alice time smearing}
    \lambda_A(t_A)=\lambda_A\delta(t_A-t_I^A)\,.
\end{equation}
However, we must consider an uncertainty on $t_I^A$ to take into account the Heisenberg principle\footnote{In appendix \ref{appendix: delta-like} we show that, if we violate the Heisenberg principle, then also the no-cloning theorem would be violated.}. To do that, we consider $t_I^A$ as a random variable in a uniform probability distribution from the values $\overline{t}_A^I-\frac{\Delta t_I^A}{2}$ to the values $\overline{t}_A^I+\frac{\Delta t_I^A}{2}$, where $\overline{t}_I^A$ is the central value of the uniform distribution (or, the mean of $t_I^A$) and $\Delta t_I^A$ is the range of values that $t_I^A$ may assume. The standard deviation of the distribution is $\Delta t_I^A/\sqrt{12}$. Then, since the minimum possible energy of Alice'detector, before the interaction, is $\omega_A/2$, the uncertainty principle implies
\begin{equation}\label{Heisenberg principle}
    \Delta t_I^A\ge\frac{\sqrt{12}}{\omega_A}\,.
\end{equation}
To simplify later calculations, we make the choice $\Delta t_I^A=\frac{2\pi}{\omega_A}$, respecting the condition \eqref{Heisenberg principle}. Notice that the condition \eqref{energy condition on the frequency} implies $\epsilon\ll\Delta t_I^A$.

Supposing that Alice and Bob have shared classical information before the protocol, Bob knows that Alice wants to send her message at the time $\overline{t}_I^A$. However, Bob has no way to predict the outcome of the random variable $t_I^A$. For this reason, in order to be sure to receive Alice's message, Bob should interact with the field in a finite time window of width $\Delta t_I^B$ including all the possible values of $t_I^A$. Bob's window should be centered around the time $\overline{t}_I^B+d(\overline{t}_I^B)$, where $\overline{t}_I^B=t_A^{-1}(\overline{t}_I^A)$ and large $\Delta t_I^B\ge\Delta t_I^A$. In particular, from Bob's perspective, Alice's minimum energy during the interaction is multiplied by a factor $\dot{t}_A(\overline{t}_I^B)\le1$. For this reason, Bob's interaction with the field should last for at least
\begin{equation}\label{window largeness relation}
    \Delta t_I^B=\frac{\Delta t_I^A}{\dot{t}_A(\overline{t}_I^A)}\,,
\end{equation}
where the upper dot, as in Sec.~\ref{sec: QLE}, refers to the first derivative w.r.t. Bob's proper time $t$. Since $\Delta t_I^B\ge \Delta t_I^A$, we have $\epsilon\ll\Delta t_I^B$ - this condition is used to perform some approximations to compute the Green function matrix elements in appendix \ref{appendix Green function matrix}.
The switching-in function of Alice is then given by Eq.~\eqref{alice time smearing}, while Bob's one reads 
\begin{equation}\label{bob time smearing}
    \lambda_B(t)=\lambda_B\frac{1}{\Delta t_I^B}\text{rect}\left(\frac{t-\overline{t}_I^B-d(\overline{t}_I^B)}{\Delta t_I^B}\right)\,,
\end{equation}
where we used the function
\begin{equation}
    \text{rect}(x)=\begin{cases}
        1\quad\text{if}\quad|2x|<1\,;\\
        0\quad\text{if}\quad|2x|>1\,.
    \end{cases}
\end{equation}
At this point, to have a rapid interaction protocol, we must impose\footnote{In Appendix \ref{appendix Green function matrix}, we show that the rapid interaction condition \eqref{rapid interaction condition} allows us to get approximated analytical results for the Green function matrix elements.}
\begin{equation}\label{rapid interaction condition}
    \Delta t_I^B/d(\overline{t}_I^B)\ll1\Rightarrow d(\overline{t}_I^B)\omega_A \dot{t}_A(\overline{t}_I^B)\gg1\,.
\end{equation}

Last, we consider the field's initial state $\ket{\Phi}$ to be the Minkowski vacuum $\ket{0}$.

\subsection{Static detectors}\label{ssec: static detectors}
We start with the simplest case where the detectors are static in a given reference frame. In this case, the proper coordinates of Alice and Bob coincide, so that we can call both of them $(t,\mathbf{x})$. 

Since the field is in a Minkowski vacuum, the dissipation kernel \eqref{dissipation kernel} can be rewritten as
\begin{align}
    \chi_{ij}&(t,s)=-2\theta(t-s)\lambda_i(t)\lambda_j(s)\nonumber\\&\times\Im\int_{\Sigma_{t}}d\mathbf{x}\int_{\Sigma_{s}}d\mathbf{x}'\tilde{f}_i(\mathbf{x})\tilde{f}_j(\mathbf{x}') W(\mathbf{x},t;\mathbf{x}',s)\,,\label{dissipation kernel element from the vacuum}
\end{align}
where $W(\mathbf{x},t;\mathbf{x}',s)$ is the Wightman function of the scalar field, i.e. $\bra{0}\hat{\Phi}(\mathbf{x},t)\hat{\Phi}(\mathbf{x}',s)\ket{0}$. The double integral in Eq.~\eqref{dissipation kernel element from the vacuum} is the two-point correlation function of the ``smeared version" of the field $\Phi$. The latter, for a Lorentzian smearing \eqref{lorentzian smearing}, was computed in Ref.~\cite{Schlicht_2004}. Using this result and making explicit $\lambda_A(t)$ and $\lambda_B(s)$, from Eqs.~\eqref{alice time smearing} and \eqref{bob time smearing} respectively, we get the elements of the dissipation kernel as
\begin{equation}\label{chiAA static}
    \chi_{AA}(t,s)=\chi_{AB}(t,s)=0\,;
\end{equation}
\begin{align}
    \chi_{BB}(t,s)=&\frac{4\lambda_B^2}{\pi^2\Delta t_I^2}\text{rect}\left(\frac{t-d-\overline{t}_I}{\Delta t_I}\right)\text{rect}\left(\frac{s-d-\overline{t}_I}{\Delta t_I}\right)\nonumber\\&\times\theta(t-s)\frac{\epsilon(t-s)}{((t-s)^2+4\epsilon^2)^2}\,;\label{chiBB static}
\end{align}
\begin{align}
    \chi_{BA}(t,s)&=\frac{4\lambda_A\lambda_B}{\pi^2\Delta t_I}\delta(s-t_I)\text{rect}\left(\frac{t-d-\overline{t}_I}{\Delta t_I}\right)\nonumber\\&\frac{\epsilon(t-t_I)}{\left((t-t_I)^2-\Delta x-4\epsilon^2\right)^2+16\epsilon^2(t-t_I)^2}\,;\label{chiBA static}
\end{align}
Similarly to the dissipation kernel in Eq.~\eqref{dissipation kernel element from the vacuum}, the elements of the noise kernel \eqref{noise kernel} can be rewritten as
\begin{align}
    \nu_{ij}&(t,s)=2\lambda_i(t)\lambda_j(s)\nonumber\\&\times\Re\int_{\Sigma_{t}}d\mathbf{x}\int_{\Sigma_{s}}d\mathbf{x}'\tilde{f}_i(\mathbf{x})\tilde{f}_j(\mathbf{x}') W(\mathbf{x},t;\mathbf{x}',s)\,,\label{noise kernel element from the vacuum}
\end{align}
and explicitly computed as
\begin{equation}
    \nu_{AA}(t,s)=\lambda_A^2\frac{\delta(t-t_I)\delta(s-t_I)}{8\pi^2\epsilon^2}\,;\label{nu AA static}
\end{equation}\small
\begin{align}
    \nu_{BB}(t,s)=&-\frac{1}{2\pi^2\Delta t_I^2}\text{rect}\left(\frac{t-d-\overline{t}_I}{\Delta t_I}\right)\text{rect}\left(\frac{s-d-\overline{t}_I}{\Delta t_I}\right)\nonumber\\&\times\frac{(t-s)^2-4\epsilon^2}{((t-s)^2+4\epsilon^2)^2}\,;\label{nu BB static}
\end{align}
\begin{align}
    \nu_{AB}(t&,s)=\nu_{BA}(s,t)=-\frac{1}{2\pi^2\Delta t_I}\delta(t-t_I)\text{rect}\left(\frac{s-d-\overline{t}_I}{\Delta t_I}\right)\nonumber\\&\times\frac{(t-s)^2-d^2-4\epsilon^2}{((t-s)^2-d^2-4\epsilon^2)^2+16\epsilon^2(t-s)^2}\,.\label{nu AB static}
\end{align}\normalsize
At this point, to calculate the transmissivity $\tau$ and the additive noise $W$ of the communication channel, we have to solve the homogeneous quantum Langevin equation \eqref{homogeneous quantum langevin equation} to obtain the elements of the Green function matrix. We report here the solution for them, leaving the detailed calculation in the Appendix \ref{appendix Green function matrix}. In particular, for $G_{AA}$ and $G_{AB}$ we have
\begin{equation}\label{GAA static}
    G_{AA}(t,s)=\frac{\sin(\omega_A(t-s))}{\omega_A}\,;
\end{equation}
\begin{equation}
    G_{AB}(t,s)=0\,.
\end{equation}
For $G_{BA}$ and $G_{BB}$ we have to distinguish three ranges of time:
\begin{itemize}
    \item $s<t<d+\overline{t}_I-\Delta t_I/2$, i.e. before detector $B$ interacts with the field;
    \item $d+\overline{t}_I-\Delta t_I/2<t<d+\overline{t}_I+\Delta t_I/2$, i.e. while detector $B$ interacts with the field;
    \item $t>d+\overline{t}_I+\Delta t_I/2$, i.e. after detector $B$ interacts with the field.
\end{itemize}
Before the interaction with the field we have
\begin{equation}\label{GBB before interaction main text}
    G_{BB}(t,s)=\frac{\sin(\omega_B(t-s))}{\omega_B}\,;
\end{equation}
\begin{equation}
    G_{BA}(t,s)=0\,.
\end{equation}
During the interaction\begin{widetext}\small
\begin{align}
    G_{BB}(\tilde{t},\tilde{s})=&\frac{e^{-\frac{B}{2}\tilde{t}}}{\omega_B\sqrt{4A-B^2}}\left((2\omega_B\cos(\omega_B\tilde{s})-B\sin(\omega_B\tilde{s}))\sin\left(\sqrt{A-\frac{B^2}{4}}\tilde{t}\right)\right.\left.-\sqrt{4A-B^2}\sin(\omega_B\tilde{s})\cos\left(\sqrt{A-\frac{B^2}{4}}\tilde{t}\right)\right)\,,\label{solution in the middle GBB main text}
\end{align}\normalsize
\begin{equation}
G_{BA}(\tilde{t},\tilde{s})=\frac{C}{A}\left(\left(1-e^{-\frac{B}{2}\tilde{t}}\cos\left(\frac{\sqrt{4A-B^2}\tilde{t}}{2}\right)\right)+\frac{BC}{A\sqrt{4A-B^2}}e^{-\frac{B}{2}\tilde{t}}\sin\left(\frac{\sqrt{4A-B^2}\tilde{t}}{2}\right)\right)\,,\label{non-simplified solution main text} 
\end{equation}\end{widetext}
where $\tilde{t}=t-(d+\overline{t}_I-\Delta t_I/2)$, $\tilde{s}=s-(d+\overline{t}_I-\Delta t_I/2)$,
\begin{equation}\label{A parameter}
  A=\left(\omega_{B}^2-\frac{\lambda_B^2}{32\pi m_B\Delta t_I^2\epsilon}\right)\, ,
\end{equation}
\begin{equation}
    B=\frac{\lambda_B^2}{32\pi m_B\Delta t_I^2}\,,
\end{equation}
and finally
\begin{equation}\label{C static}
    C=\frac{\lambda_A\lambda_B}{4\pi^2 m_B\Delta t_I}\frac{d}{\epsilon(\epsilon^2+d^2)}G_{AA}(t_I,s)\,.
\end{equation}
After the interaction with the field
\begin{align}
    G_{BB}(\tilde{t},s)&=\frac{\dot{G}_{BB}(\tilde{t}=\Delta t_I^-,s)}{\omega_B}\sin(\omega_B (\tilde{t}-\Delta t_I))\nonumber\\&+G_{BB}(\tilde{t}=\Delta t_I^-,s)\cos(\omega_B (\tilde{t}-\Delta t_I))\,.\label{GBB after interaction main text}
\end{align}
\begin{align}
    G_{BA}(\tilde{t},s)&=\frac{\dot{G}_{BA}(\tilde{t}=\Delta t_I^-,s)}{\omega_A}\sin(\omega_B (\tilde{t}-\Delta t_I))\nonumber\\&+G_{BA}(\tilde{t}=\Delta t_I^-,s)\cos(\omega_B (\tilde{t}-\Delta t_I))\,.\label{GBA after interaction static main text}
\end{align}
At this point, we can compute the parameters $\tau$ and $W$ defined in Sec.~\ref{sec: OMGC}, allowing us to calculate the capacity of the channel. In particular, we are interested in these properties after Bob has interacted with the field i.e. when $t>d+\overline{t}_I+\Delta t_I/2$.

\subsubsection{Additive noise}\label{sssec: static additive noise}
We can compute the noise by studying the determinant of the matrix $\mathbb{N}$ defined in Eq.~\eqref{noise matrix}. We start by analyzing the second term of $\mathbb{N}$, given by the matrix $N_{BB}$. The elements of this matrix, given by Eqs.~\eqref{N11}, \eqref{N12} and \eqref{N22}, could be greatly simplified with the rapid interaction protocol chosen. In fact, the elements of the noise kernel \eqref{nu AA static} and \eqref{nu AB static}, when integrated in Eqs.~\eqref{N11}, \eqref{N12} and \eqref{N22} give terms proportional to $G_{BA}(t,\overline{t}_I)$. However, since $G_{AA}(t,t)=0$ from Eq.~\eqref{GAA static} the parameter $C$ from Eq.~\eqref{C static} vanishes, making $G_{BA}(t,\overline{t}_I)=0$ for each $t$. Then, only the last integrals of Eqs.~\eqref{N11}, \eqref{N12} and \eqref{N22} are non-zero. That is\small
\begin{equation}\label{N11 simplified}
    N_{11}=\frac{1}{m_B^2}\int_s^t\int_s^tG_{BB}(t,r)\nu_{BB}(r,r')G_{BB}(t,r')drdr' \,,
\end{equation}
\begin{equation}\label{N12 simplified}
    N_{12}=\frac{1}{m_B}\int_s^t\int_s^t\dot{G}_{BB}(t,r)\nu_{BB}(r,r')G_{BB}(t,r')drdr' \,,
\end{equation}
\begin{equation}\label{N22 simplified}
    N_{22}=\int_s^t\int_s^t\dot{G}_{BB}(t,r)\nu_{BB}(r,r')\dot{G}_{BB}(t,r')drdr' \,. 
\end{equation}
\normalsize 
Regarding the first term of the noise matrix $\mathbb{N}$ in Eq.~\eqref{noise matrix}, in order to minimize the noise, we suppose Bob to prepare his oscillator in its ground state. Namely
\begin{equation}\label{Bob's ground state}
    \sigma_{BB}(s)=\frac{1}{2}\text{diag}\left((m_B\omega_B)^{-1},m_B\omega_B\right)\,.
\end{equation}
Looking at Eqs.~\eqref{N11 simplified}, \eqref{N12 simplified}, \eqref{N22 simplified} and \eqref{Bob's ground state}, we notice that the noise matrix $\mathbb{N}$ from Eq.~\eqref{noise matrix} is dependent exclusively on $G_{BB}$.
\begin{figure}
    \centering
    \includegraphics[scale=0.6]{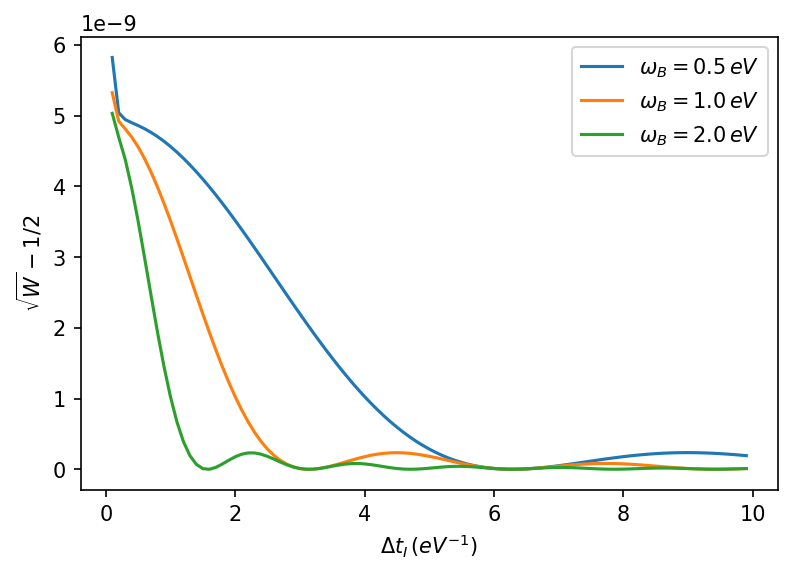}
    \caption{Plot of the parameter $\sqrt{W}-\frac{1}{2}$, quantifying the noise of the channel, in terms of the width of the window during which Bob's detector interacts with the field $\Delta t_I$, for different frequencies of Bob's detector. The values of the parameters are chosen as $\lambda_B=1$, $\epsilon=10^{-3} \,eV^{-1}$, $d=10^{5}\,eV^{-1}$ and $m_B=10^9\,eV$.}
    \label{fig:noise}
\end{figure}

We can numerically evaluate $W=\det \mathbb{N}$ from Eq.~\eqref{noise matrix}, obtaining results as in Fig.~\ref{fig:noise}. This figure shows that, apart from oscillations - due to the presence of oscillating functions in Eqs.~\eqref{GBB after interaction main text} - the noise decreases by increasing the time during which Bob interacts with the field $\Delta t_I$. In particular, numerical analyses 
have shown that the noise is maximized in the limit $\Delta t_I\to0$. This limit corresponds to the one obtained in case the interaction of Alice and Bob with the field is delta-like. This case is studied in detail in Appendix \ref{appendix: delta-like} as a limit case of the protocol described in this section, with the result
\begin{equation}\label{noise delta like main text}
    W=\frac{1}{4}+\frac{\lambda_B^2}{16\pi^2\epsilon^2\omega_Bm_B}\,.
\end{equation}
When $\Delta t$ is finite, we can consider the noise in Eq.~\eqref{noise delta like main text} as an upper bound for $W$.

Moreover, as we can see from Fig.~\ref{fig:noise}, $\sqrt{W}$ has the lower bound $1/2$, given by the first term of the matrix $\mathbb{N}$ from Eq.~\eqref{noise matrix}. This term indicates the initial state of Bob's detector, giving a noisy contribution of $1/4$ on the determinant of $\mathbb{N}$ - this contribution increases by choosing an initial state of Bob's oscillator different than the vacuum.  

The condition \eqref{energy condition} for the energy of the detector must be valid both before and after the interaction. In particular, after the interaction, each detector has absorbed energy from the field. We can calculate this energy by studying the evolution of Alice and Bob's subsystem states, represented respectively by $\sigma_{AA}(t)$ and $\sigma_{BB}(t)$. From Eq.~\eqref{complete cov matrix transformation phys basis}, the evolution of Alice's detector, in Bob's frame, reads
\begin{equation}\label{Alice's state evolution from Bob}
    \sigma_{AA}(t)=T_{AA}\sigma_{AA}(s)T_{AA}^T+N_{AA}\,.
\end{equation}
The matrix $T_{AA}$, from Eqs.~\eqref{complete transmission matrix} and \eqref{new transmission matrix}, is in general
\begin{equation}\label{TAA}
    T_{AA}=\left(\begin{matrix}
        \dot{G}_{AA}(t,s)&\frac{\dot{t}_A(s)}{m_A}G_{AA}(t,s)\\\frac{m_A}{\dot{t}_A(t)}\ddot{G}_{AA}(t,s)&\frac{\dot{t}_A(s)}{\dot{t}_A(t)}\dot{G}_{AA}(t,s)
    \end{matrix}\right)\,.
\end{equation}
In the static case, Eq.~\eqref{TAA} becomes
\begin{equation}\label{Alice to Alice transformation static case}
    T_{AA}=\left(\begin{matrix}
        \dot{G}_{AA}(t,s)&\frac{G_{AA}(t,s)}{m_A}\\
        m_A\ddot{G}_{AA}(t,s)&\dot{G}_{AA}(t,s)
    \end{matrix}\right)\,,
\end{equation}
and $N_{AA}$ is equal to $N_{BB}$ up to an exchange of the indices $A$ and $B$.

Since Alice and Bob's frames coincide, the channel \eqref{Alice's state evolution from Bob} also described the evolution of Alice's state in her own frame, which is what we have to analyze to bound the energy of Alice's detector. Recall that the interaction between Alice's detector and the field is delta-like (Eq.~\eqref{alice time smearing}), so we can use the results in appendix \ref{appendix: delta-like} to compute the elements of $N_{AA}$. Namely, we can use Eqs.~\eqref{N11 delta-like}, \eqref{N12 delta-like} and \eqref{N22 delta like} and replace the index $B$ with the index $A$. At this point, we can study the energy that Alice's detector gains by interacting with the field. Using Eq.~\eqref{oscillator energy},
\begin{equation}\label{Alice noise energy}
    \langle E_A(t=t_I^+)\rangle-\langle E_A(t=t_I^-)\rangle=\frac{\lambda_A^2}{16\pi^2\epsilon^2m_A}\,.
\end{equation}
Hence, to prevent the final energy of the detector $A$ to overcome the ultraviolet cutoff $\epsilon^{-1}$, we have to impose
\begin{equation}\label{Alice noise condition}
    \lambda_A^2\ll m_A\epsilon\,.
\end{equation}
The noise received by Bob is bounded from above by that which he would receive with a delta-like interaction \eqref{noise delta like}. Then, also the energy that Bob's detector absorbs from the interaction is bounded from above by that which it would absorb in the delta-like interaction case. The latter is computed again by using Eqs.~\eqref{N11 delta-like}, \eqref{N12 delta-like} and \eqref{N22 delta like} along with Eq.~\eqref{oscillator energy}, giving the left hand side of Eq.~\eqref{Alice noise energy} with the label $A$ replaced by $B$. Thus, to prevent Bob's oscillator from overcoming the cutoff $\epsilon^{-1}$, we must impose
\begin{equation}\label{Bob noise condition}
    \lambda_B^2\ll m_B\epsilon\,.
\end{equation}
It is worth remarking that the condition \eqref{Bob noise condition} is sufficient, but not necessary, to prevent Bob's oscillator energy to increase too much after the interaction. That is because to find the condition \eqref{Bob noise condition}, we considered an upper bound for the absorbed energy instead of its actual value. On the contrary, the condition \eqref{Alice noise condition}, preventing the same problem for Alice's detector, is both necessary and sufficient, since the absorbed energy was computed exactly in Eq.~\eqref{Alice noise energy}.

To analyze the magnitude of the upper bound of the noise in Eq.~\eqref{noise delta like main text} (called $\overline{W}$ from now on) we rewrite it as
\begin{equation}\label{noise upper bound}
    \overline{W}=\frac{1}{4}+\frac{\lambda_B^2}{16\pi^2\epsilon m_B}\frac{1}{\epsilon \omega_B}\,.
\end{equation}
From Eq.~\eqref{noise upper bound}, we notice that the condition \eqref{Bob noise condition} does not prevent the noise $\overline{W}$ from becoming large, since $\omega_B\epsilon\ll1$. In particular, the upper bound of the noise can increase arbitrarily high by decreasing $\omega_B$. From Fig.~\ref{fig:noise}, this fact seems to be true (apart from oscillations) also for the noise $W$, 
which increases as $\omega_B$ is reduced.

\subsubsection{Transmissivity}\label{sssec: transmissivity static}
\begin{figure}
    \centering
    \includegraphics[scale=0.65]{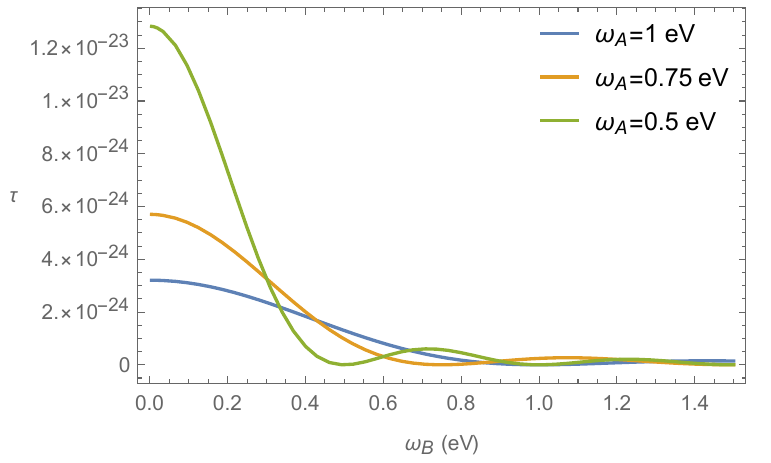}
    \caption{Plot of the transmissivity $\tau$ from Eq.~\eqref{tau realistic case simplified} in terms of the energy gap of the detector $B$, for different values of the energy gap of the detector $A$. The parameters used are $\lambda_A=\lambda_B=1$, $\epsilon=10^{-3}\,eV^{-1}$, $d=10^{5}\,eV^{-1}$ and $m_A=m_B=10^9\,eV$.}
    \label{fig:transmissivity over frequency}
\end{figure}

After the interaction time, the transmissivity $\tau=\det\mathbb{T}=\det T_{BA}$ can be computed by using Eq.~\eqref{transmissivity matrix} and the expression \eqref{GBA after interaction static main text} for $G_{BA}(t,s)$, obtaining
\begin{equation}\label{transmissivity after interaction general}
    \tau(\tilde{t}>\Delta t_I)=\frac{m_B}{m_A}\left(\dot{G}_{BA}^2(\Delta t_I,\tilde{s})+\omega_B^2G_{BA}^2(\Delta t_I,\tilde{s})\right)\,.
\end{equation}
The function $G_{BA}(\Delta t,s)$ must be computed through Eq.~\eqref{non-simplified solution main text}. By employing the condition \eqref{Bob noise condition}, one can simplify $e^{-\frac{B}{2}\tilde{t}}\sim1$, since 
\begin{equation}
    \frac{B}{2}\tilde{t}<\frac{B}{2}\Delta t_I=\frac{\lambda_B^2}{32\pi m_B\Delta t_I}\ll\frac{\lambda_B^2}{32\pi m_B\epsilon}\ll1\,.
\end{equation}
Moreover, with the same argument, we have
\begin{equation}
    4A-B^2=4\omega_B^2-\frac{\lambda_B^2}{8\pi m_B\Delta t_I^2\epsilon}\left(1+\frac{\lambda_B^2\epsilon}{128\pi m_B\Delta t_I^2}\right)\sim 4A\,,
\end{equation}
In this way the solution \eqref{non-simplified solution main text} is simplified to
\begin{equation}\label{very simplified solution main text}
    G_{BA}(\tilde{t},s)=\frac{C}{A}\left(1-\cos(\sqrt{A}\tilde{t})\right)+\frac{CB}{2A^{3/2}}\sin(\sqrt{A}\tilde{t})\,.
\end{equation}
We can finally write $\tau$ in Eq.~\eqref{transmissivity after interaction general} explicitly. Exploiting Eq.~\eqref{very simplified solution main text} for $G_{BA}$ during the interaction and taking an average over the values of $t_I^A$, we get\begin{widetext}
\begin{equation}
    \tau\sim\frac{\lambda_A^2\lambda_B^2}{128\pi^6m_Am_B\epsilon^2d^2}\frac{1}{\omega_B^2-\frac{\lambda_B^2}{128\pi^3m_B\epsilon}\omega_A^2}\left(\frac{\left(1-\cos\left(2\pi\sqrt{\frac{\omega_B^2}{\omega_A^2}-\frac{\lambda_B^2}{128\pi^3m_B\epsilon}}\right)\right)^2}{1-\frac{\lambda_B^2}{128\pi^3m_B\epsilon}\frac{\omega_A^2}{\omega_B^2}}+\sin^2\left(2\pi\sqrt{\frac{\omega_B^2}{\omega_A^2}-\frac{\lambda_B^2}{128\pi^3m_B\epsilon}}\right)\right)\label{tau realistic case simplified}
\end{equation}\end{widetext}
The transmissivity $\tau$ in Eq.~\eqref{tau realistic case simplified} is plotted in Fig.~\ref{fig:transmissivity over frequency}. The figure shows that - apart from oscillations due to trigonometric functions in Eq.~\eqref{tau realistic case simplified} - $\tau$ drops to zero as $\omega_B$ is increased. In particular, in the range $\omega_B\gg\sqrt{\frac{\lambda_B^2}{128\pi^3m_B\epsilon}}\omega_A$, Eq.~\eqref{tau realistic case simplified} can be simplified to
\begin{equation}\label{definitive tau non-accelerating}
    \tau\sim\frac{\lambda_A^2\lambda_B^2}{16\pi^4m_Am_B}\frac{1}{\epsilon^2d^2}\frac{1}{8\pi^2\omega_B^2}\left(2-2\cos\left(2\pi\frac{\omega_B}{\omega_A}\right)\right)\,.
\end{equation}
The maximum value of $\tau$ occurs in the limit $\omega_B\to0$ where \normalsize
\begin{align}
    \tau=\tau_{max}&\sim\frac{\lambda_A^2\lambda_B^2}{128\pi^6\epsilon^2m_Am_B}\frac{1}{d^2\omega_A^2}\frac{\sinh^2\left( 2\pi\sqrt{\frac{\lambda_B^2}{128\pi^3 m_B\epsilon}}\right)}{\frac{\lambda_B^2}{128\pi^3 m_B\epsilon}}\nonumber\\&\sim\frac{1}{32\pi^4}\frac{\lambda_A^2}{m_A\epsilon}\frac{\lambda_B^2}{m_B\epsilon}\frac{1}{d^2\omega_A^2}\,,\label{maximum transmissivity static}
\end{align}\normalsize
where, in the last line, we used the condition \eqref{Bob noise condition}.

\subsubsection{Quantum capacity}\label{sssec: quantum capacity static}
The maximum transmissivity the channel can have is given by $\tau_{max}$ in Eq.~\eqref{maximum transmissivity static}. Apart from the first numerical factor $(32\pi^4)^{-1}$, $\tau_{max}$ is the product of three factors, all much smaller than $1$, by the conditions \eqref{Alice noise condition} and \eqref{Bob noise condition} and \eqref{rapid interaction condition}, respectively. As a consequence $\tau_{max}\ll1$ in the static case. From Eq.~\eqref{maxcoherentinf2}, the maximized coherent information is negative and then the quantum capacity is vanishing from Eq.~\eqref{single letter quantum capacity}. This is consistent with the no-cloning theorem, since it would not be possible to communicate quantum messages to Bob if another observer, within the same distance from Alice, can achieve the same quantum message reliably (see the discussion in Appendix A.1 in Ref.~\cite{Jonsson_2018}).

\subsection{Inertial detectors}\label{ssec: inertial detectors}
We now consider Alice's detector travelling inertially with respect to Bob in a Minkowski vacuum background. In particular, we consider the detector $A$ moving in the same timelike plane as the detector $B$. Alice and Bob's coordinates are related through
\begin{equation}\label{lorentz boost}
    \begin{cases}
        t=\gamma(t_A-\beta x_A)\,;\\
        x=\gamma(x_A-\beta t_A)+d\,;\\
        y=y_A\,;\\
        z=z_A\,.
    \end{cases}
\end{equation}
In Bob's coordinates, the distance of Alice's detector from Bob is $d(t)=d-\beta t$.

Again, Alice interacts via a delta-like switching-in at a time $t_I^A$ with expected value $\overline{t}_I^A$ and uncertainty $\Delta t_I^A=\frac{2\pi}{\omega_A}$. 
To receive Alice's message, Bob interacts with the field at a time centered at $\overline{t}_I^B+d(\overline{t}_I^B)$, following again Eq.~\eqref{bob time smearing}. From Eq.~\eqref{lorentz boost}, the relation between Alice and Bob's proper time is $t_A(t)=t/\gamma$. Then, from Eq.~\eqref{window largeness relation}, the time period in which Bob should interact with the field is
\begin{equation}
    \Delta t_I^B=\gamma\Delta t_I^A=\frac{2\pi\gamma}{\omega_A}\,.
\end{equation}
We now proceed to compute the elements of the dissipation kernel. It is easy to show that, again $\chi_{AA}^A=\chi_{AB}^A=0$\,. For $\chi_{BB}^B$, we have
\begin{align}
    \chi_{BB}^B&(t,s)=\frac{4\lambda_B^2\theta(t-s)}{\pi^2\left(\Delta t_I^B\right)^2}\text{rect}\left(\frac{t-d(\overline{t}_I^B)-\overline{t}_I^B}{\Delta t_I^B}\right)\nonumber\\&\times\text{rect}\left(\frac{s-d(\overline{t}_I^B)-\overline{t}_I^B}{\Delta t_I^B}\right)\frac{\epsilon(t-s)}{((t-s)^2+4\epsilon^2)^2}\,.\label{chiBB inertial}
\end{align}
To study $\chi_{BA}^B$, we first need to study the smeared field operator $\varphi_A^B(t_B)$ from Eq.~\eqref{smeared field external observer}. To this purpose, we consider the spacetime smearing of Alice's detector and we bring it in Bob's coordinates. That is
\begin{align}
    f_A&(t_A,\mathbf{x}_A)=\frac{1}{\pi^2}\frac{\epsilon\delta(t_A-t_I^A)}{\left(\mathbf{x}_A\cdot\mathbf{x}_A+\epsilon^2\right)^2}\nonumber\\&=\frac{1}{\pi^2}\frac{\epsilon\delta(t+\beta x-\beta d-t_I^A/\gamma)}{\gamma\left(\left(\frac{x-d}{\gamma}+\beta t_I^A\right)^2+y^2+z^2+\epsilon^2\right)^2}\,.\label{Alice's smearing function}
\end{align}
Considering the variable $\tilde{x}=\frac{1}{\gamma}(x-d)+\beta t_I^A$, Alice's smearing becomes
\begin{equation}\label{Alice's smearing strange coordinates}
    f_A(t,\tilde{x},y,z)=\frac{1}{\pi^2\gamma}\frac{\epsilon\delta(t+\beta\gamma\tilde{x}-\gamma t_I^A)}{\left(\tilde{x}^2+y^2+z^2+\epsilon^2\right)^2}\,.
\end{equation}
From Eq.~\eqref{Alice's smearing strange coordinates}, $f_A$ is peaked at $\tilde{x}=0$ and drops to zero as $\left(\frac{\epsilon}{\tilde{x}}\right)^4$ outside a neighborhood of $\tilde{x}=0$ with radius $\sim\epsilon$. Since $t_I^A$ has uncertainty $\Delta t_I^A$ and $\Delta t_I^A\gg\epsilon$ we can consider the deviations of $\tilde{x}$ around zero to be negligible w.r.t. the deviations of $t_I^A$ around $\overline{t}_I^A$. For this reason, the argument of the Dirac delta in Eq.~\eqref{Alice's smearing strange coordinates} may be approximated to\footnote{In other words, we can say that the detectors are small enough to consider the relation between $t_A$ and $t$ only in the center of mass of the detectors, ignoring the fact that this relation would change along the detector's profile. This is possible as long as $\epsilon$ is negligible w.r.t. the range of times considered. In our case, $\Delta t_I^A$ represents this range of times, since it is a natural uncertainty on the interaction time.} $\sim t-\gamma t_I^A$. In this way, using Eq.~\eqref{dissipation kernel element from the vacuum}, the dissipation kernel element $\chi_{BA}^B$ becomes
\begin{widetext}
\begin{align}
    \chi_{BA}^B(t,t')=&-2\lambda_A\lambda_B\theta(t-t')\frac{\delta(t'-t_I^B)}{\Delta t_I^B}\text{rect}\left(\frac{t-\overline{t}_I^B-d(\overline{t}_I^B)}{\Delta t_I^B}\right)\nonumber\\&\times\Im\int d\mathbf{k}\frac{e^{i|\mathbf{k}|(t-t')}}{(2\pi)^3|\mathbf{k}|}\int d\mathbf{x}\frac{\epsilon e^{i\mathbf{k}\cdot\mathbf{x}}}{\pi^2\left(x^2+y^2+z^2+\epsilon^2\right)^2}\int d\mathbf{x}'\frac{\epsilon e^{-i\mathbf{k}\cdot\mathbf{x}'}}{\pi^2\gamma\left(\left(\frac{x'-d}{\gamma}+\beta t_I^A\right)^2+y'^2+z'^2+\epsilon^2\right)^2}\,.\label{element dissipation kernel inertial detectors}
\end{align}\end{widetext}
Defining $\mathbf{k}=(k_x,k_y,k_z)$ and $\mathbf{k}'\coloneqq (\gamma k_x,k_y,k_z)$, the integrals in $d\mathbf{x}$ and $d\mathbf{x}'$ in Eq.~\eqref{element dissipation kernel inertial detectors} can be evaluated to be equal to $e^{-\epsilon|\mathbf{k}|}$ and $e^{-\epsilon|\mathbf{k}'|}e^{-ik_xd(t_I^B)}$, respectively. Then, we rewrite Eq.~\eqref{element dissipation kernel inertial detectors} as
\begin{equation}
    \chi_{BA}^B=-2\lambda_A\lambda_B\frac{\delta(t'-t_I^B)}{\Delta t_I^B}\text{rect}\left(\frac{t-\overline{t}_I^B-d(\overline{t}_I^B)}{\Delta t_I^B}\right)\Im I(t)\,,
\end{equation}
where
\begin{equation}
    I(t)=\int\frac{1}{(2\pi)^3|\mathbf{k}|}e^{i|\mathbf{k}|(t-t_I^B)-ik_x d(t_I^B)-\epsilon(|\mathbf{k}|+|\mathbf{k}'|)}d\mathbf{k}\,.
\end{equation}
The integral $I(t)$ does not give an elementary function. However, since $|\mathbf{k}|<|\mathbf{k}'|<\gamma|\mathbf{k}|$ for each $\mathbf{k}$, we can use the mean value theorem to substitute $|\mathbf{k}'|$ with $\mu|\mathbf{k}|$ inside the integral, where $1<\mu<\gamma$. Then, defining $n=\frac{1+\mu}{2}$, we can evaluate $I(t)$ as
\begin{equation}
    I(t)=-\frac{1}{4\pi^2}\frac{1}{(t-t_I^B-2i\epsilon n)^2-d(t_I^B)^2}\,.
\end{equation}
The precise value of $n$ must be computed numerically. However, for the analysis we perform, it is sufficient to know that $n$ is included between $1$ (in the limit $|\mathbf{k}'|\to|\mathbf{k}|$) and $(1+\gamma)/2$ (in the limit $|\mathbf{k}'|\to\gamma|\mathbf{k}|$). 

The dissipation kernel element \eqref{element dissipation kernel inertial detectors} can be finally written as\small
\begin{align}
    &\chi_{BA}^B(t,t')=\frac{4\lambda_A\lambda_B}{\pi^2\Delta t_I^B}\delta(t'-t_I^B)\text{rect}\left(\frac{t-d-\overline{t}_I^B}{\Delta t_I}\right)\nonumber\\&\times\frac{n\epsilon(t-t_I^B)}{\left((t-t_I^B)^2-d(t_I^B)-4n^2\epsilon^2\right)^2+16 n^2\epsilon^2(t-t_I^B)^2}\,.\label{chiBA inertial}
\end{align}\normalsize
We can now compute the elements of the Green function matrix. Again, the calculations are reported in Appendix \ref{appendix Green function matrix}. Computing, we obtain $G_{AB}=0$. For $G_{AA}$ we get
\begin{equation}\label{GAA inertial main text}
    G_{AA}(t,s)=\frac{\gamma}{\omega_A}\sin\left(\frac{\omega_A}{\gamma}(t-s)\right)\,.
\end{equation}
Regarding $G_{BA}$, applying the same approximations performed in the static case, we have $G_{BA}=0$ before the interaction. During the interaction, instead, we have that $G_{BA}$ has the same behaviour of the static case, given by Eq.~\eqref{non-simplified solution main text}, but the parameter $C$ is replaced by
\begin{equation}\label{C inertial main text}
    C'=\frac{\lambda_A\lambda_B}{4\pi^2m_B\Delta t_I^B}\frac{d(\overline{t}_I^B)}{n\epsilon(n^2\epsilon^2+d(\overline{t}_I^B)^2)}G_{AA}(t_I^B,s)\,.
\end{equation}
After the interaction, $G_{BA}(t,s)$ can be expressed again as Eq.~\eqref{GBA after interaction static main text}. Finally, since $\chi_{BB}^B$ is the same of the one obtained in the static case \eqref{chiBB inertial}, $G_{BB}$ is given by Eqs.~\eqref{GBB before interaction main text}, \eqref{solution in the middle GBB main text} and \eqref{GBB after interaction main text} respectively before, during and after Bob's detector interaction with the field.

\subsubsection{Additive noise}
The elements of the noise kernel \eqref{noise kernel} can be easily computed as
\begin{equation}
    \nu_{AA}(t,s)=\lambda_A^2\frac{\delta(t-t_I^B)\delta(s-t_I^B)}{8\pi^2\gamma^2\epsilon^2}\,;\label{nu AA inertial}
\end{equation}
\begin{align}
    \nu_{BB}&(t,s)=-\frac{1}{2\pi^2\Delta t_I^2}\text{rect}\left(\frac{t-d(\overline{t}_I^B)-\overline{t}_I^B}{\Delta t_I^B}\right)\nonumber\\&\times\text{rect}\left(\frac{s-d(\overline{t}_I)-\overline{t}_I}{\Delta t_I^B}\right)\frac{(t-s)^2-4\epsilon^2}{((t-s)^2+4\epsilon^2)^2}\,;\label{nu BB inertial}
\end{align}
\begin{align}
    \nu_{AB}(t&,s)=\nu_{BA}(s,t)=-\frac{\delta(t-t_I^B)}{2\pi^2\gamma\Delta t_I}\text{rect}\left(\frac{s-d(\overline{t}_I)-\overline{t}_I}{\Delta t_I}\right)\nonumber\\&\times\frac{(t-s)^2-d(s)^2-4\epsilon^2}{((t-s)^2-d(s)^2-4\epsilon^2)^2+16\epsilon^2(t-s)^2}\,.\label{nu AB inertial}
\end{align}
Applying the same reasoning as in the static case (Sec.~\ref{sssec: static additive noise}), the elements of the noise matrix $\mathbb{N}$ simplify also here, so that for the elements of $N_{BB}$ are given by 
Eqs.~\eqref{N11 simplified}, \eqref{N12 simplified} and \eqref{N22 simplified}.

Since also $G_{BB}$ is the same of the one we had in the static case, we conclude that the additive noise received by Bob is exactly the same we computed in Sec.~\ref{sssec: static additive noise} and shown in Fig.~\ref{fig:noise}. As a consequence, the condition \eqref{Bob noise condition} is still sufficient to ensure that energy absorbed by the detectors does not overcome the detectors' cutoff.

\subsubsection{Transmissivity}\label{sssec: transmissivity inertial}
As done in Sec.~\ref{sssec: transmissivity static} we can simplify $G_{BA}(t,s)$ during the interaction as $4A-B^2\sim4A$ and $e^{-B/2\Delta t_I^B}\sim1$, obtaining Eq.~\eqref{very simplified solution main text} with $C'$ from Eq.~\eqref{C inertial main text} replacing $C$. Then, we obtain $\tau$ as done in Sec.~\ref{sssec: transmissivity inertial} considering $\Delta t_I^B=\frac{2\pi\gamma}{\omega_A}$ instead of $\Delta t_B=\frac{2\pi}{\omega_A}$ and $C'$ instead of $C$. By averaging over the possible values of $t_I^B$, we get
\begin{widetext}\footnotesize
\begin{align}
    \tau\sim \frac{\lambda_A^2\lambda_B^2}{128\pi^6m_Am_B\gamma}\left(\frac{d(\overline{t}_I^B)}{n\epsilon(n^2\epsilon^2+d(\overline{t}_I^B)^2)}\right)^2\frac{1}{\omega_B^2-\frac{\lambda_B^2\omega_A^2}{128\pi^3 m_B\epsilon\gamma^2}}\left(\frac{\left(1-\cos\left(2\pi\sqrt{\gamma^2\frac{\omega_B^2}{\omega_A^2}-\frac{\lambda_B^2}{128\pi^3m_B\epsilon}}\right)\right)^2}{1-\frac{\lambda_B^2}{128\pi^3m_B\epsilon\gamma^2}\frac{\omega_A^2}{\omega_B^2}}+\sin^2\left(2\pi\sqrt{\gamma^2\frac{\omega_B^2}{\omega_A^2}-\frac{\lambda_B^2}{128\pi^3m_B\epsilon}}\right)\right)\,.\label{transmissivity inertial}
\end{align}\normalsize
\end{widetext}
The transmissivity \eqref{transmissivity inertial} obtained when the detectors travel inertially (called $\tau_I$ from now on), is similar to the static case transmissivity \eqref{transmissivity after interaction general} (which we call $\tau_S$) apart from: the redshift of Alice's detector energy gap (namely $\omega_A$ becomes $\omega_A/\gamma$ in $\tau_I$); the factor $\left(\frac{d(\overline{t}_I^B)}{n\epsilon(n^2\epsilon^2+d(\overline{t}_I^B)^2)}\right)^2$ replacing $(\epsilon d)^{-2}$ in $\tau_I$.

In particular, by fixing the parameter $\eta_A$ equal to $\omega_A$ in the static case and equal to $\omega_A/\gamma$ in the inertial case and by taking the same distance i.e. $d=d(t_I^B)$, we have
\begin{equation}\label{transmissivity comparison}
\tau_I=\frac{\tau_S}{\gamma n}\left(\frac{d^2}{n^2\epsilon^2+d^2}\right)^2\,.    
\end{equation}
As long as $n$ is not very large, we can assume $d\gg n\epsilon$ and then we have
\begin{equation}\label{transmissivity comparison simplified}
    \tau_I\sim\frac{\tau_S}{n\gamma}\,.
\end{equation}
We can conclude that, in case two detectors move inertially w.r.t. each other with a Lorentz factor $\gamma$, the transmissivity decreases by a factor $1/(n\gamma)$ with respect to the static case. The factor $\gamma^{-1}$ comes from the presence of $\dot{t}_A(s)$ in the matrix $\mathbb{T}$, in Eq.~\eqref{transmissivity matrix}. That factor arises from the fact that the state prepared by Alice changes if seen by an external observer. This transformation is reported in Eq.~\eqref{change of coordinates} and it is extensively studied in the appendix \ref{appendix: CP map}.

The factor $n^{-1}$ comes from the fact that, despite both detectors having the same spatial smearing in their own proper frames, their smearings differ in Bob's frame. In particular, from Bob's perspective, Alice's spatial profile is contracted along the $x$ axis. This obviously affects negatively the communication properties.

\subsubsection{Quantum capacity}\label{sssec: quantum capacity inertial}
Since $\tau_I<\tau_S\ll1$, the quantum capacity is zero again from Eqs.~\eqref{single letter quantum capacity} and \eqref{maxcoherentinf2}. 

At the end of Sec.~\ref{sssec: quantum capacity static} (using Ref.~\cite{Jonsson_2018}), we discussed how, in the static case, the geometry of the protocol neglects a priori the possibility of a reliable communication of quantum messages. However, the same argument is not applicable to the pair of inertial detectors. In fact, since the detectors move inertially along the same line, there are no other observers, beside Bob, who can potentially receive the same message. Nevertheless, we showed that this argument is not sufficient to imply the possibility of a reliable quantum communication.

Summarizing, we proved not only that the quantum capacity is still zero in case the detectors travel inertially, but also that their transmissivity decreases compared with the pair of static detectors. Henceforth, since the noise achieved is the same, also the classical capacity is expected to be worse if we make the detectors travel inertially.

\subsection{Accelerating detectors}\label{ssec: accelerating detectors}
We now consider Alice undergoing a Rindler acceleration \cite{Rindler60,Rindler66} along the $x$-axis at $y=z=0$ - while Bob is static at $x=y=z=0$\footnote{The opposite situation, where Bob is Rindler accelerated and Alice is static, is considered in Refs.~\cite{Downes2013,Daiqin2014} to study the communication through Gaussian wavepackets.}. The Fermi normal coordinates of Alice are the Rindler coordinates $(t_A,x_A,y_A,z_A)$, related to Bob's coordinates $(t,x,y,x)$ through
\begin{equation}\label{rindler transformation}
    \begin{cases}
        t=\left(\frac{1}{\alpha}+x_A\right)\sinh(\alpha t_A)\,;\\
        x=\left(\frac{1}{\alpha}+x_A\right)\cosh(\alpha t_A)\,;\\
        y=y_A\,;\\
        z=z_A\,,
    \end{cases}
\end{equation}
where $\alpha$ is Alice's proper acceleration. The world-line of Alice's center of mass is $\left(\frac{1}{\alpha}\sinh(\alpha t_A),\frac{1}{\alpha}\cosh(\alpha t_A),0,0\right)$. Then, the distance between the two detectors, in Bob's frame, is given by
\begin{equation}
    d(t)=\sqrt{\frac{1}{\alpha^2}+t^2}\,.
\end{equation}
So, $1/\alpha$ represents the minimum distance between the two detectors, reached at $t_A=t=0$. To ensure that the detectors are always far from each other we need $\alpha\epsilon\ll1$. From Ref.~\cite{Perche_2022}, the condition $\alpha\epsilon\ll1$ also ensures the validity of the Fermi coordinates inside the detector, allowing to consider Alice's detector as a non-relativistic quantum system. 

We suppose that the field is initially in the Minkowski vacuum. Following the protocol described at the beginning of Sec.~\ref{sec: specific protocol}, Alice's and Bob's switching-in are given by Eqs.~\eqref{alice time smearing} and \eqref{bob time smearing}, respectively. In Bob's proper time, Alice's interaction time $t_I^A$ becomes $t_I^B=\frac{1}{\alpha}\sinh(\alpha t_I^A)$. To ensure that Bob receives Alice's message, the window of his interaction period with the field should be
\begin{equation}
    \Delta t_I^B=\frac{2\pi}{\omega_A \dot{t}_A(\overline{t}_I^B)}=\frac{2\pi\cosh(\overline{t}_I^B)}{\omega_A}\,.
\end{equation}
Regarding the elements of the dissipation kernel, we have again $\chi_{AA}^A=\chi_{AB}^A=0$. The element $\chi_{BB}^B$ is given by Eq.~\eqref{chiBB inertial}, as in the static and inertial case. To compute $\chi_{BA}^B$, we transform Alice's smearing function in Bob's coordinates using Eq.~\eqref{rindler transformation}, namely
\begin{align}
    &f_A(\mathbf{x}_A,t_A)=\frac{\epsilon\delta(t_A-t_I^A)}{(\mathbf{x}_A\cdot\mathbf{x}_A+\epsilon^2)^2}\nonumber\\&=\frac{\alpha x}{\cosh^2(\alpha t_I^A)}\frac{\epsilon\delta(t-x\tanh(\alpha t_I^A))}{\left(\left(\frac{x}{\cosh(\alpha t_I^A)}-\frac{1}{\alpha}\right)^2+y^2+z^2+\epsilon^2\right)^2}\,.\label{Alice smearing accelerating}
\end{align}\normalsize
By using the variable $\tilde{x}=\frac{x}{\cosh(\alpha t_I^A)}-\frac{1}{\alpha}$, Eq.~\eqref{Alice smearing accelerating} becomes
\begin{equation}\label{Alice smearing centered accelerating}
    f_A(\tilde{x},y,z,t)=\frac{1+\alpha\tilde{x}}{\cosh(\alpha t_I^A)}\frac{\epsilon\delta\left(t-\left(\frac{1}{\alpha}+\tilde{x}\right)\sinh(\alpha t_I^A)\right)}{(\tilde{x}^2+y^2+z^2+\epsilon^2)^2}\,.
\end{equation}
Notice that, from Eq.~\eqref{Alice smearing centered accelerating}, $f_A$ becomes negative when $\tilde{x}<-1/\alpha$. This is because, for $\tilde{x}<-1/\alpha$, the detector crosses the Rindler horizon \cite{Rindler66} and then, this region is not observable by Bob. However, from Eq.~\eqref{Alice smearing centered accelerating}, we see that the Lorentzian shape is centered at $\tilde{x}=y=z=0$ and vanishes as $(\epsilon/\tilde{x})^4$ by increasing the magnitude of $\tilde{x}$. Since $\epsilon\ll1/\alpha$, we can conclude that the portion of the Lorentzian shape crossing the Rindler horizon is negligible. Moreover, since $\epsilon\alpha\ll1$, whenever the Lorentzian shape is not-negligible (i.e. a neighborhood of radius $\sim\epsilon$), one can approximate $1+\alpha\tilde{x}\sim1$. In this way, Alice's smearing from Eq.~\eqref{Alice smearing centered accelerating} becomes
\begin{equation}
    f_A(\tilde{x},y,z,t)\sim \frac{1}{\cosh(\alpha t_I^A)}\frac{\epsilon\delta\left(t-t_I^B\right)}{(\tilde{x}^2+y^2+z^2+\epsilon^2)}\,.
\end{equation}
At this point, we can compute the dissipation kernel $\chi_{BA}^B$ in the exact same way we have done in Sec.~\ref{ssec: inertial detectors}, obtaining\small
\begin{equation}
    \chi_{BA}^B=-2\lambda_A\lambda_B\frac{\delta(t'-t_I^B)}{\Delta t_I^B}\text{rect}\left(\frac{t-\overline{t}_I^B-d(\overline{t}_I^B)}{\Delta t_I^B}\right)\Im I_a(t)\,,
\end{equation}\normalsize
where\small
\begin{equation}
    I_a(t)=\int\frac{1}{(2\pi)^3|\mathbf{k}|}e^{i|\mathbf{k}|(t-t_I^B)-ik_x d(t_I^B)-\epsilon(|\mathbf{k}|+|\mathbf{k}''|)}d\mathbf{k}\,,
\end{equation}\normalsize
and $\mathbf{k}''=(\cosh(\alpha t_I^A)k_x,k_y,k_z)$. Similarly to what we did in Sec.~\ref{ssec: inertial detectors} to evaluate Eq.~\eqref{chiBA inertial}, by using the fact that $|\mathbf{k}|<|\mathbf{k}''|<\cosh(\alpha t_I^A)|\mathbf{k}|$, we can write\small
\begin{align}
    &\chi_{BA}(t,t')=\frac{4\lambda_A\lambda_B}{\pi^2\Delta t_I^B}\delta(t'-t_I^B)\text{rect}\left(\frac{t-d-\overline{t}_I^B}{\Delta t_I}\right)\nonumber\\&\times\frac{n'\epsilon(t-t_I^B)}{\left((t-t_I^B)^2-d(t_I^B)-4n'^2\epsilon^2\right)^2+16 n'^2\epsilon^2(t-t_I^B)^2}\,,\label{cross dissipation kernel accelerating}
\end{align}\normalsize
where $n'$ is between $1$ and\footnote{To be more precise, since $t_I^A$ is a random variable bounded in the interval $\left[\overline{t}_I^A+\Delta t_I^A/2,\overline{t}_I^A-\Delta t_I^A/2\right]$, the maximum value for $n'$ is $n'=\frac{1}{2}+\frac{1}{2}\cosh\left(\alpha\left(|\overline{t}_I^A|+\frac{\pi}{\omega_A}\right)\right)$. However, since $\Delta t_I^B\ll d$, then $\alpha\ll\omega_A$ and the upper bound of $n'$ becomes $(1+\cosh(\alpha t_I^A))/2+\mathcal{O}(\alpha/\omega_A)$} $(1+\cosh(\alpha t_I^A))/2$. 

Now, we can proceed to compute the properties of the quantum channel. Using Eq.~\eqref{homogeneous quantum langevin equation}, the equation for $G_{AA}$ is, in Bob's proper time
\begin{equation}
    \ddot{G}_{AA}(t,s)+\frac{\alpha^2 t}{1+\alpha^2 t^2}\dot{G}_{AA}(t,s)+\frac{\omega_A^2}{1+\alpha^2t^2}G_{AA}(t,s)=0\,.
\end{equation}
with boundary conditions $\dot{G}_{AA}(t\to s^+,s)=1$ and $G_{AA}(t\to s^+,s)=0$. The solution can be obtained exactly as\small
\begin{equation}\label{GAA accelerating}
    G_{AA}(t,s)=\frac{\sqrt{1+\alpha^2 s^2}}{\omega_A}\sin\left(\frac{\omega_A}{\alpha}(\sinh^{-1}(\alpha t)-\sinh^{-1}(\alpha s))\right)\,.
\end{equation}\normalsize
The equation for $G_{BB}$ is the same used in the static and inertial cases (Secs.~\ref{ssec: static detectors} and \ref{ssec: inertial detectors}), giving the same solution. For $G_{BA}(t,s)$, the differential equation is similar to that for the inertial case (Eq.~\eqref{all times inertial GBA differential equation}) with $n'$ replacing $n$ and considering Eq.~\eqref{GAA accelerating} for $G_{AA}(t_I^B,s)$. In this way, the solution for $G_{BA}(t,s)$ is the same as for inertial detectors up to these substitutions.

\subsubsection{Additive noise}\label{sssec: noise accelerating}
The elements of the noise kernel \eqref{noise kernel} can be easily computed as
\begin{equation}
    \nu_{AA}(t,s)=\lambda_A^2\frac{\delta(t-t_I^B)\delta(s-t_I^B)}{8\pi^2(1+\alpha^2 (t_I^B)^2)\epsilon^2}\,;\label{nu AA accelerating}
\end{equation}
\begin{align}
    \nu_{BB}&(t,s)=-\frac{1}{2\pi^2\Delta t_I^2}\text{rect}\left(\frac{t-d(\overline{t}_I^B)-\overline{t}_I^B}{\Delta t_I^B}\right)\nonumber\\&\times\text{rect}\left(\frac{s-d(\overline{t}_I)-\overline{t}_I}{\Delta t_I^B}\right)\frac{(t-s)^2-4\epsilon^2}{((t-s)^2+4\epsilon^2)^2}\,;\label{nu BB accelerating}
\end{align}
\begin{align}
    \nu_{AB}(t&,s)=\nu_{BA}(s,t)\nonumber\\&=-\frac{\delta(t-t_I^B)}{2\pi^2\sqrt{1+\alpha^2 (t_I^B)^2}\Delta t_I}\text{rect}\left(\frac{s-d(\overline{t}_I)-\overline{t}_I}{\Delta t_I}\right)\nonumber\\&\times\frac{(t-s)^2-d(s)^2-4\epsilon^2}{((t-s)^2-d(s)^2-4\epsilon^2)^2+16\epsilon^2(t-s)^2}\,.\label{nu AB accelerating}
\end{align}
Again, putting Eqs.~\eqref{nu AA accelerating} and \eqref{nu AB accelerating} in Eqs.~\eqref{N11}, \eqref{N12} and \eqref{N22}, the Dirac deltas imply the presence of $G_{BA}(t,t_I^B)$ in the first three integrals of them. However $G_{BA}(t,t_I^B)=0$ since, from Eq.~\eqref{GAA accelerating}, $G_{AA}(t_I^B,t_I^B)=0$. In this way, Eqs.~\eqref{N11}, \eqref{N12} and \eqref{N22} reduce again to Eqs.~\eqref{N11 simplified}, \eqref{N12 simplified} and \eqref{N22 simplified}, respectively. Since both $\nu_{BB}$ and $G_{BB}$ are the same of the static case (Sec.~\ref{ssec: static detectors}) and inertial case (Sec.~\ref{ssec: inertial detectors}), also the noise is the same one computed in Sec.~\eqref{sssec: static additive noise} (and shown in Fig.~\ref{fig:noise}).

\subsubsection{Transmissivity}\label{sssec: transmissivity accelerating}
By applying the same approximations performed in Secs.~\ref{sssec: transmissivity static}, the solution for the transmissivity after Bob interacts with the field, performing an average over the values of $t_I^B$, is\begin{widetext}
\begin{align}
    \tau\sim &\frac{\theta(\overline{t}_I^B-s)\lambda_A^2\lambda_B^2}{128\pi^6m_Am_B \epsilon^2 n'^2}\frac{d^2(\overline{t}_I^B)}{(n'^2\epsilon^2+d^2(\overline{t}_I^B))^2}\frac{\sqrt{1+\alpha^2s^2}}{1+\alpha^2(\overline{t}_I^B)^2}\frac{1}{\omega_B^2-\frac{\lambda_B^2\omega_A^2}{128\pi^3 m_B\epsilon(1+\alpha^2(\overline{t}_I^B)^2)}}\nonumber\\&\times\left(\frac{\left(1-\cos\left(2\pi\sqrt{(1+\alpha^2(\overline{t}_I^B)^2)\frac{\omega_B^2}{\omega_A^2}-\frac{\lambda_B^2}{128\pi^3m_B\epsilon}}\right)\right)^2}{1-\frac{\lambda_B^2}{128\pi^3m_B\epsilon(1+\alpha^2(\overline{t}_I^B)^2)}\frac{\omega_A^2}{\omega_B^2}}+\sin^2\left(2\pi\sqrt{(1+\alpha^2(\overline{t}_I^B)^2)\frac{\omega_B^2}{\omega_A^2}-\frac{\lambda_B^2}{128\pi^3m_B\epsilon}}\right)\right)\,.\label{tau accelerating detectors}
\end{align}\end{widetext}
The behaviour of the transmissivity in Eq.~\eqref{tau accelerating detectors} (called $\tau_A$) in terms of $\omega_A/\omega_B$ is the same as the static case transmissivity \eqref{transmissivity after interaction general} up to a redshift on Alice's energy gap $\omega_A\to\omega_A/(1+\alpha^2(\overline{t}_I^B)^2)$. To compare the transmissivity $\tau_A$ with that in the static case $\tau_S$, we fix $\eta_A=\omega_A \dot{t}_A(\overline{t}_I^B)$, as we did in Sec.~\ref{sssec: transmissivity inertial} and we get
\begin{equation}\label{comparison accelerating-static case}
    \tau_A=\frac{\sqrt{1+\alpha^2 s^2}}{n'(1+\alpha^2 (\overline{t}_I^B)^2)}\left(\frac{\epsilon^2+d^2}{n'^2\epsilon^2+d^2}\right)^2\tau_S\,.
\end{equation}
From Eq.~\eqref{comparison accelerating-static case}, we see that the lowest possible value for $n'$, i.e. $1$, gives an upper bound of $\tau_A$. A lower bound for $\tau_A$ is given if $n'$ assumes its highest value $n_{max}=(1+\cosh(\alpha t_I^A))/2$.

The relation between $\tau_A$ and $\tau_S$, from Eq.~\eqref{comparison accelerating-static case}, is strongly dependent on $s$ and $\overline{t}_I^B$, being respectively: the time $s$ when both Alice and Bob prepare their initial state (since Bob's initial state is set to be the ground state of the oscillator, we refer to $s$ as the time when just Alice prepares her detector's state); the average time $\overline{t}_I^B$ when Alice interacts with the field to send her state to Bob. 

From causality, we have that both $\tau_S$ and $\tau_A$ are zero when $s>\overline{t}_I^B$. Then, depending on the value of $\overline{t}_I^B$, we can find values of $s$ such that $\tau_A<\tau_S$ or $\tau_A>\tau_S$. By using the fact that $\tau_A$ is maximized when $n'=1$ and minimized when $n'=n_{max}$, we define two time parameters $s_-$ and $s_+$, as
\begin{equation}
    s_-\coloneqq|\overline{t}_I^B|\sqrt{2+\alpha^2(\overline{t}_B^I)^2}\,;
\end{equation}
\begin{equation}
        s_+\coloneqq\frac{1}{\alpha}\sqrt{n'^2_{max}(1+\alpha^2(\overline{t}_I^B)^2)^2\left(\frac{n'^2_{max}\epsilon^2+d^2}{d^2}\right)^2-1}\,.
\end{equation}
In this way,
\begin{enumerate}
    \item if $|s|<s_-$, then $\tau_A<\tau_S$;
    \item if $|s|>s_+$, then $\tau_A>\tau_S$;
    \item if $s_-<|s|<s_+$, we need numerical calculations in order to evaluate $n'$ exactly to compare $\tau_A$ and $\tau_S$. 
\end{enumerate}
Condition $1.$ is always satisfied if $s\ge0$, so that, to improve the transmissivity of the channel w.r.t. the static case, Alice has to prepare her initial state before the time when they are at their minimum distance $1/\alpha$. 

On the contrary, for each value of $\overline{t}_I^B$, Alice can prepare her initial state with enough advance (increasing $-s$) so that the condition $2.$ is satisfied, increasing the transmissivity of the protocol where Alice accelerates. From Eq.~\eqref{comparison accelerating-static case}, $\tau_A$ can reach an arbitrarily high value by increasing $-s$, i.e. the earlier Alice decides to prepare her initial state w.r.t. the transmission of the signal $\overline{t}_I^B$. To explain why this happens, we study the evolution of Alice's state, in Bob's frame, from the time of its preparation $s$ to the time where it is sent to Bob $\overline{t}_I^B$. 

To do that, we use Eq.~\eqref{Alice's state evolution from Bob} where the matrix $T_{AA}$ is given by Eq.~\eqref{TAA} using the Green function matrix element \eqref{GAA accelerating}. The determinant of the matrix $T_{AA}$ could be seen as a transmissivity from Alice to herself, in Bob's coordinates, reading
\begin{align}
    \det T_{AA}=&\sqrt{\frac{1+\alpha^2 s^2}{1+\alpha^2(\overline{t}_I^B)^2}}+\sqrt{1+\alpha^2s^2}\frac{\alpha}{\omega_A}\frac{\alpha \overline{t}_I^B}{1+\alpha^2(\overline{t}_I^B)^2}\nonumber\\&\times\sin\left(2\omega_A(t_A(\overline{t}_I^B)-t_A(s))\right)\,.\label{Alice-alice transmissivity accelerating}
\end{align}
We can neglect the second term on the r.h.s. of Eq.~\eqref{Alice-alice transmissivity accelerating} since $\alpha\ll\omega_A$. In this way, $\det T_{AA}\sim\sqrt{\frac{1+\alpha s^2}{1+\alpha^2 t^2}}$. Then, if $|s|>|\overline{t}_I^B|$, Alice's state in Bob's frame gets amplified by a factor $\sqrt{\frac{1+\alpha s^2}{1+\alpha^2 t^2}}$ before it interacts with the field. Conversely, if $|\overline{t}_I^B|>|s|$, the input state of Alice is damped by $\tau_A$ before the interaction with the field. 
\begin{figure}
    \centering
    \includegraphics[scale=0.8]{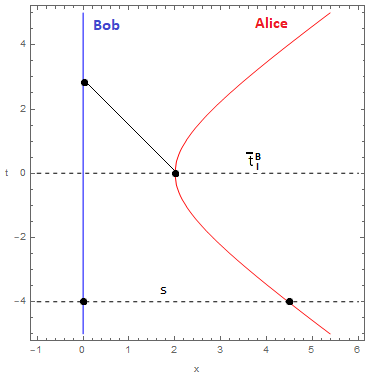}
    \caption{This image outlines the protocol used in Sec.~\ref{ssec: accelerating detectors} in Bob's coordinates, with $s=-4\,eV^{-1}$, $\overline{t}_I^B=0\,eV^{-1}$ and $\alpha=0.5\,eV$.}
    \label{fig:accelerating protocol}
\end{figure}
Henceforth, in Bob's perspective, Alice's initial state could be amplified arbitrarily, eventually overcoming the huge loss comparable to $\tau_S$ from Eq.~\eqref{tau realistic case simplified} and occurring when Alice's detector communicates with Bob's through the field.

To improve the transmissivity of the channel, the best case scenario is provided if Alice interacts with the field at the time $\overline{t}_I^B=0$, i.e. when she is at her minimum distance from Bob, as shown in Fig.~\ref{fig:accelerating protocol}. This scenario also provides an approximate analytic solution for the transmissivity, since now $n'\simeq1$. Then, Eq.~\eqref{comparison accelerating-static case} simply becomes
\begin{equation}\label{comparison simplified}
    \tau_A\sim\sqrt{1+\alpha^2s^2}\tau_S\,.
\end{equation}

\subsubsection{Quantum capacity}\label{sssec: quantum capacity accelerating}
Summarizing, in Sec.~\ref{sssec: noise accelerating}, we briefly showed how the additive noise achieved by Bob, in the accelerating case, is exactly the one of the static case (see Fig.~\ref{fig:noise}). Then, in Sec.~\ref{sssec: transmissivity accelerating} we recognized the protocol described in Fig.~\ref{fig:accelerating protocol} as the optimal one to increase the transmissivity $\tau_A$, following Eq.~\eqref{comparison simplified} where $\tau_S$ is given by Eq.~\eqref{transmissivity after interaction general}.

From Eq.~\eqref{comparison simplified}, by increasing $|s|$, it is possible to increase $\tau_A$ so that $\tau_A>1/2$ and potentially have a positive maximized coherent information \eqref{maxcoherentinf2}, leading to a quantum capacity $Q^{(1)}>0$. To analyze this possibility, we take the upper bound of the noise $\overline{W}$, from Eq.~\eqref{noise upper bound}. Since the maximized coherent information \eqref{maxcoherentinf2} decreases with $W$, then $I_c(\tau_A,\overline{W})\le I_c(\tau_A,W)$. In this way, $I_c(\tau_A,\overline{W})>0$ is sufficient to prove that the single letter quantum capacity of the channel is greater than zero.

The average number of noisy particles occurring when $W=\overline{W}$, using Eq.~\eqref{additivenoise} is
\begin{equation}\label{additive noise accelerating case}
    \overline{n}=\frac{\sqrt{\frac{1}{4}+\frac{\lambda_B^2}{16\pi^2\epsilon^2 m_B\omega_B}}}{|1-\tau_A|}-\frac{1}{2}\,.
\end{equation}
At this point, we can take a value of $|s|$ so that $\tau_A$ is sufficiently high to make $\overline{n}=0$. To this end, $\tau_A$ must read 
\begin{equation}\label{best transmissivity}
    \tau_A=\tau^\star_A\coloneqq1+\sqrt{1+\frac{\lambda_B^2}{4\pi^2\epsilon^2m_B\omega_B}}\,,
\end{equation}
leading, from Eq.~\eqref{maxcoherentinf2}, to a maximized coherent information\small
\begin{equation}\label{best coherent information}
I_c(\tau^\star_A,\overline{W})=\log\left(\frac{\tau^\star_A}{\tau^\star_A-1}\right)=\log\left(\frac{1+\sqrt{1+\frac{\lambda_B^2}{4\pi^2\epsilon^2m_B\omega_B}}}{\sqrt{1+\frac{\lambda_B^2}{4\pi^2\epsilon^2m_B\omega_B}}}\right)\,.
\end{equation}\normalsize
Since $I_c$ from Eq.~\eqref{best coherent information} is always positive, we can finally conclude that, with the protocol described in Fig.~\ref{fig:accelerating protocol}, it is possible to choose a value for $|s|$ high enough to have a quantum capacity greater than zero and communicate quantum messages reliably.

\begin{figure}
    \centering
    \includegraphics[scale=0.65]{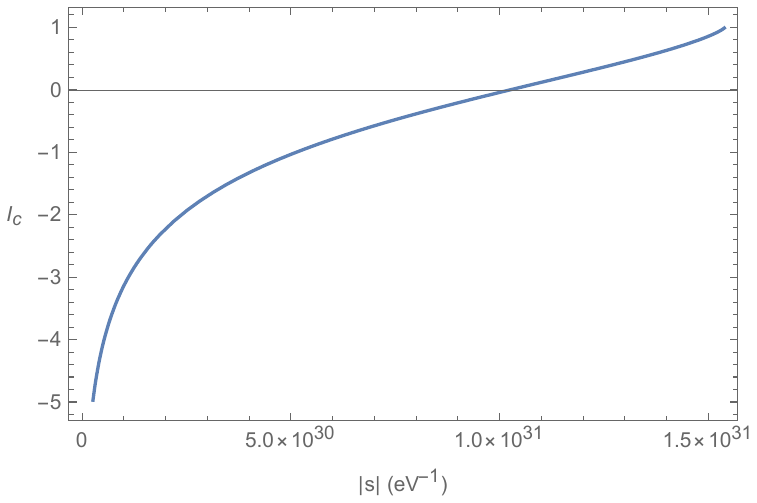}
    \caption{Maximized coherent information $I_c$ of the channel outlined in Fig.~\ref{fig:accelerating protocol} (where $\overline{t}_I^B=0$), obtained from Eq.~\eqref{maxcoherentinf2}, in terms of the time $s$ when Alice prepares her initial state (in Bob's frame). The quantum capacity of the channel corresponds to $I_c$ when $I_c>0$. The parameters chosen are $\lambda_A=\lambda_B=1$, $\omega_A=1\,eV$, $\omega_B=0.5\,eV$, $m_A=m_B=10^9\,eV$, $d=10^{5}\,eV^{-1}$, $\epsilon=10^{-3}\,eV^{-1}$.}
    \label{fig: QCaccelerating}
\end{figure}

Fig.~\ref{fig: QCaccelerating} shows how the maximized coherent information \eqref{maxcoherentinf2} grows by increasing $|s|$ from $0$ to the value $|s^\star|$, defined so that $\tau_A(s^\star)=\tau^\star_A$. 

The value of $I_c(\tau^\star,\overline{W})$ is maximized when $W$ is minimized to $1/4$, so that $\tau_A^\star=2$, and $I_c(2,1/4)=1$, implying that $Q^{(1)}\le1$, in the range $0<|s|<|s^\star|$. From Fig.~\ref{fig:noise}, we see that $W=1/4$ is reachable by choosing suitable values of $\Delta t_I$, and then of the energy gap $\omega_A$, due to the oscillations of $W$.

If we choose $s$ so that $|s|>|s^\star|$, then from Eq.~\eqref{additive noise accelerating case} $\overline{n}$ becomes negative and the channel $\mathcal{N}:\sigma_{in}\mapsto\sigma_{out}$ is no more a complete positive map, albeit Gaussian. We discuss in Appendix \ref{appendix: CP map} why this situation arises. 

Although we have shown that it is theoretically possible to communicate quantum messages reliably with this protocol, a practical realization would be really hard. The value of $|s^\star|$ (time from the preparation of the input to the communication) could be estimated considering $\omega_B=\omega_A/2$, so that
\begin{equation}\label{protocol execution time}
    |s^\star|\simeq 1.538\cdot 10^{4} \cdot \frac{m_Am_B\epsilon^2\omega_A^2}{\lambda_A^2\lambda_B^2\alpha^3}\,.
\end{equation}
The quantities chosen in Fig.~\ref{fig: QCaccelerating} for the parameters $\omega_i$, $m_i$, $\epsilon$, $d$ emulate an atomic scale detector. A quantum capacity greater than zero occurs if $|s|\simeq 10^{31}\,eV^{-1}$, comparable with the age of the universe ($\simeq6.611\cdot10^{32}\,eV^{-1}$). 

Moreover, the minimum distance $d=1/\alpha\simeq10^{5}\,eV^{-1}$ from the two detectors was chosen to have an acceleration $\alpha$ generating an Unruh thermal radiation of $\sim1\,K$, far from being reachable by modern experimental setups. However, it was recently shown how detectors with a circular motion could provide an analog Unruh effect and potentially reach high accelerations in a limited space \cite{Biermann2020,Bunney:2023ipk,D_Bunney_2023}. To this perspective, we notice from Eq.~\eqref{protocol execution time} that the protocol execution time $\sim|s^\star|$ scales as $\alpha^{-3}$. So, if one finds a way to increase the acceleration of a detector, the execution time could drastically decrease to reach reasonable values. 

To decrease $|s^\star|$, the couplings $\lambda_{B}$ could be increased as well. Indeed, the condition \eqref{Bob noise condition} guarantees that Bob's detector cannot absorb energy beyond its limits. For this condition, we considered the upper bound for the additive noise \eqref{noise upper bound}, obtained in the limit $\Delta t_I\to0$. However, as we see from Fig.~\ref{fig:noise}, the noise achieved by Bob could drastically decrease when $\Delta t_I$ is finite. Then, to limit the energy of the detector after the interaction with the field, one can choose a more permissible condition instead of the condition \eqref{Bob noise condition}, giving the possibility to increase $\lambda_{B}$ and reduce the execution time $|s^\star|$.

\section{Final discussions and perspectives}\label{sec: conclusions}
For particle detector models in a $(3+1)D$ spacetime, the possibility to send quantum messages reliably via an isotropic interaction with the field is usually prevented by the no-cloning theorem \cite{Jonsson_2018}. In this paper, we showed that this problem can be circumvented by taking the two detectors in motion with respect to each other.

To do that, we use the method presented in Ref.~\cite{Lapponi_2023} to study the communication of Bosonic signals in non-perturbative regimes. In Secs. \ref{sec: Hamiltonian}, \ref{sec: QLE} and \ref{sec communication protocol}, we generalized this method for whatever spacetime smearing of the detectors and for whatever background spacetime they move in. Although the expressions could be very complicated - implying most of the times non-exact solutions - the freedom on the detectors' smearing could be used to simplify those expressions. In this work, we exploited this possibility to study some protocols involving a rapid interaction between field and detector. However, as a future perspective, particular spacetime smearing for the detectors could allow an analytic study of the communication properties between two detectors with a wider variety of trajectories or background spacetime. Obviously, those smearing functions should satisfy the Fermi bound discussed in Ref.~\cite{Perche_2022}, to ensure that the detectors can be considered non-relativistic quantum systems. 

For example, in appendix \ref{appendix: delta-like}, we see how a delta-like switching-in of the detectors can drastically simplify the properties of the channel and its quantum capacity (defined in Sec.~\ref{sec: OMGC}). Despite this simplicity, we showed that the no-cloning theorem is violated in this case, since a quantum capacity greater than zero is possible also in a Minkowski spacetime, contradicting the geometric argument presented in Ref.~\cite{Jonsson_2018}. In this context, it is interesting to observe how the violation of the no-cloning theorem is related to the violation of the uncertainty principle. Indeed, in Sec.~\ref{sec: specific protocol}, we presented a protocol similar to the one in appendix \ref{appendix: delta-like}, but ensuring that the Heisenberg principle is respected. Although the results are similar, the violation of the no-cloning theorem is prevented in this case, due to an infrared cutoff on the energy gap of Alice's detector.

Then, we considered a static receiver on a Minkowski background. The sender was considered in $3$ different situations: static w.r.t. the receiver (Sec.~\ref{ssec: static detectors}); travelling inertially w.r.t. the receiver (\ref{ssec: inertial detectors}); undergoing a Rindler acceleration (\ref{ssec: accelerating detectors}). The noise received by the receiver (Bob) is always the same, as a consequence of the fact that the motion of Bob is the same and that Alice's interaction with the field is delta-like (even if her interaction time presents an uncertainty).

When the two detectors are static, the transmissivity of the channel is so low that each possibility of reliable quantum communication is prevented (as expected from the no-cloning theorem). This result is comparable numerically to the one obtained in Ref.~\cite{Lapponi_2023} - where the switching-in function is considered to be a Heaviside theta - in case of low coupling. Moreover, in Ref.~\cite{Tjoa_2022}, it is shown how the classical capacity of the channel \cite{Lupo_2011,Pilyavets_2012} - built with a pair of two-level detectors interacting rapidly - increases by increasing the ratio $\lambda_A/\lambda_B$. In our scenario, a similar behaviour is suggested by the fact that the transmissivity \eqref{transmissivity after interaction general} is proportional to $\lambda_A^2\lambda_B^2$ while the upper bound of the noise is $\propto \lambda_B^2$, from Eq.~\eqref{noise upper bound}. Then, the study of the classical capacity of the channel deserves further investigations in the future. If the result of Ref.~\cite{Tjoa_2022} is confirmed, then the increasing of the classical capacity with the ratio $\lambda_A/\lambda_B$ could be a typical feature of the ``rapid interaction" between field and detectors. Indeed, for long interaction periods, the classical capacity drastically drops if we consider detectors with a different coupling with the field \cite{Lapponi_2023}.

By considering two detectors travelling inertially in the same line, the geometric argument for the no-cloning theorem \cite{Jonsson_2018} is no more applicable. However, in Sec.~\ref{ssec: inertial detectors} we proved not only that the quantum capacity is always zero again, but also that the classical capacity decreases with respect to the static case. This can be ascribed to two factors: the length contraction of the detector; the fact that Alice's state is seen by an observer in a different frame.

These results suggest that, to seek for a quantum capacity greater than zero, one has to look for a situation where the two observers are not inertial w.r.t. each other. For this reason, we investigated the case where Alice is Rindler accelerated with respect to Bob (Sec.~\ref{ssec: accelerating detectors}) as shown in Fig.~\ref{fig:accelerating protocol}. In this case, we proved that a non-zero quantum capacity is possible if Alice prepares her state in the remote past w.r.t. the time when Alice and Bob are at their minimum distance. This is due to the strong redshift of the energy gap of Alice, which can circumvent the infrared cutoff arising in the static case. Moreover, one can study the evolution of Alice's state in Bob's frame (Eq.~\eqref{Alice-alice transmissivity accelerating}) to see how Alice's state undergoes an amplification which can compensate the loss occurring during the interaction with the field (the same obtained in the static case \ref{ssec: static detectors}).

In general, a non-inertial detector undergoes quantum effects (e.g. the Unruh effect) as predicted by quantum field theory in curved spacetimes. These effects could play a crucial role on the possibility to achieve a reliable quantum communication. To this prospect, also entanglement harvesting could be pivotal \cite{Reznik_2003,Perche_2023}. Indeed, an arbitrary high entanglement between the detectors always lead to an assisted quantum capacity greater than zero \cite{Landulfo_2016}. Future works would try to see in more detail how quantum effects, given by the detectors' non-inertial motion, could affect the possibility of achieving a quantum capacity greater than zero. 

Notice that, while Alice waits before sending her signal, no entanglement harvesting occurs, since the two detectors do not interact with the field. However, it is interesting to see how the communication scheme behaves as entanglement harvesting occurs. Indeed, if entanglement between the detectors is harvested from the time $-s$ to the time $t_I=0$, it makes sense that the entanglement between the two detectors could eventually assist the communication of quantum messages, making the quantum capacity greater than zero \cite{Landulfo_2016}. 

The analogy with the entanglement harvesting occurs also in the quantitative analysis performed at the end of Sec.~\ref{sssec: quantum capacity accelerating}. The latter indicates that a practical realization of this protocol would require a huge amount of time, making the protocol practically impossible to achieve with today's means. One can relate the difficulty to realize this protocol with the difficulty to see the Unruh effect (or entanglement harvesting) in a laboratory. Indeed, Eq.~\eqref{protocol execution time} shows that the protocol execution time scales as $\alpha^{-3}$. In this way, if we find a way to create a detector achieving a considerable Unruh temperature, the required time could drastically drop to make the protocol realizable. 

The possibility to achieve a non-negligible Unruh temperature is recently being theoretically investigated with detectors undergoing a stationary motion in a finite space, e.g. a circular motion \cite{Biermann2020,Bunney:2023ipk,D_Bunney_2023}. In a future work, we investigate if those stationary motion setups could allow a non-zero quantum capacity as well. Moreover, it's worth investigating what could be the role of a curved background on the communication of quantum messages. To this perspective, it is known that the spacetime curvature decreases the communication capabilities of single-mode signals \cite{Mancini_2014,Good_2021}. We wonder if the same occurs in the communication of Bosonic states through particle detectors.

\section*{Acknowledgements}
A.L. is grateful to the Gravity Laboratory Group of the University of Nottingham for the support and the interesting discussions on the main topic of the work. Among people there, A.L. especially thanks Leo Parry, Adam Wilkinson and Cisco Gooding for their active help to solve the problems faced in this study. 
We thank an anonymous referee for helpful comments. 
The work of J.L. was supported by United Kingdom Research and Innovation Science and Technology Facilities Council [grant number ST/S002227/1]. S.M. acknowledges financial support from Italian Ministry of Universities and Research through ``PNRR MUR project PE0000023-NQSTI". For the purpose of open access, the authors have applied a CC BY public copyright licence to any Author Accepted Manuscript version arising.

\bibliography{main}

\begin{thebibliography}{10}

\bibitem{MannRQI}
R.~Mann and T.~Ralph.
\newblock Relativistic quantum information.
\newblock {\em Classical and Quantum Gravity}, 29(22):220301, November 2012.

\bibitem{ManciniBook}
S.~Mancini and A.~Winter.
\newblock {\em A Quantum Leap in Information Theory}.
\newblock World Scientific, 2020.

\bibitem{Wald_1999}
Robert~M Wald.
\newblock Gravitation, thermodynamics and quantum theory.
\newblock {\em Classical and Quantum Gravity}, 16(12A):A177, December 1999.

\bibitem{Belenchia_2022}
A.~Belenchia et~Al.
\newblock Quantum physics in space.
\newblock {\em Physics Reports}, 951:1--70, March 2022.

\bibitem{KRELINA2023101563}
M.~Krelina.
\newblock The prospect of quantum technologies in space for defence and security.
\newblock {\em Space Policy}, 65:101563, August 2023.

\bibitem{Radchenko_2014}
I.~V. Radchenko, K.~S. Kravtsov, S.~P. Kulik, and S.~N. Molotkov.
\newblock Relativistic quantum cryptography.
\newblock {\em Laser Physics Letters}, 11(6):065203, April 2014.

\bibitem{Birrell:1982ix}
N.~D. Birrell and P.~C.~W. Davies.
\newblock {\em {Quantum Fields in Curved Space}}.
\newblock Cambridge Monographs on Mathematical Physics. Cambridge Univ. Press, Cambridge, UK, February 1984.

\bibitem{Hollands_2015}
S.~Hollands and R.~M. Wald.
\newblock Quantum fields in curved spacetime.
\newblock {\em Physics Reports}, 574:1–35, April 2015.

\bibitem{PCWDavies_1975}
P.~C.~W. Davies.
\newblock Scalar production in {Schwarzschild} and {Rindler} metrics.
\newblock {\em Journal of Physics A: Mathematical and General}, 8(4):609--616, April 1975.

\bibitem{Unruh1976}
W.~G. Unruh.
\newblock Notes on black-hole evaporation.
\newblock {\em Phys.~Rev.~D}, 14(4):870--892, August 1976.

\bibitem{Hawking:1975vcx}
S.~W. Hawking.
\newblock {Particle Creation by Black Holes}.
\newblock {\em Commun.~Math.~Phys.}, 43:199--220, 1975.
\newblock [Erratum: Commun.~Math.~Phys.~46, 206 (1976)].

\bibitem{Ford1987}
L.~H. Ford.
\newblock Gravitational particle creation and inflation.
\newblock {\em Phys.~Rev.~D}, 35(10):2955, May 1987.

\bibitem{Mancini_2014}
S.~Mancini, R.~Pierini, and M.~M. Wilde.
\newblock Preserving information from the beginning to the end of time in a {Robertson}{\textendash}{Walker} spacetime.
\newblock {\em New Journal of Physics}, 16(12):123049, December 2014.

\bibitem{Good_2021}
M.~R.~R. Good, A.~Lapponi, O.~Luongo, and S.~Mancini.
\newblock Quantum communication through a partially reflecting accelerating mirror.
\newblock {\em Phys.~Rev.~D}, 104(10):105020, November 2021.

\bibitem{Unruh1984}
W.~G. Unruh and R.~M. Wald.
\newblock What happens when an accelerating observer detects a {Rindler} particle.
\newblock {\em Phys.~Rev.~D}, 29(6):1047, March 1984.

\bibitem{Hu_2012}
B.~L. Hu, S.~Lin, and J.~Louko.
\newblock Relativistic quantum information in detectors{\textendash}field interactions.
\newblock {\em Classical and Quantum Gravity}, 29(22):224005, October 2012.

\bibitem{1991RSPSA.435..205R}
D.~J. {Raine}, D.~W. {Sciama}, and P.~G. {Grove}.
\newblock {Does a uniformly accelerated quantum oscillator radiate?}
\newblock {\em Proceedings of the Royal Society of London Series A}, 435(1893):205--215, October 1991.

\bibitem{Gibbons1977}
G.~W. Gibbons and S.~W. Hawking.
\newblock Cosmological event horizons, thermodynamics, and particle creation.
\newblock {\em Phys.~Rev.~D}, 15(10):2728, May 1977.

\bibitem{Mart_n_Mart_nez_2014}
N.~C. Martín-Martínez, E.~and~Menicucci.
\newblock Entanglement in curved spacetimes and cosmology.
\newblock {\em Classical and Quantum Gravity}, 31(21):214001, October 2014.

\bibitem{Hodkinson2014}
L.~Hodgkinson, J.~Louko, and A.~C. Ottewill.
\newblock Static detectors and circular-geodesic detectors on the {Schwarzschild} black hole.
\newblock {\em Phys.~Rev.~D}, 89(10):104002, May 2014.

\bibitem{NG2014}
K.~K. Ng, L.~Hodgkinson, J.~Louko, R.~B. Mann, and E.~Mart\'{\i}n-Mart\'{\i}nez.
\newblock Unruh-{DeWitt} detector response along static and circular-geodesic trajectories for {Schwarzschild}--anti-de {Sitter} black holes.
\newblock {\em Phys.~Rev.~D}, 90(6):064003, September 2014.

\bibitem{Bueley_2022}
K.~Bueley, L.~Huang, K.~Gallock-Yoshimura, and R.~B. Mann.
\newblock Harvesting mutual information from {BTZ} black hole spacetime.
\newblock {\em Phys.~Rev.~D}, 106(2):025010, July 2022.

\bibitem{Reznik_2003}
B.~Reznik.
\newblock Entanglement from the vacuum.
\newblock {\em Foundations of Physics}, 33(1):167--176, 2003.

\bibitem{Reznik_2005}
B.~Reznik, A.~Retzker, and J.~Silman.
\newblock Violating {Bell}'s inequalities in vacuum.
\newblock {\em Phys.~Rev.~A}, 71(4):042104, April 2005.

\bibitem{Salton_2015}
G.~Salton, R.~B. Mann, and N.~C. Menicucci.
\newblock Acceleration-assisted entanglement harvesting and rangefinding.
\newblock {\em New Journal of Physics}, 17(3):035001, March 2015.

\bibitem{Cliche2010}
M.~Cliche and A.~Kempf.
\newblock Relativistic quantum channel of communication through field quanta.
\newblock {\em Phys.~Rev.~A}, 81(1):012330, January 2010.

\bibitem{Jonsson_2018}
R.~H. Jonsson, K.~Ried, E.~Martín-Martínez, and A.~Kempf.
\newblock Transmitting qubits through relativistic fields.
\newblock {\em Journal of Physics A: Mathematical and Theoretical}, 51(48):485301, October 2018.

\bibitem{Lapponi_2023}
A.~Lapponi, D.~Moustos, D.~E. Bruschi, and S.~Mancini.
\newblock Relativistic quantum communication between harmonic oscillator detectors.
\newblock {\em Phys.~Rev.~D}, 107(12):125010, June 2023.

\bibitem{Brown_2013}
E.~G. Brown, E.~Martìn-Martìnez, N.~C. Menicucci, and R.~B. Mann.
\newblock Detectors for probing relativistic quantum physics beyond perturbation theory.
\newblock {\em Phys.~Rev.~D}, 87(8):084062, April 2013.

\bibitem{Simidjia2020}
P.~Simidzija, A.~Ahmadzadegan, A.~Kempf, and E.~Mart\'{\i}n-Mart\'{\i}nez.
\newblock Transmission of quantum information through quantum fields.
\newblock {\em Phys.~Rev.~D}, 101(3):036014, February 2020.

\bibitem{Landulfo_2016}
A.~G.~S. Landulfo.
\newblock Nonperturbative approach to relativistic quantum communication channels.
\newblock {\em Phys.~Rev.~D}, 93(10):104019, May 2016.

\bibitem{Tjoa_2022}
E.~Tjoa and K.~Gallock-Yoshimura.
\newblock Channel capacity of relativistic quantum communication with rapid interaction.
\newblock {\em Phys.~Rev.~D}, 105(8):085011, April 2022.

\bibitem{Barcellos2021}
I.~B. Barcellos and A.~G.~S. Landulfo.
\newblock Relativistic quantum communication: Energy cost and channel capacities.
\newblock {\em Phys.~Rev.~D}, 104(10):105018, Nov 2021.

\bibitem{chen2008capacity}
X.~Chen.
\newblock The capacity of transmitting atomic qubit with light.
\newblock {\em Quantum Information Processing}, 9:451--462, November 2008.

\bibitem{Bruschi_2013}
D.~E. Bruschi, A.~R. Lee, and I.~Fuentes.
\newblock Time evolution techniques for detectors in relativistic quantum information.
\newblock {\em Journal of Physics A: Mathematical and Theoretical}, 46(16):165303, April 2013.

\bibitem{Mart_n_Mart_nez_2021}
E.~Martín-Martínez, T.~R. Perche, and B.~S.~L. Torres.
\newblock Broken covariance of particle detector models in relativistic quantum information.
\newblock {\em Phys.~Rev.~D}, 103(2):025007, January 2021.

\bibitem{Perche_2021}
T.~R. Perche and E.~Martín-Martínez.
\newblock Antiparticle detector models in {QFT}.
\newblock {\em Phys.~Rev.~D}, 104(10):105021, November 2021.

\bibitem{PhysRevD.101.045017}
E.~Mart\'{\i}n-Mart\'{\i}nez, T.~R. Perche, and B.~de~S.~L.~Torres.
\newblock General relativistic quantum optics: Finite-size particle detector models in curved spacetimes.
\newblock {\em Phys.~Rev.~D}, 101(4):045017, February 2020.

\bibitem{Perche_2022}
T.~R. Perche.
\newblock Localized nonrelativistic quantum systems in curved spacetimes: A general characterization of particle detector models.
\newblock {\em Phys.~Rev.~D}, 106(2):025018, July 2022.

\bibitem{CALDEIRA1983587}
A.~O. Caldeira and A.~J. Leggett.
\newblock Path integral approach to quantum {Brownian} motion.
\newblock {\em Physica A: Statistical Mechanics and its Applications}, 121(3):587--616, 1983.

\bibitem{Breuer2006}
H.P. Breuer and F.~Petruccione.
\newblock {\em The Theory of Open Quantum Systems}.
\newblock Oxford University Press, 2002.

\bibitem{serafini2017quantum}
A.~Serafini.
\newblock {\em Quantum Continuous Variables: A Primer of Theoretical Methods}.
\newblock CRC Press, Boca Raton, FL, 2017.

\bibitem{Caruso_2006}
F.~Caruso, V.~Giovannetti, and A.~S. Holevo.
\newblock One-mode bosonic {Gaussian} channels: a full weak-degradability classification.
\newblock {\em New Journal of Physics}, 8(12):310, December 2006.

\bibitem{Giovannetti_2014}
V.~Giovannetti, A.~S. Holevo, and R.~García-Patrón.
\newblock A solution of {Gaussian} optimizer conjecture for quantum channels.
\newblock {\em Communications in Mathematical Physics}, 334(3):1553--1571, August 2014.

\bibitem{Holevo2015}
A.~Holevo.
\newblock Gaussian optimizers and the additivity problem in quantum information theory.
\newblock {\em Russian Mathematical Surveys}, 70:331--367, April 2015.

\bibitem{Lupo_2011}
C.~Lupo, S.~Pirandola, P.~Aniello, and S.~Mancini.
\newblock On the classical capacity of quantum {Gaussian} channels.
\newblock {\em Physica Scripta}, T143:014016, February 2011.

\bibitem{Pilyavets_2012}
O.~V. Pilyavets, C.~Lupo, and S.~Mancini.
\newblock Methods for estimating capacities and rates of {Gaussian} quantum channels.
\newblock {\em {IEEE} Transactions on Information Theory}, 58(9):6126--6164, September 2012.

\bibitem{Cubitt_2015}
T.~Cubitt, D.~Elkouss, W.~Matthews, M.~Ozols, D.~Pérez-García, and S.~Strelchuk.
\newblock Unbounded number of channel uses may be required to detect quantum capacity.
\newblock {\em Nature Communications}, 6(1):6739, March 2015.

\bibitem{Oskouei_2018}
S.~K. Oskouei and S.~Mancini.
\newblock Algorithmic complexity of quantum capacity.
\newblock {\em Quantum Information Processing}, 17(4):94, March 2018.

\bibitem{HolevoQC}
A.~Holevo.
\newblock One-mode quantum {Gaussian} channels: Structure and quantum capacity.
\newblock {\em Problems of Information Transmission}, 43:1--11, March 2007.

\bibitem{Br_dler_2015}
K.~Bràdler.
\newblock Coherent information of one-mode {Gaussian} channels{\textemdash}the general case of non-zero added classical noise.
\newblock {\em Journal of Physics A: Mathematical and Theoretical}, 48(12):125301, March 2015.

\bibitem{NoCloning}
V.~Bu\ifmmode~\check{z}\else \v{z}\fi{}ek and M.~Hillery.
\newblock Quantum copying: Beyond the no-cloning theorem.
\newblock {\em Phys.~Rev.~A}, 54(3):1844, September 1996.

\bibitem{Schlicht_2004}
S.~Schlicht.
\newblock Considerations on the {Unruh} effect: causality and regularization.
\newblock {\em Classical and Quantum Gravity}, 21(19):4647, September 2004.

\bibitem{Rindler60}
W.~Rindler.
\newblock Hyperbolic motion in curved space time.
\newblock {\em Phys. Rev.}, 119(6):2082, September 1960.

\bibitem{Rindler66}
W.~Rindler.
\newblock {Kruskal Space and the Uniformly Accelerated Frame}.
\newblock {\em American Journal of Physics}, 34(12):1174--1178, December 1966.

\bibitem{Downes2013}
T.~G. Downes, T.~C. Ralph, and N.~Walk.
\newblock Quantum communication with an accelerated partner.
\newblock {\em Phys.~Rev.~A}, 87(1):012327, January 2013.

\bibitem{Daiqin2014}
D.~Su and T.~C. Ralph.
\newblock Quantum communication in the presence of a horizon.
\newblock {\em Phys.~Rev.~D}, 90(8):084022, October 2014.

\bibitem{Biermann2020}
S.~Biermann, S.~Erne, C.~Gooding, J.~Louko, J.~Schmiedmayer, W.~G. Unruh, and S.~Weinfurtner.
\newblock Unruh and analogue {Unruh} temperatures for circular motion in $3+1$ and $2+1$ dimensions.
\newblock {\em Phys.~Rev.~D}, 102(8):085006, October 2020.

\bibitem{Bunney:2023ipk}
C.~R.~D. Bunney, L.~Parry, T.~R. Perche, and J.~Louko.
\newblock Ambient temperature versus ambient acceleration in the circular motion {Unruh} effect.
\newblock {\em Phys. Rev. D}, 109(6):065001, Mar 2024.

\bibitem{D_Bunney_2023}
C.~R.~D. Bunney and J.~Louko.
\newblock Circular motion analogue {Unruh} effect in a 2+1 thermal bath: robbing from the rich and giving to the poor.
\newblock {\em Classical and Quantum Gravity}, 40(15):155001, June 2023.

\bibitem{Perche_2023}
T.~R. Perche, B.~Ragula, and E.~Martín-Martínez.
\newblock Harvesting entanglement from the gravitational vacuum.
\newblock {\em Phys.~Rev.~D}, 108(8):085025, October 2023.

\end{thebibliography}

\bibliographystyle{unsrt}

\newpage

\onecolumngrid

\appendix

\section{Calculation of the Green function matrix elements}\label{appendix Green function matrix}
In this appendix we report the detailed calculations for the elements of the Green function matrix, obtained from the homogeneous version of the quantum Langevin equation \eqref{vectorial form of langevin equation}, i.e\small
\begin{equation}\label{homogeneous quantum langevin equation}
    \left(\begin{matrix}
        \frac{d^2}{dt_B^2}-\frac{\ddot{t}_A}{\dot{t}_A}\frac{d}{dt_B}+\dot{t}_A^2\omega_A^2&0\\0&\frac{d^2}{dt_B^2}+\omega_B^2
    \end{matrix}\right)\mathbb{G}(t_B,s_B)-\int_{-\infty}^{t_B}\left(\begin{matrix}
        \frac{\dot{t}_A(t_B)^2\dot{t}_A(r_B)}{m_A}&0\\0&\frac{1}{m_B}
    \end{matrix}\right)\left(\begin{matrix}
        \chi_{AA}^A(t_B,r_B)&\chi_{AB}^A(t_B,r_B)\\\chi_{BA}^B(t_B,r_B)&\chi_{BB}^B(t_B,r_B)
    \end{matrix}\right)\mathbb{G}(r_B,s_B)dr_B=\delta(t_B-s_B)\mathbb{I}\,.
\end{equation}\normalsize

When the two detectors are static, Eq.~\eqref{homogeneous quantum langevin equation} becomes
\begin{equation}\label{explicit hom langevin equation}
    \begin{cases}
    \ddot{G}_{AA}(t,s)+\omega_A^2G_{AA}(t,s)-\frac{1}{m_A}\int_{-\infty}^{t}\chi_{AA}(t,r)G_{AA}(r,s)dr-\frac{1}{m_A}\int_{-\infty}^t\chi_{AB}(t,r)G_{BA}(r,s)dr=0\,;\\
    \ddot{G}_{BA}(t,s)+\omega_B^2G_{BA}(t,s)-\frac{1}{m_B}\int_{-\infty}^t\chi_{BA}(t,r)G_{AA}(r,s)dr-\frac{1}{m_B}\int_{-\infty}^t\chi_{BB}(t,r)G_{BA}(r,s)dr=0\,;\\
    \ddot{G}_{AB}(t,s)+\omega_A^2G_{AB}(t,s)-\frac{1}{m_A}\int_{-\infty}^t\chi_{AA}(t,r)G_{AB}(r,s)dr-\frac{1}{m_A}\int_{-\infty}^t\chi_{AB}(t,r)G_{BB}(r,s)dr=0\,;\\
    \ddot{G}_{BB}(t,s)+\omega_B^2G_{BB}(t,s)-\frac{1}{m_B}\int_{-\infty}^t\chi_{BA}(t,r)G_{AB}(r,s)dr-\frac{1}{m_B}\int_{-\infty}^t\chi_{BB}(t,r)G_{BB}(r,s)dr=0\,,
    \end{cases}
\end{equation}
with boundary conditions $G_{ij}(t\to s^+,s)=0$ and $\dot{G}_{ij}(t\to s^+,s)=\delta_{ij}$.

Using the dissipation kernel elements \eqref{chiAA static}, \eqref{chiBB static} and \eqref{chiBA static}, the first and third equations of the system \eqref{explicit hom langevin equation} read respectively
\begin{equation}
    \ddot{G}_{AA}(t,s)+\omega_A^2G_{AA}(t,s)=0\,;
\end{equation}
\begin{equation}
    \ddot{G}_{AB}(t,s)+\omega_A^2G_{AB}(t,s)=0\,.
\end{equation}
The solutions are, respectively,
\begin{equation}\label{GAA static appendix}
    G_{AA}(t,s)=\frac{\sin(\omega_A(t-s))}{\omega_A}\,.
\end{equation}
\begin{equation}
    G_{AB}(t,s)=0\,.
\end{equation}
The fourth of Eq.~\eqref{explicit hom langevin equation} for $G_{BB}(t,s)$ becomes 
\begin{equation}
    \ddot{G}_{BB}(t,s)+\omega_B^2G_{BB}(t,r)-\frac{\lambda_B^2}{4\pi^2 m_B\Delta t_I^2}\text{rect}\left(\frac{t-d-\overline{t}_I}{\Delta t_I}\right)\int_{d+\overline{t}_I-\frac{\Delta t_I}{2}}^{d+\overline{t}_I+\frac{\Delta t_I}{2}}\theta(t-r)\frac{\epsilon(t-r)}{((t-r)^2+4\epsilon^2)^2}G_{BB}(r,s)dr=0\,.\label{GBBdiffeqComplete}
\end{equation}
Before the interaction, i.e. when $s<t<d+\overline{t}_I-\Delta t_I/2$, the solution of $G_{BB}$ is trivially
\begin{equation}\label{GBB before}
    G_{BB}(t-s)=\frac{\sin(\omega_B(t-s))}{\omega_B}\,.
\end{equation}
To find an approximate solution for $G_{BB}(t,s)$ during the interaction, i.e. in the interval $d+\overline{t}_I-\Delta t_I/2<t<d+\overline{t}_I+\Delta t_I/2$, we study the third term on the left-hand side of Eq.~\eqref{GBBdiffeqComplete}. By integrating the integral by parts we get
\begin{align}
    \int_{d+\overline{t}_I-\frac{\Delta t_I}{2}}^{d+\overline{t}_I+\frac{\Delta t_I}{2}}\theta(t-r)\frac{\epsilon(t-r)}{((t-r)^2+4\epsilon^2)^2}G_{BB}(r,s)dr=&\frac{G_{BB}(t,s)}{8\epsilon}-\frac{1}{2}\frac{\epsilon G_{BB}(d+\overline{t}_I-\Delta t_I/2,s)}{4\epsilon^2+\left(t-d-\overline{t}_I+\frac{\Delta t_I}{2}\right)^2}\nonumber\\&-\frac{1}{2}\int_{d+\overline{t}_I-\frac{\Delta t_I}{2}}^{t}\frac{\epsilon}{(t-r)^2+4\epsilon^2}\dot{G}_{BB}(r,s)dr\,.\label{third integral BB}
\end{align}
We start by analyzing the third term on the r.h.s. of Eq.~\eqref{third integral BB}. As long as $t-d-\overline{t}_I+\Delta t_I/2\gg\epsilon$, the integrand can be approximated by the Dirac delta $\frac{1}{2}\delta(t-r)$. Taking into account that the upper bound of the integral lies on the peak of the Dirac delta, we have when $t-d-\overline{t}_I+\Delta t_I/2\gg\epsilon$
\begin{equation}\label{dirac delta in the integral}
    -\frac{1}{2}\int_{d+\overline{t}_I-\frac{\Delta t_I}{2}}^{t}\frac{\epsilon}{(t-r)^2+4\epsilon^2}\dot{G}_{BB}(r,s)dr=-\frac{\pi}{8}\dot{G}_{BB}(t,s)\,.
\end{equation}
When $t-d-\overline{t}_I+\Delta t_I/2\sim\epsilon$, the approximation in Eq.~\eqref{dirac delta in the integral} cannot be performed. However, one can prove that, by increasing $t$ starting from $d+\overline{t}_I-\Delta t_I/2$, the left hand side of Eq.~\eqref{dirac delta in the integral} increases its magnitude from zero until to reach the value $-\frac{\pi}{8}\dot{G}_{BB}(t,s)$ for $t\gtrsim\epsilon$.

The second term on the r.h.s. of Eq.~\eqref{third integral BB} is negligible w.r.t. the first term except for times $t$ such that $t-d-\overline{t}+\Delta t_I/2\sim\epsilon$. Since $\omega_B\epsilon\ll1$, from the condition \eqref{energy condition on the frequency}, the function $G_{BB}$, given Eq.~\eqref{GBB before}, is expected to change by a negligible amount in the interval $\left(d+\overline{t}-\Delta t_I/2-\epsilon,d+\overline{t}-\Delta t_I/2+\epsilon\right)$. Then, we can make the following approximation
\begin{align}
    \frac{G_{BB}(t,s)}{8\epsilon}-\frac{1}{2}\frac{\epsilon G_{BB}(d+\overline{t}_I-\Delta t_I/2,s)}{4\epsilon^2+\left(t-d-\overline{t}_I+\frac{\Delta t_I}{2}\right)^2}&\sim\frac{G_{BB}(t,s)}{8\epsilon}-\frac{1}{2}\frac{\epsilon G_{BB}(t,s)}{4\epsilon^2+\left(t-d-\overline{t}_I+\frac{\Delta t_I}{2}\right)^2}\nonumber\\&=\frac{G_{BB}(t,s)}{8\epsilon}\left(\frac{(t-d-\overline{t}_I+\Delta t_I/2)^2}{4\epsilon^2+(t-d-\overline{t}_I+\Delta t_I/2)^2}\right)\,.\label{Approximating the first two terms}
\end{align}
From Eq.~\eqref{Approximating the first two terms}, the first two terms on the r.h.s. of Eq.~\eqref{third integral BB} give $0$ at the beginning of the interaction time $t=d+\overline{t}_I-\Delta t_I/2$ and grow to $G_{BB}(t,s)/(8\epsilon)$ after a time comparable to $\epsilon$. We can conclude that the integral \eqref{third integral BB}, from being zero at time $t=d+\overline{t}_I-\Delta t_I/2$, increases its magnitude to become, after a time of the order $\epsilon$, the following
\begin{align}
    \int_{d+\overline{t}_I-\frac{\Delta t_I}{2}}^{d+\overline{t}_I+\frac{\Delta t_I}{2}}\theta(t-r)\frac{\epsilon(t-r)}{((t-r)^2+4\epsilon^2)^2}G_{BB}(r,s)dr\sim&\frac{G_{BB}(t,s)}{8\epsilon}-\frac{\pi}{8}\dot{G}_{BB}(r,s)dr\,.\label{third integral BB approximated}
\end{align}
If we have $\Delta t_I\gg\epsilon$, then the range of time where the approximation \eqref{third integral BB approximated} is not valid is very small with respect to the entire interaction time. In other words, if we consider the approximation in Eq.~\eqref{third integral BB approximated} to be valid during the entire interaction time, the error expected on the value of $G_{BB}(t,s)$ after the interaction would be of an order $\mathcal{O}(\epsilon/\Delta t_I)$ and then negligible.

Thus, for our purposes, we can rewrite Eq.~\eqref{GBBdiffeqComplete} as
\begin{equation}\label{GBB complete equation}
    \ddot{G}_{BB}(t,s)+\frac{\lambda_B^2}{32\pi m_B\Delta t_I^2}\text{rect}\left(\frac{t-d-\overline{t}_I}{\Delta t_I}\right)\dot{G}_{BB}(t,s)+\left(\omega_{B}^2-\frac{\lambda_B^2}{32\pi m_B\Delta t_I^2\epsilon}\text{rect}\left(\frac{t-d-\overline{t}_I}{\Delta t_I}\right)\right)G_{BB}(t,s)=0\,.
\end{equation}
The solution of Eq.~\eqref{GBB complete equation} during the interaction ($d+\overline{t}_I-\Delta t_I/2\le t<d+\overline{t}_I+\Delta t_I/2$, i.e. $0<\tilde{t}<\Delta t_I$), by matching $G_{BB}$ and its derivative with the solution \eqref{GBB before} at $t=d+\overline{t}_I-\Delta t_I/2$ is
\begin{align}
    G_{BB}(\tilde{t},\tilde{s})=&\frac{e^{-\frac{B}{2}\tilde{t}}}{\omega_B\sqrt{4A-B^2}}\left((2\omega_B\cos(\omega_B\tilde{s})-B\sin(\omega_B\tilde{s}))\sin\left(\sqrt{A-\frac{B^2}{4}}\tilde{t}\right)\right.\nonumber\\&\left.-\sqrt{4A-B^2}\sin(\omega_B\tilde{s})\cos\left(\sqrt{A-\frac{B^2}{4}}\tilde{t}\right)\right)\,,\label{solution in the middle GBB}
\end{align}
where, for simplicity, we defined $\tilde{t}\coloneqq t-(d+\overline{t}_I-\Delta t_I/2)$, $\tilde{s}\coloneqq s-(d+\overline{t}_I-\Delta t_I/2)$, $A\coloneqq\left(\omega_{B}^2-\frac{\lambda_B^2}{32\pi m_B\Delta t_I^2\epsilon}\right)$ and $B\coloneqq\frac{\lambda_B^2}{32\pi m_B\Delta t_I^2}$. 

After the interaction ($\tilde{t}\ge\Delta t$), Eq.~\eqref{GBBdiffeqComplete} returns the one of a simple harmonic oscillator. By matching $G_{BB}$ and its derivative with the solution \eqref{solution in the middle GBB} at $\tilde{t}=\Delta t_I$, we get
\begin{align}\label{GBB after}
    G_{BB}(\tilde{t},s)=\frac{\dot{G}_{BB}(\tilde{t}=\Delta t_I^-,s)}{\omega_B}\sin(\omega_B (\tilde{t}-\Delta t_I))+G_{BB}(\tilde{t}=\Delta t_I^-,s)\cos(\omega_B (\tilde{t}-\Delta t_I))\,.
\end{align}
Eqs.~\eqref{GBB before}, \eqref{solution in the middle GBB} and \eqref{GBB after} give then the complete solution for $G_{BB}(t,s)$ before, during and after the interaction, respectively.\\

The second equation of the system \eqref{explicit hom langevin equation}, using the dissipation kernel elements \eqref{chiBB static} and \eqref{chiBA static}, reads
\begin{align}
    &\ddot{G}_{BA}(t,s)+\omega_B^2G_{BA}(t,r)-\frac{\lambda_B^2}{4\pi^2 m_B\Delta t_I^2}\text{rect}\left(\frac{t-d-\overline{t}_I}{\Delta t_I}\right)\int_{d+\overline{t}_I-\frac{\Delta t_I}{2}}^{d+\overline{t}_I+\frac{\Delta t_I}{2}}\theta(t-r)\frac{\epsilon(t-r)}{((t-r)^2+4\epsilon^2)^2}G_{BA}(r,s)dr\nonumber\\&=\frac{4\lambda_A\lambda_B}{\pi^2\Delta t_I}\text{rect}\left(\frac{t-d-\overline{t}_I}{\Delta t_I}\right)G_{AA}(t_I,s)\frac{\epsilon(t-t_I)}{\left((t-t_I)^2-d^2-4\epsilon^2\right)^2+16\epsilon^2(t-t_I)^2}\,.\label{GBAdiffeqCorrected}
\end{align}
Regarding the third term on the left-hand side of Eq.~\eqref{GBAdiffeqCorrected} we can simplify it using Eq.~\eqref{third integral BB approximated} - with $G_{BA}$ instead of $G_{BB}$ and using the same argument explained before. Concerning the r.h.s. of Eq.~\eqref{GBAdiffeqCorrected}, we study the factor $\frac{\epsilon(t-t_I)}{\left((t-t_I)^2-d^2-4\epsilon^2\right)^2+16\epsilon^2(t-t_I)^2}$ in the range where $\text{rect}\left(\frac{t-d-\overline{t}_I}{\Delta t_I}\right)\ne0$, i.e. when $d-\overline{t}_I-\Delta t_I/2<t<d-\overline{t}_I+\Delta t_I/2$. Since $|t_I-\overline{t}_I|<\Delta t_I/2$, we can prove that
\begin{equation}\label{approximation r.h.s. term GBA diffeq}
    \frac{\epsilon(t-t_I)}{\left((t-t_I)^2-d^2-4\epsilon^2\right)^2+16\epsilon^2(t-t_I)^2}=\frac{1}{16}\frac{d}{\epsilon(\epsilon^2+d^2)}\left(1+\mathcal{O}\left(\frac{\Delta t_I}{d}\right)\right)\,.
\end{equation}
Then, from the rapid interaction condition \eqref{rapid interaction condition}, the second term of the latter can be neglected. Using these approximations, the equation \eqref{GBAdiffeqCorrected} becomes:
\begin{align}
    &\ddot{G}_{BA}(t,s)+\frac{\lambda_B^2}{32\pi m_B\Delta t_I^2}\text{rect}\left(\frac{t-d-\overline{t}_I}{\Delta t_I}\right)\dot{G}_{BA}(t,s)+\left(\omega_{B}^2-\frac{\lambda_B^2}{32\pi m_B\Delta t_I^2\epsilon}\text{rect}\left(\frac{t-d-\overline{t}_I}{\Delta t_I}\right)\right)G_{BA}(t,s)\nonumber\\&=\frac{\lambda_A\lambda_B}{4\pi^2m_B\Delta t_I}\text{rect}\left(\frac{t-d-\overline{t}_I}{\Delta t_I}\right)\frac{d}{\epsilon(\epsilon^2+d^2)}G_{AA}(t_I,s)\label{all times simplified GBA differential equation}\,.
\end{align}
Before the interaction, i.e. when $s<t<d+\overline{t}_I-\frac{\Delta t_I}{2}$, given the boundary conditions $G_{BA}(t=s,s)=\dot{G}_{BA}(t=s,s)=0$, we have $G_{BA}(t,s)=0$. During the interaction ($d+\overline{t}_I-\Delta t_I/2<t<d+\overline{t}_I+\Delta t_I/2$), Eq.~\eqref{all times simplified GBA differential equation} becomes
\begin{equation}\label{interesting time GBA differential equation}
    \ddot{G}_{BA}(t,s)+\frac{\lambda_B^2}{32\pi m_B\Delta t_I^2}\dot{G}_{BA}(t,s)+\left(\omega_{B}^2-\frac{\lambda_B^2}{32\pi m_B\Delta t_I^2\epsilon}\right)G_{BA}(t,s)=\frac{\lambda_A\lambda_B}{4\pi^2 m_B\Delta t_I}\frac{d}{\epsilon(\epsilon^2+d^2)}G_{AA}(t_I,s)\,.
\end{equation}
Setting
\begin{equation}
    C=\frac{\lambda_A\lambda_B}{4\pi^2 m_B\Delta t_I}\frac{d}{\epsilon(\epsilon^2+d^2)}G_{AA}(t_I,s)\sim\frac{\lambda_A\lambda_B}{4\pi^2 m_B\Delta t_I}\frac{1}{\epsilon d}G_{AA}(t_I,s)\,,
\end{equation}
the solution of Eq.~\eqref{interesting time GBA differential equation} reads
\begin{equation}\label{non-simplified solution}
G_{BA}(\tilde{t},s)=\frac{C}{A}\left(\left(1-e^{-\frac{B}{2}\tilde{t}}\cos\left(\frac{\sqrt{4A-B^2}\tilde{t}}{2}\right)\right)+\frac{BC}{A\sqrt{4A-B^2}}e^{-\frac{B}{2}\tilde{t}}\sin\left(\frac{\sqrt{4A-B^2}\tilde{t}}{2}\right)\right)\,,    
\end{equation}
where $\tilde{t}$, $A$ and $B$ are defined below Eq.~\eqref{solution in the middle GBB}. Finally, when $\tilde{t}>\Delta t_I$, Eq.~\eqref{all times simplified GBA differential equation} becomes $\ddot{G}_{BA}+\omega_B^2G_{BA}=0$ whose solution is 
\begin{equation}\label{GBA after interaction static}
    G_{BA}(\tilde{t},s)=\frac{\dot{G}_{BA}(\tilde{t}=\Delta t_I^-,s)}{\omega_A}\sin(\omega_B (\tilde{t}-\Delta t_I))+G_{BA}(\tilde{t}=\Delta t_I^-,s)\cos(\omega_B (\tilde{t}-\Delta t_I))\,.
\end{equation}
Then, $G_{BA}(t,s)=0$ before the interaction, while during and after it, $G_{BA}$ follows Eqs.~\eqref{non-simplified solution} and \eqref{GBA after interaction static}, respectively.\\

In case the two detectors travel inertially w.r.t. each other (as in Sec.~\ref{ssec: inertial detectors}), for the elements of the dissipation kernel we have $\chi_{AA}^A=\chi_{AB}^A=0$ and $\chi_{BB}^B$ and $\chi_{BA}^B$ given by Eqs.~\eqref{chiBB inertial} and \eqref{chiBA inertial}. Using these, the homogenous Langevin equation \eqref{homogeneous quantum langevin equation} yields the following equations for the elements of the Green function matrix
\begin{equation}\label{eq GAA inertial}
    \ddot{G}_{AA}(t,s)+\frac{\omega_A^2}{\gamma^2}G_{AA}(t,s)=0\,;
\end{equation}
\begin{equation}\label{eq GAB inertial}
    \ddot{G}_{AB}(t,s)+\frac{\omega_A^2}{\gamma^2}G_{AB}(t,s)=0\,;
\end{equation}
\begin{align}
    &\ddot{G}_{BA}(t,s)+\frac{\lambda_B^2}{32\pi m_B(\Delta t_I^B)^2}\text{rect}\left(\frac{t-d(\overline{t}_I^B)-\overline{t}_I^B}{\Delta t_I^B}\right)\dot{G}_{BA}(t,s)+\left(\omega_{B}^2-\frac{\lambda_B^2}{32\pi m_B(\Delta t_I^B)^2\epsilon}\text{rect}\left(\frac{t-d(\overline{t}_I^B)-\overline{t}_I^B}{\Delta t_I^B}\right)\right)G_{BA}(t,s)\nonumber\\&=\frac{\lambda_A\lambda_B}{4\pi^2 m_B\Delta t_I^B}\text{rect}\left(\frac{t-d(\overline{t}_I^B)-\overline{t}_I^B}{\Delta t_I^B}\right)\frac{d(\overline{t}_I^B)}{n\epsilon(n^2\epsilon^2+d(\overline{t}_I^B)^2)}G_{AA}(t_I^B,s)\label{all times inertial GBA differential equation}\,;
\end{align}
while for $G_{BB}$ we have again Eq.~\eqref{GBBdiffeqComplete}. On the r.h.s. of Eq.~\eqref{all times inertial GBA differential equation}, we took $\Delta t_I\ll d$ to approximate $d(t_I^B)\sim d(\overline{t}_I^B)$, i.e., we considered the distance $d(t)$ to change by a negligible amount in the support of $\lambda_B(t)$ from Eq.~\eqref{bob time smearing}.\\

Using the boundary conditions $G_{ij}(t\to s^+,s)=0$ and $\dot{G}_{ij}(t\to s^+,s)=\delta_{ij}$, the solutions of Eqs.~\eqref{eq GAA inertial} and \eqref{eq GAB inertial} are respectively
\begin{equation}\label{GAA inertial}
    G_{AA}(t,s)=\frac{\gamma}{\omega_A}\sin\left(\frac{\omega_A}{\gamma}(t-s)\right)\,;
\end{equation}
\begin{equation}
    G_{AB}(t,s)=0\,.
\end{equation}
The Green function element $G_{BA}(t,s)$ is zero again at times $s<t<d(t_I^B)+\overline{t}_I^B-\Delta t_I^B/2$. In the range $d(t_I^B)+\overline{t}_I^B-\Delta t_I^B/2<t<d(t_I^B)+\overline{t}_I^B+\Delta t_I^B/2$ we have again Eq.~\eqref{non-simplified solution} considering $\tilde{t}=t-(d(t_I^B)+\overline{t}_I^B-\Delta t_I^B/2)$, where $C$ is replaced by
\begin{equation}\label{C inertial}
    C'=\frac{\lambda_A\lambda_B}{4\pi^2m_B\Delta t_I^B}\frac{d(\overline{t}_I^B)}{n\epsilon(n^2\epsilon^2+d(\overline{t}_I^B)^2)}G_{AA}(t_I^B,s)\,.
\end{equation}
The equation for $G_{BB}(t,s)$ is the same as that computed for static detectors, so that also the solution does not change. That is, the solution for $G_{BB}$ before, during and after the interaction is given respectively by Eqs.~\eqref{GBB before}, \eqref{solution in the middle GBB} and \eqref{GBB after}.

In case Alice's detector undergoes a Rindler acceleration and Bob is static, as described in Fig.~\ref{fig:accelerating protocol}, the computation of the Green function matrix is similar to the one performed for inertially travelling detectors. 
This computation is explained directly in the main text (see Sec.~\ref{ssec: accelerating detectors}).

\section{Delta-like interaction}\label{appendix: delta-like}
In this appendix, we study the case where both Alice and and Bob interact with the field with a delta-like interaction. Namely, $\lambda_i(t_i)=\lambda_i\delta(t_i-t_I^i)$. This protocol could be seen as a limit case of the protocol studied in Sec.~\ref{sec: specific protocol}, obtained by neglecting the uncertainty on Alice's interaction time, so that $\Delta t_I\sim0$. 

In the literature this model often provides exact results for the response function of the detector and for the capacities of the communication channel between two detectors \cite{Simidjia2020,Barcellos2021,Tjoa_2022}. However, we show here that, unless we impose an infrared cutoff for the energy gap of Alice's detector $\omega_A$, the delta-like interaction potentially leads to a violation of the no-cloning theorem. Hence, despite its simplicity, this interaction model could be controversial when studying the communication of Bosonic states. 

To show this, we consider the case where the two detectors are static. Taking a distance $d$ between the two, we have $\lambda_A(t)=\lambda_A\delta(t-t_I)$ and $\lambda_B(t)=\lambda_B\delta(t-t_I-d)$. The dissipation and noise kernel elements become (from Eqs.~\eqref{dissipation kernel element from the vacuum} and \eqref{noise kernel} and considering Ref.~\cite{Schlicht_2004} for the Lorentzian smeared Wightman function)
\begin{equation}\label{disskernelAA}
    \chi_{AA}(t,s)=\chi_{BB}(t,s)=\chi_{AB}(t,s)=0\,;
\end{equation}
\begin{equation}\label{disskernelBA}
    \chi_{BA}(t,s)=\frac{\lambda_A\lambda_B}{4\pi^2}\delta(t-t_I-d)\delta(s-t_I)\frac{d}{\epsilon(d^2+\epsilon^2)}\,;
\end{equation}
\begin{equation}\label{noisekernelAA}
    \nu_{AA}(t,s)=\lambda_A^2\frac{\delta(t-t_I)\delta(s-t_I)}{8\pi^2\epsilon^2}\,;
\end{equation}
\begin{equation}\label{noisekernelBB}
    \nu_{BB}(t,s)=\lambda_B^2\frac{\delta(t-t_I-d)\delta(s-t_I-d)}{8\pi^2\epsilon^2}\,;
\end{equation}
\begin{equation}\label{noisekernelAB}
    \nu_{AB}(t,s)=\nu_{BA}(s,t)=\frac{\lambda_A\lambda_B}{16\pi^2}\frac{\delta(t-t_I)\delta(s-t_I-d)}{\epsilon^2+d^2}\,.
\end{equation}
Using Eq.~\eqref{disskernelAA} and \eqref{disskernelBA}, the homogeneous quantum Langevin equation \eqref{explicit hom langevin equation} results:
\begin{equation}\label{hom Langevin eq minkowski delta like}
    \begin{cases}
    \ddot{G}_{AA}(t,s)+\omega_A^2G_{AA}(t,s)=0\,;\\
    \ddot{G}_{BA}(t,s)+\omega_B^2G_{BA}(t,s)=\frac{\lambda_A\lambda_B}{4\pi^2m_B}\frac{d}{\epsilon(d^2+\epsilon^2)}\delta(t-d-t_I)G_{AA}(t_I,s)\,;\\
    \ddot{G}_{AB}(t,s)+\omega_A^2G_{AB}(t,s)=0\,;\\
    \ddot{G}_{BB}(t,s)+\omega_B^2G_{BB}(t,s)=0\,,
    \end{cases}
\end{equation}
with boundary conditions $G_{ij}(t\to s^+,s)=0$ and $\dot{G}_{ij}(t\to s^+,s)=\delta_{ij}$. 

The solution for $G_{ii}$ with $i=A,B$, following Eq.~\eqref{hom Langevin eq minkowski delta like}, is 
\begin{equation}\label{same detector green delta case}
    G_{ii}(t,s)=\frac{\sin(\omega_i(t-s))}{\omega_i}\,.
\end{equation}
The solution for $G_{BA}(t,s)$, following the second equation of the system \eqref{hom Langevin eq minkowski delta like}, is instead
\begin{equation}
    G_{BA}(t,s)=\theta(t-d-t_I)\frac{\sin(\omega_B(t-d-t_I))}{\omega_B}\frac{\lambda_A\lambda_B}{4\pi^2m_B}\frac{d}{\epsilon(d^2+\epsilon^2)}\frac{\sin(\omega_A(t_I-s))}{\omega_A}\,.
\end{equation}
The transmissivity of the channel $\tau$ could be immediately calculated through the determinant of the matrix $\mathbb{T}$ from Eq.~\eqref{transmissivity matrix}, as
\begin{equation}\label{tau delta like}
    \tau=\theta(t-d-t_I)\frac{m_B}{m_A}\left(\dot{G}_{BA}^2-G_{BA}\ddot{G}_{BA}\right)=\theta(t-d-t_I)\frac{\lambda_A^2\lambda_B^2}{16\pi^4m_Am_B}\frac{d^2}{\epsilon^2(d^2+\epsilon^2)^2}\frac{\sin^2(\omega_A(t_I-s))}{\omega_A^2}\,.
\end{equation}
The latter is obviously $0$ when $t<t_I+d$, since the detectors are not causally connected.

Regarding the noise, we proceed to compute the matrix \eqref{noise matrix} by supposing that Bob's detector is prepared in its ground state \eqref{Bob's ground state}. The first term of the matrix $\mathbb{N}$ can be easily computed using Eq.~\eqref{same detector green delta case}. For the second term we can compute $N_{11}$, $N_{12}$ and $N_{22}$ from Eqs.~\eqref{N11}, \eqref{N12} and \eqref{N22}. Namely, using \eqref{noisekernelAA}, \eqref{noisekernelBB} and \eqref{noisekernelAB}, we get
\begin{equation}\label{N11 delta-like}
    N_{11}=\frac{\lambda_B^2}{8\pi^2\epsilon^2 m_B^2\omega_B^2}\sin^2(\omega_B(t-t_I-d))\,;
\end{equation}
\begin{equation}\label{N12 delta-like}
    N_{12}=\frac{\lambda_B^2}{8\pi^2\epsilon^2 m_B\omega_B}\sin(\omega_B(t-t_I-d))\cos(\omega_B(t-t_I-d))\,;
\end{equation}
\begin{equation}\label{N22 delta like}
    N_{22}=\frac{\lambda_B^2}{8\pi^2\epsilon}\cos^2(\omega_B(t-t_I-d))\,.
\end{equation}
Then, the quantity $W\coloneqq\det\mathbb{N}$ can be computed exactly as
\begin{equation}\label{noise delta like}
    W=\frac{1}{4}+\frac{\lambda_B^2}{16\pi^2\epsilon^2\omega_Bm_B}\,.
\end{equation}
The noise \eqref{noise delta like} corresponds to the upper bound of the noise \eqref{noise upper bound} achieved in the protocol described in Sec.~\ref{sec: specific protocol}.

There are some caveats in the expression \eqref{tau delta like} for the transmissivity $\tau$. In fact, Alice could decrease the energy gap of her detector $\omega_A$ arbitrarily so that the transmissivity of the channel becomes, in the limit $\omega_A\to0$,
\begin{equation}\label{transmissivity low energy gap}
    \tau\sim\frac{\lambda_A^2\lambda_B^2}{16\pi^4m_Am_B}\frac{d^2}{\epsilon^2(d^2+\epsilon^2)^2}(t_I-s)^2\,.
\end{equation}
In this case, the transmissivity of the channel is proportional to $t_I-s$, i.e. the time Alice waits after the preparation of the state to interact with the field. This proportionality is similar to the one we obtain in case Alice is Rindler accelerated and Bob is static, as studied in Sec.~\ref{ssec: accelerating detectors}. However, in that context, we showed how, in Bob's frame, Alice's state changes from the time $s$ to the time $t_I$, undergoing an amplification (see Sec.~\eqref{sssec: transmissivity accelerating}). Instead, here the evolution of Alice's covariance matrix from $s$ to $t_I$ is provided by Eq.~\eqref{Alice's state evolution from Bob} where $N_{AA}=0$ and $T_{AA}$ is given by Eq.~\eqref{Alice to Alice transformation static case}. Therefore, we see that $T_{AA}$ is a symplectic matrix, so that Alice's substate evolves with a unitary transformation from $s$ to $t_I$.

As a consequence, the fact that the transmissivity of the protocol in the static case depends on $t_I-s$ is not expected, because effectively Alice's state does not change from the time $s$ to the time $t_I$. Moreover, the expression \eqref{transmissivity low energy gap} for the transmissivity would allow $\tau$ to be arbitrarily high by increasing $t_I-s$, eventually reaching a situation where the quantum capacity becomes greater than zero. However, this violates the no-cloning theorem, because of the isotropy of the spacetime considered. As explained in detail in Ref.~\cite{Jonsson_2018}, if Alice is able to communicate a quantum message to Bob, Alice would be able to communicate a copy of this quantum message to every third detector whose distance from Alice's is $d$. 

As we proved in Sec.~\ref{ssec: static detectors}, all these problems are solved if we consider the uncertainty principle on the time $t_I$. Indeed, the uncertainty on $t_I$ imposes a natural cutoff on Alice's energy gap so that $\omega_A\ll1/d$ and the approximation leading to Eq.~\eqref{transmissivity low energy gap} is prevented. Moreover, in this case, the dependence of $\tau$ on $t_I-s$ disappears since $t_I$ becomes as a random variable with uncertainty $\sim1/\omega_A$.

Concluding, this appendix shows that, despite its simplicity, the delta-like interaction between field and detectors should be taken with caution. Namely, this kind of interaction could be considered as an approximating case valid whenever the period of interaction is very small with respect to the period of time needed for the detectors to be causally connected (i.e. $d$ in the static case). However, this approximation should not overcome the physical limits imposed by causality or by the uncertainty principle, to prevent unphysical results.

\section{Analysis of the complete positivity of the channel}\label{appendix: CP map}
In Sec.~\ref{sssec: quantum capacity accelerating}, we studied the quantum capacity of the protocol described in Fig.~\ref{fig:accelerating protocol}, when Alice undergoes a Rindler acceleration and Bob is static. In this context, we have seen how, if $|s|>|s^\star|$, then the average number of noisy particles \eqref{additive noise accelerating case} is negative. From the discussion at the end of Sec.~\ref{ssec: canonical form}, this means that the map describing the channel does not satisfy the complete positivity condition. Thus, when $|s|>|s^\star|$ it is no more true that each Gaussian input $\sigma_{in}$ is mapped into a valid, observable Gaussian output $\sigma_{out}$.

By going back to the general case in Sec.~\ref{sec communication protocol}, looking at the map \eqref{quantum channel mapping} together with Eq.~\eqref{output in terms of the input}, there is no way to guarantee that this map is always complete positive. However, from Eq.~\eqref{vectorial form of langevin equation} it is clear that the evolution of the moment operator $\hat{q}_i$ of the detectors (and of its canonically conjugate $\hat{p}_i$) is linear. As a consequence, each input Gaussian state is always mapped to an output Gaussian state, so that the channel is always Gaussian. Then, the eventual lack of complete positivity is not caused by a disrupting of the Gaussian form of the output. Instead, we show that this problem originates from the coordinate transformation of Alice's input state from Alice's frame to Bob's frame, which is in general a non-CP map.

The map $\mathcal{N}$ described by Eq.~\eqref{quantum channel mapping} could be rewritten as the composition between three different maps, namely $\mathcal{N}_1$, $\mathcal{N}_2$ and $\mathcal{N}_3$, where:
\begin{itemize}
    \item $\mathcal{N}_1$ maps the input state in Alice's frame into the input state in Bob's frame, called $\sigma_{AA}^B$. Namely:
    \begin{equation}\label{change of coordinates}
        \mathcal{N}_1:\sigma_{in}(s)\mapsto\sigma_{AA}^B(s)=\text{diag}(1,\dot{t}_A)\sigma_{in}(s)\text{diag}(1,\dot{t}_A)\,.
    \end{equation}
    \item $\mathcal{N}_2$ represents the time evolution of Alice's state in Bob's frame from $s$ to $t_I$, i.e.
    \begin{equation}\label{channel N2}
        \mathcal{N}_2:\sigma_{AA}^B(s)\mapsto \sigma_{AA}^B(t_I)\,.
    \end{equation}
    The evolution of Alice's state specified in Eq.~\eqref{Alice's state evolution from Bob} is the composite map $\mathcal{N}_2\circ\mathcal{N}_1:\sigma_{in}\to T_{AA}\sigma_{in}T_{AA}^T$\,.
    \item The map $\mathcal{N}_3:\sigma_{AA}^B(t_I)\mapsto\sigma_{out}$ maps Alice's state in Bob's frame at the time $t_I$ to Bob's state after a defined time. This is the channel occurring when Alice and Bob communicate. 
\end{itemize}
Each one of these maps $\mathcal{N}_i$ could be written as Eq.~\eqref{quantum channel mapping} with matrices $\mathbb{T}_i$ and $\mathbb{N}_i$. Then, $\mathcal{N}_i$ can be characterized by the parameters $\tau_i=\det\mathbb{T}_i$ and $W=\det\mathbb{N}_i$. However, the map $\mathcal{N}$ is a valid one-mode Gaussian channel if, given their $\tau_i$ and $W_i$, the relative $\overline{n}$ from Eq.~\eqref{additivenoise}, called $\overline{n}_i$, is positive. In the protocols studied in Sec.~\eqref{sec: specific protocol}, the map $\mathcal{N}_3$ always satisfy this property, because $W_3\ge1/4$ and $\tau_3\ll1$. Then, we have $\overline{n}_3\ge0$ as long as $\tau_3\le2$.

However, for the maps $\mathcal{N}_1$ and $\mathcal{N}_2$, we have $W_1=W_2=0$, implying $\overline{n}_1=\overline{n}_2=-1/2$ and making them non-complete positive maps. The problems that may arise in the complete channel $\mathcal{N}=\mathcal{N}_3\circ\mathcal{N}_2\circ\mathcal{N}_1$ are then caused by the applications of $\mathcal{N}_1$ and $\mathcal{N}_2$.

In particular, we now see that the main problem resides in $\mathcal{N}_1$ and we show that, by solving it, we automatically remove the possibility to have $\overline{n}<0$.

This analysis requires further tools on Bosonic Gaussian states that we are going to introduce now. Namely, the covariance matrix representing a one-mode Gaussian state is defined in Eq.~\eqref{one-mode Gaussian state} when $i=j$. The canonical variables $\hat{q}_i$ and $\hat{p}_i$ defining the covariance matrix $\sigma_{ii}$ must satisfy the uncertainty principle, consequence of the algebra described in Eq.~\eqref{algebra}. Mathematically speaking, the covariance matrix $\sigma_{ii}$ must satisfy
\begin{equation}\label{uncertainty principle covariance matrix}
    \sigma_{ii}+\frac{1}{2}\left(\begin{matrix}
        0&\rm{i}\\-\rm{i}&0
    \end{matrix}\right)\ge0
\end{equation}
In general, following the commutation relation \eqref{algebra}, a covariance matrix $\sigma_{ii}$ representing a one-mode Gaussian state can be written as 
\begin{equation}\label{covariance matrix n and m}
    \sigma_{ii}=\left(\begin{matrix}
        \frac{1}{2}+n_i+\Re m_i&\Im m_i\\\Im m_i&\frac{1}{2}+n_i-\Re m_i
    \end{matrix}\right)\,,
\end{equation}
where $n_i\coloneqq\langle a^\dagger_ia_i\rangle$, i.e., the average number of particles in the mode $i$, and $m_i\coloneqq\langle a_ia_i\rangle$. To satisfy Eq.~\eqref{uncertainty principle covariance matrix}, one needs $|m_i|\le n_i+n_i^2$. The entropy of the one-mode Gaussian state represented by $\sigma_{ii}$ is given by $h(\sqrt{\det\sigma_{ii}})$, where $h$ is defined in Eq.~\eqref{entropy function}. Since $h$ is not defined when its argument is less than $1/2$, we have $\det\sigma_{ii}\ge1/4$. This condition is equivalent to \eqref{uncertainty principle covariance matrix}.

Looking at the action of the channel $\mathcal{N}_1$ from Eq.~\eqref{change of coordinates}, it is clear that, starting with an input $\sigma_{in}$ whose determinant is greater than $1/4$, the output of $\mathcal{N}_1$ does not always satisfy Eq.~\eqref{uncertainty principle covariance matrix}. For example, starting with the unsqueezed vacuum state $\sigma_{in}=\frac{1}{2}\mathbb{I}$, using Eq.~\eqref{change of coordinates}, we have 
\begin{equation}
    \mathcal{N}_1\left(\frac{1}{2}\mathbb{I}\right)=\left(\begin{matrix}
        \frac{1}{2}&0\\0&\frac{\dot{t}_A^2}{2}
    \end{matrix}\right)\,,
\end{equation}
whose determinant is $\frac{\dot{t}_A^2}{4}<\frac{1}{4}$. In this case, the output of the channel is not an observable state. In particular, the input of the channel $\mathcal{N}_1$ must satisfy certain conditions to have an observable output. Namely, by applying the channel $\mathcal{N}_1$ to the general input state \eqref{covariance matrix n and m}, which we call $\sigma_{in}$, we need
\begin{equation}\label{condition for observability}
    \det(\mathcal{N}_1(\sigma_{in}))=\dot{t}_A^2\left(\frac{1}{4}+n_i+n_i^2-|m_i|^2\right)\ge\frac{1}{4}\,.
\end{equation}
A generic covariance matrix $\sigma_{in}$ satisfying the condition \eqref{condition for observability} can be decomposed as $\sigma_{in}=\sigma_{in}'+\mathbb{N}_0$, where $\sigma_{in}'$ could be a whatever one-mode Gaussian state and $\mathbb{N}_0$ is a matrix whose determinant, to ensure that Eq.~\eqref{condition for observability} is satisfied, must be
\begin{equation}\label{condition general for N0}
    \sqrt{\det\mathbb{N}_0}\ge\frac{1}{2}\left(\frac{1}{\dot{t}_A}-1\right)\,.
\end{equation}
At this point, the channel $\mathcal{N}_1$ can be considered a complete positive one-mode Gaussian channel where the input is $\sigma_{in}'$ and the matrix $\mathbb{N}_0$ plays the role of an additive noise, namely
\begin{equation}\label{map decomposing}
    \mathcal{N}_1(\sigma_{in})=\text{diag}(1,\dot{t}_A)\sigma_{in}'\text{diag}(1,\dot{t}_A)+\text{diag}(1,\dot{t}_A)\mathbb{N}_0\text{diag}(1,\dot{t}_A)\,.
\end{equation}
By reducing the channel \eqref{map decomposing} to its canonical form (see Sec.~\ref{ssec: canonical form}), we now have a transmissivity $\tau_1=\dot{t}_A$ and a noise $W_1=\dot{t}_A^2\det(\mathbb{N}_0)$. Taking the minimum $\det\mathbb{N}_0$ possible, from Eq.~\eqref{condition general for N0}, we have $W_0=\frac{1}{4}(1-\dot{t}_A)^2$. producing a number of noisy particles $\overline{n}_1=0$. Then, if we apply the channel $\mathcal{N}_2$ to the r.h.s. of Eq.~\eqref{map decomposing}, we end up again with an output with a null number of noisy particles. Since $\mathcal{N}_3$ was always recognized as a one-mode Gaussian channel, we can conclude that the lack of complete positiveness of the channel $\mathcal{N}$ is due by the channel $\mathcal{N}_1$. Hence, by ensuring $\mathcal{N}_1$ is complete positive by taking input states satisfying Eq.~\eqref{condition for observability}, the channel $\mathcal{N}$ is always complete positive as well.

\end{document}